# The State of AI Ethics
## October 2020

MAIEI

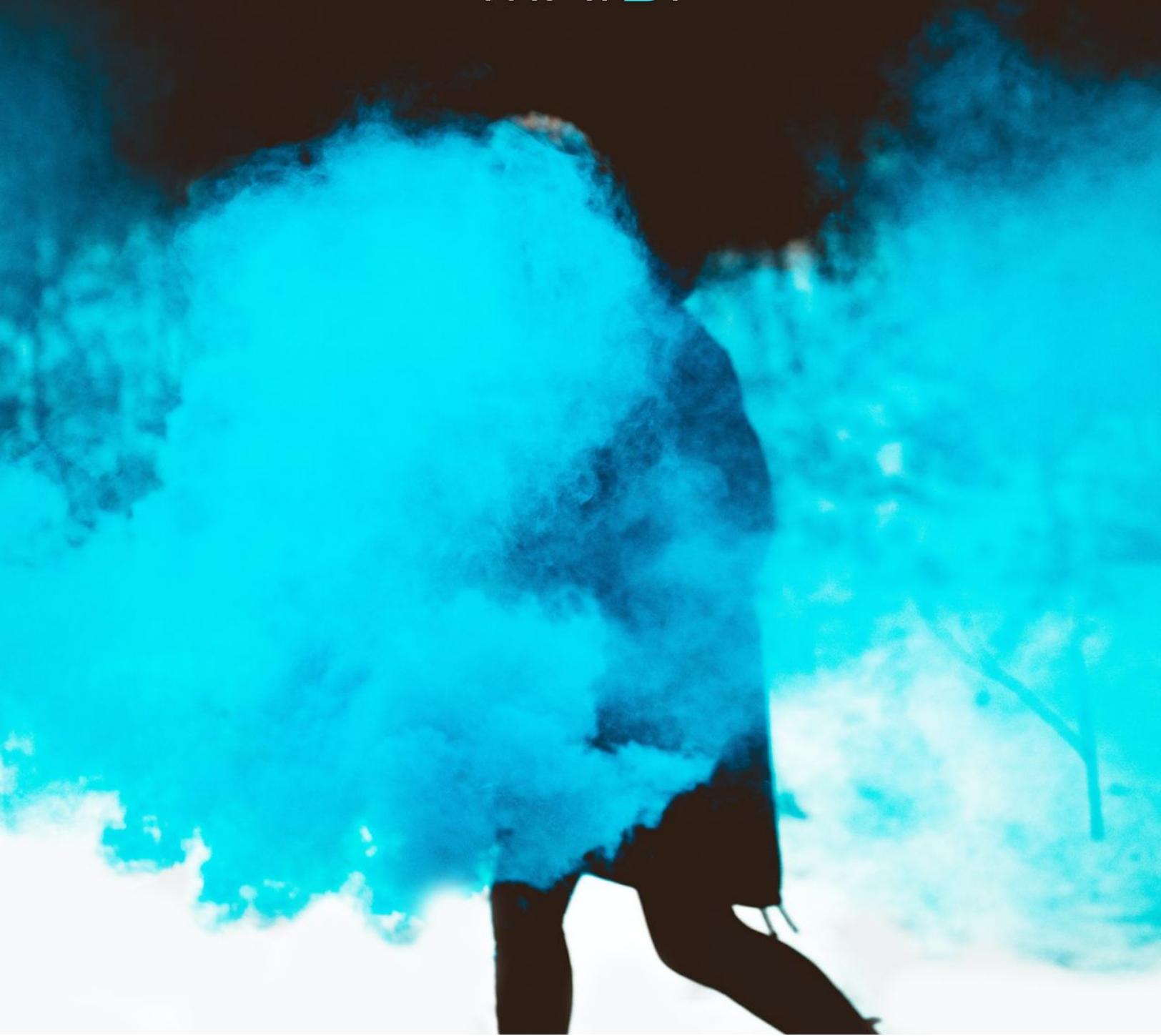

This report was prepared by the **Montreal AI Ethics Institute** (MAIEI) — an international non-profit research institute helping people understand the societal impacts of AI and equipping them to take action. **Learn more on our website or subscribe to our newsletter.**

This work is licensed under a Creative Commons Attribution 4.0 International License.

Primary contact for the report: **Abhishek Gupta (abhishek@montrealethics.ai)**

Full team behind the report:

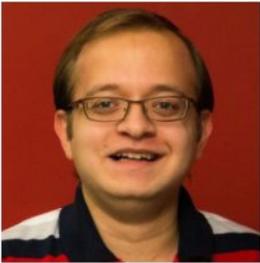
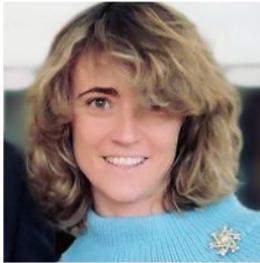
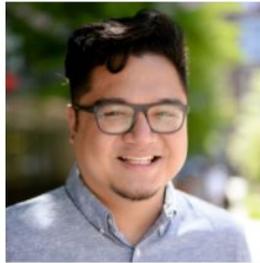
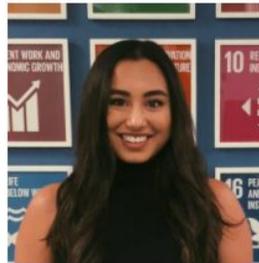
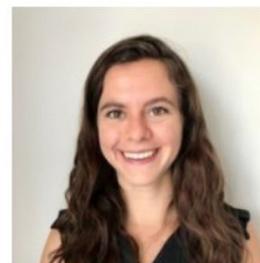

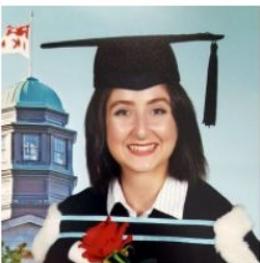
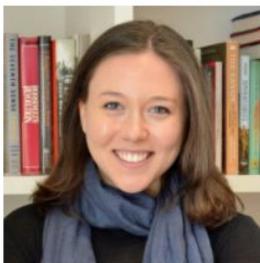
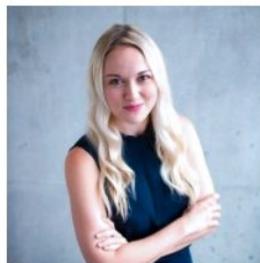
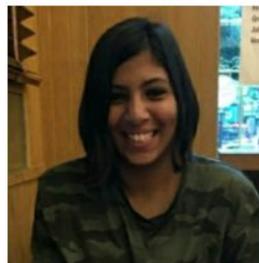
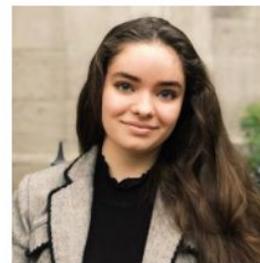

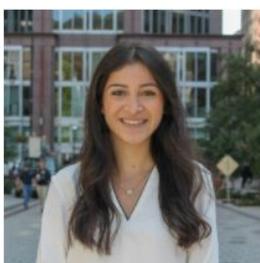
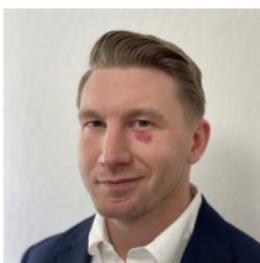
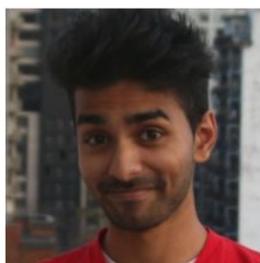
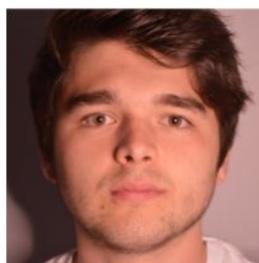
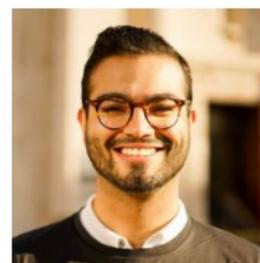

Special thanks to: Khaulat Ayomide, Brooke Criswell, Samuel Curtis, Pablo Nazé, Ameen Jauhar.

# Table of Contents

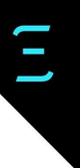

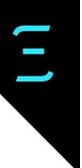

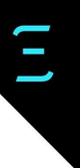

*Note that the original sources are linked under the title of each piece. The work in the following pages combines summarization of the material supplemented with insights from the research staff.

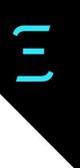

# Foreword by Danit Gal (Technology Advisor, The United Nations)

I'll start by congratulating the MAIEI team for yet another thorough overview of developments in AI ethics. These reports are an incredibly valuable resource for which I am personally and professionally grateful. Swimming in a seemingly endless ocean of AI ethics reports and overviews, three things set MAIEI's reports apart:

**1) They cut through the hype.**

We have reached a pivotal moment where most people, from world leaders to private sector CEOs and AI trailblazers, recognize that ethical considerations are inseparable from the development, use, and regulation of AI. This is a pivotal moment: AI ethics now centers many important conversations and actions. But this progress notwithstanding, much of the attention (often superficial) that this field receives results in "ethics washing" at best, and "ethics dumping" at worst. The MAIEI reports cut through the hype. They consistently go beyond merely writing about what others have to say on the topic, and instead ask what is actually being done.

**2) They focus on implementation.**

In the same vein, MAIEI's reports focus on substance across a wide variety of developments and implementation pathways. With over 180 available sets of (largely repetitive) AI ethics principles, we have clear consensus on some key principles. How we move forward with actually implementing these commonly-supported principles, however, is a different story. Today, many of my conversations are ripe with stories of entities who designed their policies and guidelines around these principles but don't have a clear understanding of what more they should and could do with them. This is where critical discussions on the many attempted implementations and applications of these principles become the fuel for meaningful progress.

**3) They are open-access and freely available.**

Open-access has become rare at a time when AI ethics is increasingly being sold as a service, often with little regard for how these technologies actually work, or for the consequences when they don't. Aside from helping readers sort through the constant stream of news and publications, MAIEI's reports lower the AI ethics literacy threshold. Their deep-dives, summaries, and analyses serve as an accessible

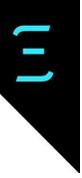

and engaging means to interact with this field and its exciting and consequential developments.

MAIEI's reports, however, also share a glaring challenge with most AI ethics initiatives: they are mostly Western-focused and facing. This is a bias of which we are all keenly aware, but not very well equipped to address. This is not a novel observation, but the problem will haunt and taunt us as long as we fail to resolve it. Because while many of us have moved from principles to action, many more are still struggling to understand what AI ethics mean. This gap is fueled by digital divides, colonial technology exports, and closed-door conversations and collaborations. The gap may appear negligible to some, but AI ethics washing and dumping will come back to bite us all with a vengeance.

This globally-shared challenge is at the heart of the United Nations Secretary-General's call for the establishment of a multi-stakeholder advisory body on global AI cooperation, in his Roadmap for Digital Cooperation. This is a mission I've been especially privileged to help shape and support, and one that enjoys the participation of MAIEI, alongside an increasingly globally-diverse assembly of member states, companies, universities, AI initiatives, international and civil society organizations, and UN entities leading global AI work like ITU, UNESCO, and UNICEF.

This is not a challenge one entity can solve alone. Our ability to achieve the goal of true global participation in developing AI that benefits the whole of humanity and leaves no one behind, depends on our capacity for cooperation. For that reason, I'm especially grateful for MAIEI's reports, which help us build common interest and understanding. I hope you find this series as helpful as I have. If you do, please consider joining us at the United Nations and MAIEI in creating a diverse, inclusive, and informed global AI cooperation network.

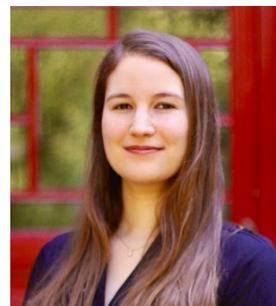

My very best wishes,
Danit Gal (**@DanitGal**)
Technology Advisor, The United Nations

Danit Gal is Technology Advisor at the United Nations, leading work on AI in the implementation of the United Nations Secretary-General's Roadmap for Digital Cooperation. Danit serves as the former chair and vice chair of the P7009 IEEE standard on the Fail-Safe Design of Autonomous and Semi-Autonomous Systems, and the executive committee of The IEEE Global Initiative on Ethics of Autonomous and Intelligent Systems. She is an Associate Fellow at the Leverhulme Centre for the Future of Intelligence at the University of Cambridge, an Affiliate at the Center for Information Technology Policy at Princeton University.

# Introduction

Since the publication of the [previous iteration of this report](#), the world has both accelerated and proceeded at a glacial pace. Accelerated in the sense of having ever more misuses of technology, violating human rights, subverting privacy, spreading disinformation, sowing discontent, and praying on our vulnerabilities to deepen and widen chasms in a world that is fragmenting every second every day. Yet, even with the realizations that we have all these problems, it seems that the front of solution-development has been glacial and we are spinning around in circles trying to resurface problems that we know too well. Discussions have also had a very deep emphasis and focus on the Western world and the problems that it faces. [Noticeably, the solutions have also then been Western-centric](#).

But, there is a glimmer of hope: a rising tide of demands to move to solutions and a realization that we might not be able to solve all the problems all at once but that shouldn't dissuade us from trying.

I realize that the field has so much information coming out on a weekly basis across a variety of subfields in the domain of AI ethics that it's nearly impossible to keep up. This report is our team's humble attempt at curating what we found to be the most insightful. It will share some items that you might have come across yourself, others that might have slipped past in the deluge of information online. While it is not a comprehensive overview by any means, we have spent many days working hard to assimilate and present information and highlights in this report linking them in a manner that will help you move past the obvious within each of the subfields of AI ethics.

These are challenging times certainly for all of us, and my hope is that you find this report an opportunity to reflect critically on the developments in this domain over the past quarter and bring a more nuanced conversation to your communities in discussions about AI.

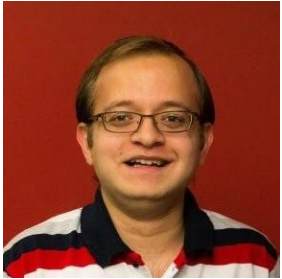

Abhishek Gupta ([**@atg_abhishek**](#))
Founder & Principal Researcher, Montreal AI Ethics Institute

Abhishek Gupta is the founder and principal researcher at the Montreal AI Ethics Institute, seeking to define humanity's place in a world increasingly characterized and driven by algorithms. He is also a machine learning engineer at Microsoft, where he serves on the CSE Responsible AI Board. His book '[Actionable AI Ethics](#)' will be published by Manning in 2021.

# 1. AI & Society

**Opening Remarks** by Adam Murray (U.S. Diplomat, and Chair of the OECD Network of Experts on AI)

The potential of Artificial Intelligence is truly transformative.  In the midst of the COVID-19 pandemic, we are seeing how AI can speed the detection and diagnosis of disease and the development of treatments, including a vaccine.  AI promises to deliver better educational outcomes, smarter and more sustainable manufacturing, new ways of connecting and communicating with others, and much more.  It seems like not a day goes by without the announcement of another innovative product or service that will change the way we live.  Yet, for AI to fulfill that potential, our societies must believe that it is trustworthy, safe, and working for their benefit.

Two years ago, the Organization for Economic Cooperation and Development began a multi-stakeholder process to develop a set of principles for the responsible stewardship of trustworthy AI.  Adopted in 2019 by the 36 OECD member governments and eight non-members, the Principles represent the first intergovernmental agreement on a vision for AI in society.  Now, the OECD's Network of Experts on AI is helping move from principle to practice by identifying use cases and tools that demonstrate responsible AI in action.

The chapter before you shows how all of us have a role in building trust in AI.  On the research and design front, we read about some great examples of efforts to encode human norms and values into AI models and algorithms.  Can we push the technical boundaries to peer inside the "black box" and make AI systems more transparent, robust, and trustworthy?  Could a "differential privacy" approach deliver the same outcomes without exposing sensitive information in datasets?  At the same time, the "wicked" problems vexing society today may not always have a "silver bullet" technical solution.  Policymakers need to be knowledgeable of AI – its potential and its limits – and adopt the right technical and policy approaches to the problem they seek to address.

We are also reminded of the importance of broadening our aperture to consider how other disciplines may have dealt with similar problems.  Humans have been keeping records since antiquity, for example, and some good practices from archive management may help make data collection for AI/ML more accurate and representative.  It is a useful illustration of the benefits of a multidisciplinary and multi-stakeholder approach to the design, development, and deployment of AI systems.

Our chapter ends with a study of why it is essential that we get this right.  Sadly, we already have examples of how some bad actors are using AI systems to trample on human rights.  But responsible actors can also use AI technologies to identify human rights abuses, helping to hold those bad actors to account and hopefully even preventing those abuses from happening in the first place.

Bringing about the full potential of AI is up to all of us.  We must continue to innovate, invest in research and development, and encourage the adoption of AI in our societies.  We must also ensure that AI systems are used in ways that reflect our core values and respects human rights and civil liberties.  In doing so, we can make sure that AI remains a force for positive change in our lives.

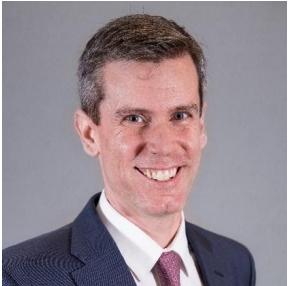

Adam Murray
U.S. diplomat working on technology policy,
Chair of the OECD Network of Experts on AI

*[The views expressed are his own and not necessarily those of the U.S. government]*

Adam Murray is a career U.S. diplomat in the Division of International Communications and Information Policy at the Department of State, where he covers digital economy policy and emerging technologies. He represents the United States at the Organization for Economic Cooperation and Development (OECD) Committee on Digital Economy Policy and the Asia Pacific Economic Cooperation (APEC) Telecommunications and Information Working Group.  He chairs the OECD Network of Experts on AI, which is developing implementation guidance for the OECD Recommendation on AI.

# Go Deep: Research Summaries

## What Does It Mean for ML to Be Trustworthy?
([Original presentation](#) by Nicolas Papernot)

This video provides Nicolas Papernot's presentation of his involvement in a project on making ML more trustworthy. This included appeals to LP norms, differential privacy, admission control at test time, model governance, and deepfakes. These mentions will be dealt with as sections of the summary, ending with Papernot's conclusions from his research.

**Trustworthy ML:**

In order to determine how to make ML more trustworthy, the group needed to determine what that would look like. To do this, they sought to include how an ML model could be robust against the threat of adversarial examples. Here, LP norms were utilized, placing the model as a constant predictor inside an LP ball, and thus making it less sensitive to perturbations. The LP norms then allowed for a new way of detecting adverse examples to be proposed. They were able to be detected more clearly through their excessive exploitation of the excessive invariance that the LP norms had coded into the model. Hence, the question is asked whether such a method can be used to better detect threats through this resultant excessive exploitation, and then help train models to be robust against such threats in the future.

This is answered through framing the question in terms of an AI arms race. Traditional computer systems have treated cybersecurity in terms of 'locking up a house.' They 'lock the door' in order to prevent intruders, and weigh this up against additionally 'locking the window' in case a bear was to break through. If this were to happen, the windows would then be locked and weighed up against the possibility of a hawk descending through the chimney. If this were to happen, then the chimney would be locked etc. creating an AI arms race to try and defend against these threats. Instead, using LP norms to increase the robustness of the model to better detect these intruders, bears and hawks could be a way forward that saves time, money, and increases trust in ML.

**Privacy:**

One way to increase the trust using this model is in terms of privacy. ML lives on data, and thus it is sometimes at risk of the data subject wanting to privatize such data, even once it has helped train the model. Thus, the group leant towards a definition of privacy as "differential privacy" through their Private Aggregation of Teacher Ensembles (PATE) method. Here, instead of training the model on a whole data set, the group split the data up and assigned "teachers" to each data partition. These partitions aren't related, so each teacher is trained independently to solve the same task of the model, whereby the group can aggregate the predictions made by the model, and then add some noise to the data in order to make the predictions more private. Hence, the data is only included in one of the data sets and only influences one of the teachers, meaning that if there's a prediction being made, your data is very unlikely to influence said prediction as you will only impact one of the predictions made. Resultantly, the data can be privatised more effectively through having less of an impact on the output, helping to align ML's version of privacy to the human norms that fall under the same term.

The video then splits into two main topic areas: admission control at test time, and model governance, which I shall now introduce in turn.

**Admission control at test time:**

One way of abstaining from making a prediction within a model to help aid admission control is to promote a measure of uncertainty within the model, which the group linked back to the training data. The group then implemented the method of Deep K-Nearest Neighbours, which allowed them to open up the "black box" of the model and see what was occurring in each layer (focusing on individual problems, rather than the model itself). Here, they looked at how each layer represents their test input, and in each of the layers' representation space, they performed a nearest neighbour search. This is done until each layer is completed, whereby the nearest neighbour search will reveal how the labels of the nearest neighbour are inconsistent, and thus point to the problem.

One question then arises as to whether there's a possibility of defining objectives at the level of each layer to make models more amenable to test-time predictions. The group looked at this question in their use of 'soft nearest neighbour loss' in 2019's ICML conference. They asked whether it was better for the model to learn representation spaces that separate the data from classes with a large margin (like a support vector machine), or whether it's better to entangle different classes together within the layers of representations.

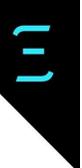

They found that the latter was better for the Deep K-Nearest Neighbours method to estimate the uncertainty of the model. So, they then introduced the 'soft nearest-neighbour loss' into the model to encourage it to entangle points from different classes in the layers of representations. The soft loss method will then encourage the model to co-opt features between different classes and the lower layers (which are able to be recognized using the same lower level features). This then helped them identify uncertainty when they have a test point that doesn't fall in any of the clusters, meaning they would've had to have guessed whom their nearest neighbours are. Instead, there's now support when doing these searches as the different classes are subsequently entangled.

**Model governance:**

The group explored the problem of how to return data to a data subject that no longer wants their data to be utilized, despite it already being used to train the model, requiring a form of "machine-unlearning". The question then becomes: is the differential privacy proposed enough to prevent such time-consuming machine-unlearning from occurring? As in, can the data being subjected to only affecting the outcome of one teacher be enough to satisfy what machine-unlearning would achieve?

The group decided that this probably wouldn't be the case. The model's data points (like in stochastic gradient descent) would still be influenced by the initial data point of the data subject, thus making it hard to remove their data entirely. Hence, Shared Isolated Sliced Aggregate training is then explored. Here, the method involves splitting the model into shards, and those shards into slices, meaning that the data point will only be in one shard and in one slice of that shard, whereby only one shard will need to be retrained rather than the whole model. Such retraining will then be quicker, for each shard completion contains a checkpoint before moving onto the next shard, proving a launching pad for retraining the model.

**Deepfakes:**

The notion of ML's role in deepfakes is then considered, with progress in ML accelerating the progress in digital alteration. The group considered three approaches as to how to combat this:

- Detect artifacts within the altered image (such as detecting imperfections, like imperfect body movements).

- Reveal content provenance (secure record of all entities and systems that manipulate a particular piece of content).

- Advocate for a notion of total accountability (record every minute of your life).

- The group believed that none of these three methods would achieve total coverage of all the problematic areas of deepfakes, so the supplementation of policy on areas such as predictive policing and feedback loops is required.

**Conclusion:**

The group concluded that research is needed in order to align ML with human norms. Once this is done, trustworthy ML is an opportunity to make ML better, and a cause that provides much food for thought for the future.

## Roles for Computing in Social Change
([Original paper](#) by Rediet Abebe, Solon Barocas, Jon Kleinberg, Karen Levy, Manish Raghavan, David G. Robinson)

With the impact that technology has on altering societal dynamics, it has led to questions on what role computing should play in social change. In particular, what degree of attention we should pay to use technology as a lever compared to addressing the underlying issues at a social level.

A lot of technical work in incorporating definitions of fairness, bias, discrimination, privacy, and other endeavours have met with concerns if the way this work happens is the best use of our efforts in addressing these challenges. The authors of this paper raise the point that while technology isn't a silver bullet in solving problems of social injustice in society, they can still serve as a useful lever. Technical methods can surface evidence and act as a diagnostic tool to better address social issues.

As an example, such methods have showcased the pervasiveness, depth, and scope of social problems like bias in society against minorities. By quantifying the extent of the problem, it provides us with a prioritized roadmap in addressing issues where there is the most significant impact first. The framing by the authors that these methods don't absolve practitioners of their responsibilities; instead, it offers them the requisite diagnostic abilities to begin addressing the challenges is one that needs to be adopted industry-wide.

Science and technology studies (STS) already have existing tools to interrogate sociotechnical phenomena, and the authors advocate the use of a blend of computation and STS methods together to come up with holistic solutions. Often the problem with diagnostic results is that the results from those exercises become targets themselves. The authors caution practitioners to craft narratives around those results that drive results and action rather than focussing solely on the results and fall prey to Goodhart's Law.

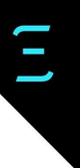

Computation is a formalization mechanism that has the potential to posit clearly what the expectations of actors are while moving away from abstract and vague delegations where actors might apply disparate standards to address social challenges within their systems. Since there is an explicit statement of inputs, outputs, and the rules associated with the system as a part of the formalization process, it presents an opportunity to lay bare the stakes of that system from a social perspective. More importantly, it offers the stakeholders the opportunity to scrutinize and contest the design of the system if it is unjust.

Advocacy work that calls forth participation from those who are affected usually focuses on how rules are made but not what the rules actually are. The formalized, computational approach offers an intermediate step between high-level calls for transparency and concrete action in the design and development of the systems. This isn't without challenges. In situations where there is limited data available, the practitioners are constrained in utilizing what they have at hand which might pose harder to solve difficulties. Yet, this might also be the opportunity to call for the collection of alternate or more representative data so that the problem becomes more tractable and solves the issue at hand.

This discussion can also serve to highlight the limitations of technical methods and the subsequent dependence and limitations of the policies premised on the outputs from these systems. Specifically, a critical examination can drive policymakers to reflect on their decisions such that they don't exacerbate social issues. While computational methods can highlight how to better use limited resources, it can also shed light on non-computational interventions that repurpose existing resources or advocate for higher resource allocation in the first place rather than a mere optimization of current, limited resources.

A risk with this approach that the authors identify is that this can shift the discussion away from policy fundamentals to purely technology focussed discussions that look at how to improve technology to better address the problem rather than changing fundamental dynamics in society to address the root cause of the problem. As an example, the alternative to an ill-conceived algorithm might not be a better algorithm, but perhaps no algorithm at all. Computational approaches can act as a synecdoche bringing social issues forward in a new light. For example, the authors point to how inequities in society have gained a boost from the attention that framing some of those challenges from a technological perspective has brought with it.

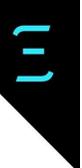

With short attention spans and the inherent multivalent nature of large societal problems, traditional policy-making takes the approach of chipping away gradually from different angles. A technological spotlight brings in more actors who now attack the problem for several dimensions, lead to greater resource allocation and potentially quicker mitigation. The synecdochal approach treads a fine line between overemphasis and shining a new light. Current society's obsession with technology can be utilized in a positive manner to drive concrete action on beginning to address the fundamental challenges we face in creating a more just society for all.

## Lessons from Archives: Strategies for Collecting Sociocultural Data in Machine Learning
([Original paper](#) by Eun Seo Jo, Timnit Gebru)

It's no secret that there are significant issues related to the collection and annotation of data in machine learning (ML). Many of the ethical issues that are discussed today in ML systems result from the lack of best practices and guidelines for the collection and use of data to train these systems. For example, Professor Eun Seo Jo (Stanford University) and Timnit Gebru (Google) write, "Haphazardly categorizing people in the data used to train ML models can harm vulnerable groups and propagate societal biases."

In this article, Seo Jo and Gebru set out to examine how ML can apply the methodologies for data collection and annotation utilized for decades by archives: the "oldest human attempt to gather sociocultural data." They argue that ML should create an "interdisciplinary subfield" focused on "data gathering, sharing, annotation, ethics monitoring, and record-keeping processes." In particular, they explore how archives have worked to resolve issues in data collection related to consent, power, inclusivity, transparency, and ethics & privacy—and how these lessons can be applied to ML, specifically to subfields that use large, unstructured datasets (e.g. Natural Language Processing and Computer Vision).

The authors argue that ML should adapt what archives have implemented in their data collection work, including an institutional mission statement, full-time curators, codes of conduct/ethics, standardized forms of documentation, community-based activism, and data consortia for sharing data. These implementations follow decades of research and work done by archives to address "issues of concern in sociocultural material collection."

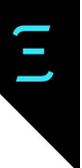

There are important differences to note between archival and ML datasets, including the level of intervention and supervision. In general, data collection in ML is done without "following a rigorous procedure or set of guidelines," and is often done without critiquing the origins of data, as well as the motivations behind the collection, and the potential impacts on society. Archives, on the other hand, are heavily supervised and have several layers of intervention that help archivists determine whether certain documents or sources should be added to a collection. Seo Jo and Gebru point out another important difference between ML and archival datasets: their motivations and objectives. For the most part, ML datasets are built to further train a system to make it more accurate, while archival datasets are built to preserve cultural heritage and educate society, with particular attention to "authenticity, privacy, inclusivity, and rarity of sources."

The authors argue that there should be a more interventionist approach to data collection in ML, similar to what is done by archives. This is due to the fact that from the very beginning, historical bias and representational bias infect data. Historical bias refers to the "structural, empirical inequities inherent to society that is reflected in the data," and representational bias comes from the "divergence between the true distribution and digitized input space." The best way to mitigate these biases is to implement what archives have put into place in their data collection practices, which includes:

1. Drafting an institutional mission statement that prioritizes "fair representation or diversity" rather than "tasks or convenience." This will prevent collection methods or even research questions from being driven solely by the accessibility and availability of datasets, which can replicate bias. It also encourages researchers to publicly explain their collection processes and allows for feedback from the public.

2. Ensuring consent through community and participatory approaches. This is especially crucial for ML researchers who are building datasets based on demographic factors. "ML researchers without sufficient domain knowledge of minority groups," write Seo Jo and Gebru, "frequently miscategorize data, imposing undesirable or even detrimental labels onto groups." Archives have attempted to solve similar issues by creating community archives where collections are built and essentially "owned" by the community being represented. These archives are open to public input and contributions, often enabling minority groups to "consent to and define their own categorization."

3. Creating data consortia to increase "parity in data ownership." Archives, alongside libraries, have created a consortia model through institutional frameworks that allow them to "gain economies of scale" by sharing resources and preventing redundant collections. This model has been adopted by the Open Data Institute, for example, to share data among researchers in ML. However, issues around the links between profit and data may prevent widespread adoption by ML companies and organizations.

4. Encourage transparency by creating appraisal records and committee-based data collection practices. Archives follow rigorous record-keeping standards, including 1) data content standards, 2) data structure standards, and 3) data value standards that pass through several layers of supervision. They also record the process of their data collection to ensure even more transparency. ML should build and maintain similar standards in its data collection practices to address issues emanating from the public (and other researchers) about ML systems.

5. Building overlapping "layers of codes on professional conduct" that guide and enforce decisions regarding ethical concerns. For archives, these codes are maintained and enforced by international groups (e.g. International Council on Archives), and because many archivists are employed as professional data collectors, they are held to specific standards that are enforced by ethics panels or committees. ML could benefit immensely from creating similar mechanisms that ensure accountability, transparency, and ethical responsibility.

Of course, there are limitations to the ML field's ability to adopt the measures outlined above. In particular, the authors argue, the sheer amount of data in ML datasets is much larger than many archives and the resources needed to implement these measures may be beyond what many ML-focused companies, researchers, etc. are willing to commit to. This is especially due to the fact that their motivations are primarily profit-focused. However, the ML community must contend with and end its current, problematic data collection practices—and a "multi-layered" and "multi-person" intervention system informed by systems put into place by archives would be a good place to start.

## Mass Incarceration and the Future of AI
(**[Original *Harvard University* discussion paper](#)** **by Eun Teresa Y. Hodge, Laurin Leonard**)

The integration of AI decision-making in the US criminal justice system and its accompanying biases have sparked numerous controversies. Automated bias in the courts, policing, and law enforcement agencies will impact hundreds of millions of people's lives. In the US, one in three individuals lives with a criminal record. The over-incarceration of the country's population has fueled government agencies' need to amass and track citizens' data from aggregated sources. Arrest and/or conviction records, made readily available for background screening tools under public safety guise, are lifelong shackles that restrict social mobility and access to employment, housing, and educational opportunities. Hodge & Leonard are hoping to start a discussion on the line between an individual's human rights and information that is necessary for society's well-being.

Hodge and Leonard, a mother-daughter duo, speak from personal experience. Hodge served a 78-month sentence in Federal prison before becoming a criminal justice advocate. Drawing on Weber's social theory, they point to the creation of an underclass in American society, where people living with criminal records cannot fully compete in the open labour market and face open stigmatization offline and online. The rise in criminal background checks due to a fear of terrorism, the gig economy's growth, and the need to digitize government records serve to accentuate the divide between this underclass and the rest of American society. Individuals with criminal records are unable to dispute privacy-infringing information spread and purchased online. They are also not given any indication of what types of information will prop up in online searches. With nearly half of the FBI's background checks failing to indicate the outcome of a case after an arrest, individuals with dismissed charges and no convictions face unjust prejudice.

The paper works to stimulate a public debate on access to data, individual rights of privacy and dignity, and setting quality standards for the source of data shared. It touches on building mechanisms to ensure that these standards are being met. The modernization of government databases to online records happened before parameters of access to low-level data and the integrity of source data were put in place. People living with criminal records face around 50 000 known sanctions and restrictions. Unchecked background screening tools built on incomplete and sometimes inaccurate data will only serve to increase that number.

Equality, opportunity, human dignity and respect for privacy are often cited as the core values of American society, yet automated online background checks impede these rights and freedoms. Individuals with criminal records need to have their data rights applied in practice and hold shared ownership over their personal information. Without these protections, background checks will continue to obstruct upward mobility and perpetuate the "life sentence" suffered by individuals with records. Policymakers tackling how data is collected, stored, and shared must include the voices of vulnerable populations in their decision.

## PolicyKit: Building Governance in Online Communities
([Original paper](#) by Amy X. Zhang, Grant Hugh, Michael S. Bernstein)

Online community governance models are often incapable of evolving; if they are to change, the methods to do so are quite labour-intensive. Online communities are defined as gatherings on platforms such as Slack, Reddit subreddits, Facebook groups, and mailing lists. Governance for these online communities includes roles and permissions. A role and permission based governance model dictates who can join the group and guides decisions about the type of content that can be broadcasted. This model of roles and permission is based on UNIX file permission models on almost all big community platforms. However, these governance models are not flexible to alternative forms of governance. Because of the lack of flexibility, communities are forced to solve their problems with moderated communities, which can lead to moderators facing burn out, new communities' members being overwhelmed, and reduced legitimacy. This exact problem is what the software "PolicyKit" solves.

The goal of PolicyKit is for online communities to develop and deploy their governance models that include flexible governance policy code that is inspired, borrowed, or altered from other online communities. This can be accomplished form the two main functions of PolicyKit's infrastructure:

1. Software library for users to learn how to write their policies in code.

2. A server that can process and execute policies "against actions" within the community.

3. Integration of PolicyKit on the platform to know when actions have been performed in the past.

4. And finally, access to a website where community members can propose new actions to change the governance model and create new policies in the code editor.

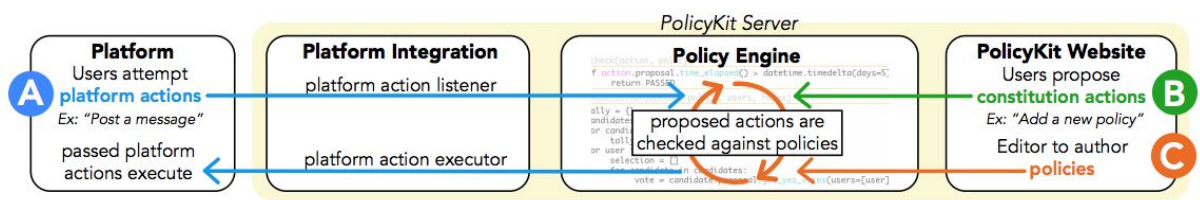

As mentioned previously, a significant issue that this platform solves is to change governance from permissions to procedures, where procedures can allow for alternative forms of governance such as participatory and democratic models. Furthermore, since these actions require a short amount of code that can be inputted by community members, new policies and governance models can evolve faster. Changing the platform's governance system is possible through two main abstractions that PolicyKit provides, which are actions and policies. Actions are a one-time occurrence within a community that is proposed by a community member. In comparison, policy governs a user's experience and can govern more than one action. Before a user submits an action, the policy must approve the action itself.

Thus, the platform can function in the following process: a) there must be a platform integration for the community home platform, b) every community hosted on that platform is now able to use PolicyKit, c) once PolicyKit is installed to the community platform, an initial governance system is installed- this governance system is a constitution policy model, d) community members can now propose actions, e) the policy engine continues to review actions to see if they can or cannot occur, f) when the action passes, it is deployed via the PolicyKit server. Finally, PolicyKit has a detailed and secure infrastructure that attempts to secure the accuracy of the policies and test policies before deployment to reduce the risk that can be posed to the community.

# Go Wide: Article Summaries

### Don't Ask If Artificial Intelligence Is Good or Fair, Ask How It Shifts Power
(Original *Nature* article by Pratyusha Kalluri)

This article sheds light on the power dynamics that exist in the field of AI and how they help to reinforce the status quo in society at large, rendering harm on those who are already marginalized. Instead of empowering data subjects to have greater control over how these systems make decisions about them, the systems abstract away that power into even more concentrated interests. The author rightly points out some of the concerns around the definitions of "fair", "transparent" and calls attention to the potential problem of "ethics-washing" that might be causing more pollution in the ecosystem in terms of legitimate vs. superficial initiatives that can help to address some of the problems with AI systems today.

Concluding the article by mentioning the following, we believe that this is increasingly important for everyone working on technology solutions that integrate and interact with people in a societal context to take into consideration: "When the field of AI believes it is neutral, it both fails to notice biased data and builds systems that sanctify the status quo and advance the interests of the powerful. What is needed is a field that exposes and critiques systems that concentrate power, while co-creating new systems with impacted communities: AI by and for the people."

### The Cost of Training Machines Is Becoming a Problem
(Original *The Economist* article)

AI presents an opportunity to act as a democratizing force. But, with recent advances in large-scale models accompanied by massive data requirements, the field has been skewed towards those who have the resources to participate in this increasingly competitive ecosystem. The most recent GPT-3 model with 175 billion parameters is not something that you could train on a few GPU instances that you spin up in the cloud. It requires access to heavy computing and the dollars to pay for it.

In some of the work by the Montreal AI Ethics Institute, we have proposed the requirement for evaluating these inequities more holistically so that AI can live up to its potential of benefitting all.

Innovations at both the software and the hardware level aim to better leverage the potential of AI. Specifically, novel fabrication approaches, chip architecture optimizations tailored for the kind of computations used in AI, and tinkering with the fidelity of the values used in computations, all have the potential to squeeze more from existing and new hardware.

Quantum computing is another avenue that can have a massive impact on how AI development happens. Some researchers even advocate for neuromorphic approaches, ones that mimic how the human brain works, as a methodology for achieving higher computational performance for the same levels of energy consumption.

## Human Rights Activists Want to Use AI to Help Prove War Crimes in Court
([Original *MIT Tech Review* article](#) by Karen Hao)

Just as is the case with content moderation that hurts the human agents who have to review it and are subsequently traumatized by it, analyzing footage from conflict zones to identify potential human rights violations poses similar challenges. Machine learning-enabled solutions that can automatically parse footage and identify different violations can spare humans from having to look at this kind of footage. However, and more importantly, these techniques have the benefit of being able to process orders of magnitude more content, which can be tremendously useful in making more substantive cases in international courts against authoritative regimes and other violators of human rights.

Given that instances of some kinds of prohibited weapons and techniques are rare, to aid the machine learning process, researchers are utilizing synthetic data crafted from limited information sets to supplement the process. What would have taken years for teams of humans working around the clock to analyze all the accumulated footage, the machine learning system can do in a few days. It makes the case made by humanitarian organizations that much more robust since they can demonstrate a systematic abuse of human rights by pulling out several examples through the automated analyses of the accumulated footage. It might just usher in a new era where more malicious actors are held liable for their war crimes, and help bring justice to those who don't have the resources to express their problems.

## Can Killing Cookies Save Journalism?
([Original *Wired* article](#) by Gilad Edelman)

Cookies help to track users and their behaviour across the internet and have long been heralded as the saving grace for publishers and advertisers as the mechanism that will help them both gain more value from the dollars that are allocated to marketing budgets. Yet, a recent experiment by a major publisher in the Netherlands (NPO) shows otherwise. NPO switched from microtargeting using cookies to non-tracked contextual targeting for ads and found that the revenues that they got actually jumped, even after factoring in for the recessionary effects of the pandemic raging at the moment.

Microtargeting is when users are bucketed into categories based on their demographics and other data that is supposed to allow advertisers to fine-tune who their ads are shown to. Contextual advertising, on the other hand, relies on the content on the web page and displays the ad based on that rather than using information about the user.

A few reasons have led to better performance of contextual advertising compared to microtargeting, according to NPO. The move away from platforms like Google Ads that provide this service means that publishers get to keep a much larger chunk of the incoming revenues. Secondly, a more privacy-conscious audience is inclined to visit websites that don't explicitly track them. And lastly, perhaps the strongest claim, whether or not someone clicks on an ad selling them pizza, is more influenced by whether they are hungry and reading up recipes than factors like their age group or where they live. This might just be the future of how journalism operates with a restoration of privacy of users while boosting revenues that will make digital journalism sustainable in the long run.

# Closing Remarks by Brent Barron (Director of Strategic Projects & Knowledge Mobilization, CIFAR)

The past four months have seen a startling change in the world around us. Since the publication of the June State of AI Ethics report, we have attempted a societal reckoning with the legacy of enslavement and colonialism, witnessed green shoots of hope wither to the second wave of COVID-19, and experienced increasingly acrimonious politics around the world.

Despite—or perhaps in response to—these uncertain times, researchers in the AI and Society community continue to develop new insights, models, and practices that can help ensure that we understand the effects of AI on the world around us, and channel its power towards the common good.

In this edition of the State of AI Ethics, MAIEI reports on research that spans the gamut from technical, as we see in Papernot's work on trust and differential privacy, to implementation-focused, as we see in the creation of PolicyKit to improve the responsiveness, flexibility, and efficacy of governance in online communities. For those of us who have seen our attention spans reduced to that of a goldfish by the modern internet, the report provides quick-hit summaries of very timely work on improving the health of online journalism, the challenges of the growth in model size and training cost, using AI to prosecute war crimes, and a suggestion to briefly pause unending discussions of AI fairness and goodness, and instead talk about how it expands, reduces, and shifts power.

As the world around us feels increasingly dark, it is even more important that we recognize members of our community who are working tirelessly to better understand how technology is shaping and is shaped by, the world around us. We're also seeing an increasing amount of collaboration between computer science and the social sciences and humanities, as well as between the academic, private, public, and civil society sectors to produce new tools, technologies, and practices. This work is helping to improve the lived experiences of humans around the world, albeit still unevenly, and that should give us all a glimmer of hope.

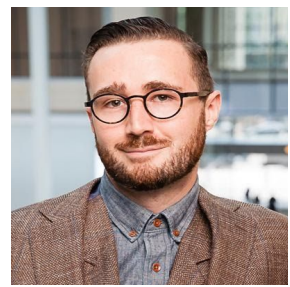

Brent Barron
Director of Strategic Projects & Knowledge Mobilization, CIFAR

Brent Barron is Director, Strategic Projects, Knowledge Mobilization at CIFAR where he is responsible for engaging the policy community around cutting edge science. He played an important role in the development of the Pan-Canadian Artificial Intelligence Strategy, and oversaw its AI & Society program.

# 2. Bias and Algorithmic Injustice

**Opening Remarks** by Abhishek Gupta (Founder, Montreal AI Ethics Institute)

Problems seem to continue to abound since the publication of the last iteration of this report. The fight to combat bias and algorithmic injustice is on and this chapter gives a glimpse into the myriad efforts that are being spearheaded by activists and researchers alike in the domain. My overall takeaway has been to empower the voices of those who are on-the-ground and who have lived experiences, sharing the spotlight when we have a chance and actively promoting these conversations at every stage of the design, development, and deployment of the AI lifecycle rather than just as an afterthought.

One of the things that caught my attention this past quarter was the emphasis on decolonial theory as an instrument in better understanding the challenges and power dynamics when it comes to the deployment of AI systems. This is supplemented by the problems that can arise when we merge datasets that can create "the whole is bigger than the sum of its parts" kind of challenges exacerbating the problems faced by those who are disenfranchised in our world driven by the hunger to amass more data. At the same time, it seems that even continued agglomeration of data doesn't yet solve the problems when it comes to mitigating biases in speech recognition systems. Something that is evident from the research that is highlighted in this chapter.

Technical approaches like the one mentioned in this chapter that helps to assess fairness outcomes on multiple sensitive attributes can help to adhere a bit more to notions of intersectionality rather than treating each sensitive attribute in isolation.

But, there are larger systemic changes afoot, like how platform algorithms can promote a certain kind of behavior, as is the case with Instagram favoring nudity over other forms of content and shaping our social media behaviour in the process of seeking the ever-elusive virality on these platforms.

I don't see an end to us bending to algorithmic injustice until we question critically all aspects of algorithmic design, whether that is data collection, algorithm design, or measurement of what success looks like in the deployment of these automated systems. Listening to scholars in this domain who have been engaging critically with this work for many years now is essential and learning from their work will help us move away from rediscovering problems to solutions. I hope that we will have more solutions and progress to report in this chapter next quarter.

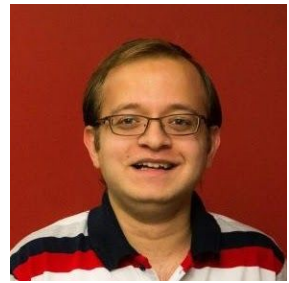

Abhishek Gupta (**@atg_abhishek**)
Founder & Principal Researcher, Montreal AI Ethics Institute
ML Engineer & CSE Responsible AI Board Member, Microsoft

Abhishek Gupta is the founder and principal researcher at the Montreal AI Ethics Institute, seeking to define humanity's place in a world increasingly characterized and driven by algorithms. He is also a machine learning engineer at Microsoft, where he serves on the CSE Responsible AI Board. His book 'Actionable AI Ethics' will be published by Manning in 2021.

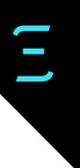

# **Go Deep: Research Summaries**

### **Decolonial AI: Decolonial Theory as Sociotechnical Foresight in Artificial Intelligence**
([Original paper](#) by Shakir Mohamed, Marie-Therese Png, William Isaac)

After repeated cases of algorithms gone awry, AI's potential, and the possibilities for its misuse, has come under the scrutiny of governments, industries and members of civil society. When evaluating the aims and applications of AI, the authors make clear that we often fail to recognize and question the asymmetrical power dynamics that underlie both the technology and the systems of networks and institutions to which it is linked. Our failure to acknowledge these power relations undermines our ability to identify and prevent future harms arising out of these systems. Traditional ethical standards for human-subject research in the sciences often do not consider structural inequities, such as systematic racism. This oversight is particularly alarming as AI can ingest, perpetuate and legitimize inequalities to a scale and scope no technology has ever done before. As follows, we must rethink our tools for the evaluation and creation of socially beneficial technologies.

For Mohamed, Png & Isaac, one way forward is to adopt a critical science-based approach, grounded in decolonial theory, to unmask the values, cultures and power dynamics at play between stakeholders and AI technologies. Critical science and decolonial theory, when used in combination, can pinpoint the limitations of an AI system and its potential ethical and social ramifications. These approaches offer a "sociotechnical foresight tool" for the development of ethical AI.

AI, like all technologies, did not emerge out of an ahistorical and isolated scientific bubble. The power dynamics between the world's advantaged and disadvantaged, instilled during the colonial era, continue to resurface in the contemporary design, development and use of AI technologies. The authors point to numerous instances where colonial practices of oppression, exploitation and dispossession are present in AI systems. They refer to these cases as algorithmic coloniality.

The examples brought up by the authors touch on the use of AI systems, its labour market, and its testing locations. The authors point to the biases against certain groups in algorithmic decision-making systems in US law enforcement. They refer to the unethical working conditions of ghost workers who do data labelling and annotation. The beta-testing and fine-tuning of AI systems are also part of a phenomenon known as "ethics dumping." AI developers purposely test and deploy their technologies in countries with weaker data protection laws. Finally, the geopolitical imbalance in AI governance policies and the paternalism of technological-focused international social development projects entrench global dependency patterns. All in all, it is made evident that AI is both shaped and supported by colonial structures of power.

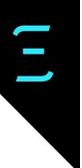

Adopting a decolonial framework would allow for an analysis of AI technologies within a socio-political and global context and address these types of abuses. Recognizing the bigger context can contribute to the design of a more inclusive and well-adapted mechanism of oversight for AI systems. The authors list three tactics for future decolonial AI design, being a critical technical practice of AI, the establishment of reciprocal engagements and reverse pedagogies, and the renewal of affective and political communities.

Critical technical practices are a middle-ground between the technical work of developing new AI algorithms and the critical work of questioning taken-for-granted assumptions. The topics of AI fairness, safety, equity, decision-making, and resistance aim to create more context-aware technological development. Reciprocal engagements and reverse pedagogies address the possibilities of knowledge exchange between AI researchers and stakeholders. They can take place through the form of intercultural dialogue, data documentation and meaningful community-engaged research design. The renewal of affective and political communities refers to the creation of new types of solidarity-based communities that have the power to address, contest and redress emerging challenges in tech.

As AI will have far-reaching impacts across the social strata, a diversity of intellectual perspectives must be a part of its development and application. Critical science and decolonial theory, along with its associated tactics, are useful tools for identifying and predicting the more nefarious uses of AI. Historical hindsight is always beneficial to technological foresight. The challenge remains of finding concrete avenues for marginalized groups to have a real influence in the decision-making process. Those who have the most to lose from AI systems are often readily aware of the social inequities they live and face. While it is sometimes tempting to get lost in semantics and academic jargon when discussing AI ethics issues, we must focus our efforts on making AI development, and mechanisms for its criticism, truly legible and accessible to all members of society. It includes rendering the decolonial theory and its associated concepts comprehensible to AI developers and members of the tech industry.

## Fairness in Clustering with Multiple Sensitive Attributes
([Original paper](#) by Savitha Sam Abraham, Deepak P., Sowmya S Sundaram)

For a quick definition, clustering is a statistical technique that involves grouping similar points or objects in a dataset. With AI systems, part of an algorithm's task is to sort through this set of objects (which can refer to individuals) and form these similar clusters. For data analysts, clustering is a necessary task given the infeasibility of conducting manual per-object assessment or appreciation in each dataset, especially when those objects number in the thousands. In a hiring scenario with a high volume of applicants, large corporations may design an algorithm to group and rank candidates based on their resumes. Those in the top cluster, with similar desired attributes, will be sent out an email informing them of the shortlisting decision. Already in this example, a few ethics-minded eyebrows might be raised.

As we have learned, algorithms, unless explicitly instructed to, will not consider individual or group fairness principles when categorizing objects. Group fairness is related to protecting people who share sensitive attributes, such as age, gender, ethnicity, relationship status and so on. If left unchecked, algorithms can create highly skewed and homogenous clusters that do not represent the demographics of the dataset. These bias clusters may serve to reinforce societal stereotypes. Even if data identifiers, such as gender, are removed, statistical correlations can still lead to gender-homogenous clusters. Gender is just but one among the plurality of sensitive attributes to consider within analytics pipelines to avoid undue discrimination. Data scientists have tried to find ways to balance one "sensitive" attribute at a time or to account for multiple binary-only (ex: citizen or non-citizen) attributes.

The authors offer a new and fairer clustering method called Fair K-Means (Fair KM). Their Fair K-means is a pioneering statistical technique in that it can consider multiple multi-valued or numeric sensitive attributes in clustering for various scenarios. For the layperson, in k-means clustering, you identify a target number k, which represents the number of groups formed with the data. Each data point is assigned to a K group based on shared similar features. With every k group, the centroid (the mean position of all the data points) in the cluster pinpoints what type of population each group represents. The Fair K-Means, when assigning points to a cluster, will not only consider similarities to the centroid but also which cluster it will skew the least in terms of sensitive attributes. To be more precise, the clusters formed will proportionally respect the demographic characteristics of the overall dataset, hence allowing for representational fairness in clustering. In their application of Fair KM, the algorithm was able to cluster individuals according to nine attributes, while balancing five sensitive ones, such as relationship status and country of origin.

The clusters formed with the FairKM method scored better on both clustering quality and fair representation of sensitive attribute groups than other prominent clustering methods. The Fair K-Means can also be fed-back into algorithmic training to repeatedly improve clustering performances. One limitation of the approach was the original dataset's quality, a census of 15 000 individuals, used in the method tests. It will be necessary to test whether this approach can hold up for extremely skewed data points.

With Fair KM, we are inching closer to forming statistical methods for decision-making algorithms that abide by principles of equality, equity, justice, and respect for diversity, the core tenets of democratic societies. Although the perfect balance may be a statistical ideal, it remains far from a lived reality. There are societal limitations that continue to prevent total gender parity or social equality. Those factors are outside the immediate control of data scientists but should be acknowledged when designing predictive decision-making algorithms.

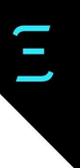

# **Go Wide: Article Summaries**

**Undress or Fail: Instagram's Algorithm Strong-arms Users into Showing Skin**
(Original *AlgorithmWatch* article by Judith Duportail, Nicolas Kayser-Bril, Kira Schacht and Édouard Richard, in partnership with the European Data Journalism Network)

The degree to which platforms govern the ecosystem and the impact that they have in shaping the wider ecosystem is evident in this experiment that was carried out by the team at AlgorithmWatch on Instagram users. In trying to ascertain if the platform as a whole had a slant towards a particular kind of content, the team from AlgorithmWatch requested volunteers to install a browser plug-in that monitored the content that was shown to the users in their feed. The team then analyzed the distribution of the types of content in the feeds of these users and found that while personalization based on tastes of users, as expressed by their interactions with content on the platform and past behaviour, factored into what was shown, there was a skew towards content that favored the showing of skin.

Overall, this has had a negative impact on the community for those whose values don't align with this and those whose businesses have nothing to do with things like swimwear, etc. and from anecdotal evidence collected by the team, they found that creators who ran accounts for things like food blogs also had to resort to these tactics to have their content feature prominently in the feeds of people. The AlgorithmWatch team reached out to the Facebook (owns Instagram) team for comment and they replied saying that the methodology used by the AlgorithmWatch team was flawed.

One of the patents filed in this space by them specifically identified "gender, ethnicity, and "state of undress" could be used in computing the engagement metric which would be used to present items in a person's feed. Additionally, the items are presented not just based on the prior actions of that user but all users which can lead to a collective shaping of behaviour on the platform. Bias can naturally creep in when supervised computer vision techniques are used to train systems to automatically categorize content and then present it to the users, for example, in using popular crowdsourced platforms to label training data, there is a risk that categorization is done in the coarsest possible manner because of the low rates of pay to users which lowers the amount of effort that workers potentially exert in finding more nuanced categories.

A problem with content creators speaking out is that they fear the "shadow-ban", a practice exercised by platform owners that pushes down the content from those facing that ban into the depths of an ethereal abyss such that their content doesn't show up prominently in people's feeds even when that piece of content might be relevant to that user.

## Speech Recognition Tech Is Yet Another Example of Bias
(Original *Scientific American* article by Claudia Lopez Lloreda)

In numerous talks given by our founder Abhishek Gupta, he has pointed out how speech recognition technologies create nudges that are reshaping our conversation patterns such that we, as humans, have to adopt our speech patterns to meet the needs of the machines rather than have the machines adopt their structures so that they can accommodate human diversity. For those who use English as their second language or don't have a "mainstream" accent, chances of having high failure rates by speech recognition systems are pretty high. The study linked in the article points to how black people in the US face a disproportionate burden of such speech-related discrimination that imposes an unnecessary choice for them to either abandon their identity or abandon the use of those devices.

In human speech, we make accommodations for each other on a daily basis to be able to negotiate and navigate different accents and dialects. But, with machines, there is no such negotiation, and it is a very binary outcome that marginalizes those who don't conform to what the machines have been trained on. In cases of people with speech disabilities, this is even more problematic since they might rely on such systems to go about their daily lives.

In addition, language is highly contextual and depending on the speaker and the context within which words and phrases are being used determine to a large degree the meaning and response which is something that is still not quite possible with current NLP systems. While adding in more diversity in training datasets is one option, a lot of it will also depend on more inclusive design practices. As a starting point, having more testing being done with those who are going to be the users of these systems will go a long way in making these systems better. Some researchers at the Montreal AI Ethics Institute have advocated for participatory design approaches as a way of making technological solutions more context- and culture-sensitive.

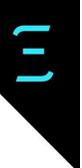

**Merging Public Data Sets Has Implications for Racial Equity**
(**[Original *Built In* article](#)** by Stephen Gossett)

The article begins by pointing out an example where Detroit PD pushed to separate the racial impacts of facial recognition technology (FRT) from the city's video surveillance program. The thrust of their argument is that FRT is one among many tools used by detectives, and hence it doesn't have a huge impact. But, on every step in the data lifecycle, there are pitfalls in terms of exacerbating bias, especially when data gets integrated across multiple agencies and creates richer profiles.

The racial equity toolkit created by the researchers in the article advocates for the interrogation of these issues in the planning, data collection, data access, algorithms and statistical tools selection, data analysis, and reporting phases of the life cycle.

Data integration is a turbocharged version of data sharing whereby more invasive analysis can be carried out on the resulting, merged datasets. The researchers point to a case in North Carolina whereupon combining data from multiple agencies, citizens who utilized several government aid programs in parallel were targeted and subsequently lost access as they were seen to be high-risk individuals. Involving a data custodian — someone who owes a responsibility to the data subjects, is a great way to ensure that the rights of citizens are respected. Even at the conception stage, there needs to be a reflection on whether the project is the best way to achieve a social outcome or if it is just being carried out because there are grant dollars available. Augmenting existing data with qualitative analysis and surveys can unearth potential problems with the quantitative data on hand.

Disaggregating data presents a promising approach, yet it shouldn't be relied upon solely as the means to address problems with racial equity. Finally, the researchers say that ignorance is no longer a defensible position, the use of a toolkit like this forces difficult conversations that will guide people in building inclusive AI systems.

# 3. Disinformation

**Opening Remarks** by **Abhishek Gupta (Founder, Montreal AI Ethics Institute)**

At the time of writing this segment, the election in the US is only a few days away and the information ecosystem has been awash with disinformation that has the potential to sway how we vote and bring into question the integrity of some of the fundamental tenets of our democratic processes. This chapter presents a broader view on how we can navigate this landmine and put our best foot forward not only as participants in this ecosystem but also as researchers who are attempting to steer the design, development, and deployment of everyday data-driven technologies.

Taking a quick stroll through private Facebook groups, it is evident that they hold tremendous influence in steering conversational direction on politically charged topics. Being opaque to external study, they pose a serious risk to the integrity of our information ecosystems. Speed of action on combating misinformation on Facebook can be quite slow and lead to damage even before the content is flagged and taken down. For example, with climate change misinformation, often the burden is disproportionate on scientists who try to debunk such information but face even higher standards to prove the veracity of their posts vs. those who are spreading misinformation in the first place. In the case of Black Lives Matter, organizations such as the IRA have played both sides of the movement to further widen the chasm between ideological extremes.

Multimodal information also presents a new war front in this theatre of disinformation, making it particularly challenging for automated systems to tackle. Researchers showcased in this chapter point out the potency of text-based disinformation as well, particularly in skewing conversations subtly over a long period of time, evading our scrutiny. These abilities have been supercharged with generating more and more believable deepfakes through the use of technologies like GPT-3. Quick punchy visuals also exacerbate the problem, as highlighted in the piece in this chapter that talks about memetic warfare that draws on our inclination to react to memes.

A report looking beyond just the information ecosystem in the US helps gain a broader perspective on how information operations in different countries take place and perhaps we can use them as a learning opportunity to implement safeguards based on different capabilities of domestic actors and their motivations. For example, the widespread use of Whatsapp in places like Brazil and India and how the spread of disinformation can be curbed there can serve as a cautionary tale to the increasing use of other platforms in addition to the ones that are discussed more frequently, like Facebook and YouTube. Some places have also experienced the spread of disinformation under the guise of real news and media outlets who have been broken into and have had stories planted into legitimate outlets causing widespread harm before these compromises are detected.

We also need more nuanced vocabulary when it comes to the ecosystem of problematic information and relying solely on a limited set of terms like fake news and disinformation doesn't do justice to the complexity of the entire ecosystem and the tailoring of the measures required to combat different actors and the different kinds of problematic information proliferating on social platforms. The segment on Lexicon of Lies helps to provide that grounding and I recommend it as a must-read in this chapter.

Whether you are a concerned citizen or an active researcher in this space, I believe that this chapter provides a wide-ranging selection of articles and research that can help you expose yourself to ideas behind those that are most frequently covered in popular media. I hope that you have a chance to share this with other colleagues and friends within your network so that they can add nuance to their own work and engage more critically with the problems in this domain.

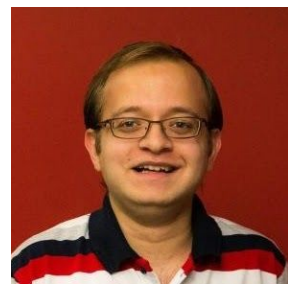

Abhishek Gupta (**@atg_abhishek**)
Founder & Principal Researcher, Montreal AI Ethics Institute
ML Engineer & CSE Responsible AI Board Member, Microsoft

Abhishek Gupta is the founder and principal researcher at the Montreal AI Ethics Institute, seeking to define humanity's place in a world increasingly characterized and driven by algorithms. He is also a machine learning engineer at Microsoft, where he serves on the CSE Responsible AI Board. His book 'Actionable AI Ethics' will be published by Manning in 2021.

# Go Deep: Research Summaries

### A Picture Paints a Thousand Lies? The Effects and Mechanisms of Multimodal Disinformation and Rebuttals Disseminated via Social Media
([Original paper](#) by Hameleers, Thomas E. Powell)

In the current information environment, fake news and disinformation are spreading, and solutions are needed to contrast the effects of the dissemination of inaccurate news and information. In particular, many worry that online disinformation – intended as the intentional dissemination of false information through social media – is becoming a powerful, persuasive tool to influence and manipulate users' political views and decisions.

Whereas so far research on disinformation has mostly focused on only textual input, this paper taps into a new line of research by focusing on multimedia types of disinformation which include both text and images. Visual tools may represent a new frontier for the spread of misinformation because they are likely to be perceived as more 'direct' representations of reality. Accordingly, the current hypothesis is that multimedia information will be more readily accepted and believed than merely textual inputs. And since now images can be easily manipulated, the worry that animates this research is that they will constitute a very powerful tool in future disinformation-campaigns. Therefore, the primary goals of this paper are (1) to investigate the persuasive power of multimedia online disinformation in the US and (2) to study the effects of journalistic debunking tools against multimedia disinformation.

In the experimental study conducted in this paper, subjects were all shown false tweets concerning two highly politicized topics: school shootings and refugees. The tweets were either only textual input or text + image, and they would come from established new sources (aka CNN) or from ordinary citizens. In some cases, subjects were shown corrective information: a rebuttal tweet (text + image) from PolitiFact (a popular fact checking software) that disproves the fake tweet content. Subjects were then asked to rate the initial tweets' credibility and truthfulness. The political and ideological views of the participants were also tracked to establish whether they would influence the participants' reactions to multimedia disinformation and subsequent debunking strategies.

The outcomes of this study are the following:

- The empirical results partially show that multimodal tools are rated as slightly more trustworthy than solely textual inputs. This is likely due to the fact that words are abstract indicators, whereas images provide a seemingly direct representation of reality. So multimedia tweets may appear more truthful and believable than mere textual inputs.

- The results indicate that fact checkers constitute useful debunking tools to contrast misinformation and disinformation. The positive effects of fact checking were stronger for those whose political and ideological beliefs aligned with the debunked content. That means that users who would typically agree with the content of the false tweets were more affected by the corrective tweets from PolitiFact. This result is in opposition to the expectation of the so-called 'backfire effect' (i.e. that contrary evidence not only does not change the mind of partisan users but actually reinforces their preexisting political views). It is however still an open question whether multimodal fact checkers are more effective than simply textual corrective information.

- The results indicate that the source of the information does not matter: in the study subjects assessed the credibility of the news inputs independently of whether they were from established journalistic sources such as CNN or from ordinary citizens. This result paints a discouraging picture of users' media literacy skills because it reveals that they are unable to distinguish between reliable news sources from unreliable ones.

The paper concludes with two recommendations. First, fact checkers should be widely used as journalistic tools as they are effective ways to debunk false information online. What's more, the paper highlights the importance of media literacy in fostering citizens' ability to spot misinformation and in educating them to rely on established, reliable news sources.

## Lexicon of Lies: Terms for Problematic Information
(**[Original *Data & Society* report](#)** **by Caroline Jack**)

This article seeks to explain the terms used to describe problematic information. They could be inaccurate, misleading, or altogether fabricated. The terms we use in describing information would impact how information spreads, who spreads it, and who receives it as this choice is based solely on the perspective of the descriptor. This makes the labelling of information complex, inconsistent and imprecise.

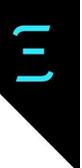

The 2 major divisions of problematic information are *misinformation* and *disinformation*:

Misinformation is when the incorrectness/ inaccuracy of information is not intentional but due to mistakes. This is caused by the failure to independently verify a source's claims or the rush to pass information across- for example in the case of journalists trying to win the competition of being the first to report.

Disinformation, on the other hand, is when information is deliberately intended to mislead.

Social media has facilitated the spread of both forms of problematic information. With computational systems that push for the spread of "trending topics", it allows these topics to reach more and more people.

Whether a given story or piece of content is labelled as misinformation or disinformation can depend as much on a speaker's intent as on the professional standards of who is evaluating it.

Automated systems with bugs and sites created just for profit without concern for the accuracy of content posted on it are another source of misinformation. The interaction between news content and entertainment content contribute to the complexity of information interpretation. Generally, intentions behind contents on social media are usually not clear.

Extra caution is taken by journalists before labelling information as misinformation or disinformation because misrepresentations can lead to reputational damage, professional sanctions, and legal repercussions.

*Publicity* and *propaganda* are persuasive information campaigns that try to link brands, people, products, or nations with certain feelings, ideas, and attitudes. Both focus on reaching a large crowd. While the former tries to get information – which may be accurate information, misinformation, disinformation, or a mix of all three, the later, is a deliberate attempt to deceive or manipulate. The existence of both forms in the same space makes their difference not very obvious.

In practice, the lines separating *advertising*, *public relations*, and *public diplomacy* (terms often regarded as neutral) from the pejorative term propaganda (which usually implies deliberate intent to manipulate or deceive) can be hard to discern.

The source of campaigns could help us understand the category in which a piece of information belongs – whether propaganda or publicity. Some information, however, does not have an obvious source. Information like advertising, public relations, or public diplomacy have obvious sources while other information might come from information operations – a term which originated from the military and referred to the strategic use of technological, operational, and psychological resources to disrupt the enemy's informational capacities and protect friendly forces. Today, this term is used to describe the deliberate and systematic attempts by unidentified actors to steer public opinion using inauthentic accounts and inaccurate information.

The differences in languages also make categorizing information difficult. In Spanish, for example, *la propaganda* can refer to political communications, advertising, and even junk mail which does not follow the standard definition mentioned above.

Propaganda can sometimes be a deliberate act to cultivate attitudes and/or provoke action. When it takes this form, it is termed agitprop, however, this term is rarely used as all forms of propaganda are just called propaganda. At some time – around the 20th century, propaganda was classified into 3 groups; white, grey, or black, depending on the information's accuracy, the channel of distribution and also the formality of it. White propaganda is accurate and from accurately identified sources, whereas black propaganda is inaccurate or deceptive and their source is misrepresented. Grey propaganda combines accurate and inaccurate content.

Far more pressing issues beyond differentiating publicity from propaganda is when the goal of an information campaign is not to promote support for an idea but to confuse people by spreading uncertainty and starting debates that are most likely to divert.

One form of confusion is *gaslighting* – a situation in which a person misinterprets and changes the narrative of an event in a deceptive and inaccurate manner to the extent that their victim stops trusting their own judgments and perceptions.

*Dezinformatsiya* is a coordinated state effort to share false or misleading information to the media in targeted countries or regions. This involves taking active measures in spreading disinformation, especially with the goals of widening existing rifts, stoking existing tensions, and destabilizing other states' relations with their publics and one another.

At the moment, there is no obvious solution to the spread of problematic information. Media literacy is necessary, but not sufficient for understanding today's problematic information flows. With the increasing number of fact-checking and rumour-debunking news stories, there seems to be no headway in restoring the authority of the press or social institutions, as the parties involved in creating disruption, destabilization, distraction and derailing information seem to have the upper hand.

One method adopted to solve false information spread is *xuanchuan*, a Chinese term, to describe a misdirection strategy on social media in which coordinated posts don't spread false information, but instead flood conversational spaces with positive messages or attempts to change the subject. This act is perceived positively even though it is still a form of propaganda. This further demonstrates the ambiguous boundary between publicity and propaganda to be a cross-cultural phenomenon as seen in the Spanish example above.

As useful as these cross-cultural terms like dezinformatsiya, and xuanchuan are, they should be used with care because of the cultural associations they can raise. There is a risk of forming certain assumptions of the cultures involved.

Other terms come from using rhetorical means to describe issues of society. These means could be playful, humorous, or ironic which are all not intended to be perceived seriously. Examples of these forms of information are *satire, parody, culture jamming*, and *hoaxing*.

Satire uses exaggeration, irony, and absurdity to amuse the audience while calling attention to, and critiquing perceived wrongdoing.

Parody is a form of satire that exaggerates the notable features of a public figure, artist, or genre. Culture jamming turns the tools of parody against advertising culture, ironically repurposing the logos and conventions of advertising in order to critique corporate culture.

Hoax is a deliberate deception that plays on people's willingness to believe.

Hoaxes depend, at least initially, on some people taking them seriously. They are a means of challenging authority, custom, or the status quo and also could be motivated by self-interest. An example of a hoax is the April Fools' Day, a day with a lot of misinformation flying around. This might not go well with everyone as information in any form has the ability to attract and retain the attention of fickle audiences.

The networked nature of social media often contributes to changing the context of the original information as it is passed on. This makes it difficult to judge whether a piece of content is serious or sarcastic in nature.

The interconnection of disinformation and misinformation and that of serious and sarcastic information makes it easy for people to take advantage of the ambiguity and instrumentalize it. Those who give ideas and perform actions far outside the mainstream that end up being criticized automatically label their actions or ideas as sarcastic. The defence of "it was just a joke!" mobilizes plausible deniability and frames anyone who objects as intolerant of free speech.

In today's information environment, we may need to modify and qualify the terms we have, or find new metaphors and models that acknowledge the complexity and ambiguity of today's problematic information.

In conclusion, the term chosen to describe an information campaign conveys information about who is running that campaign and the goals they might have in running it. It also reveals information about the writer – namely, how she assesses the accuracy, validity, and potential consequences of the information campaign.

Misinformation and disinformation should be discussed with care; writers must be mindful that their representations of problematic information in today's world can bolster assumptions that may be inaccurate and can reestablish social divisions.

## Acting the Part: Examining Information Operations Within #BlackLivesMatter Discourse
([Original paper](#) by Arif, A., Stewart, L. G., & Starbird, K.)

The researchers at the University of Washington analyzed Twitter activity on the #BlackLivesMatter movement and police-related shootings in the United States during 2016 to better understand how information campaigns manipulate the social media discussions taking place. They focused on publicly suspended accounts that were affiliated with the Internet Research Agency, a Russian organization. This organization supports full-time employees to do "professional propaganda" on social media (Arif, Stewart, & Starbird, 2018).

Social media has become a platform for information operations, especially for foreign invaders, to alter the infrastructure and spread disinformation (Arif, Stewart, & Starbird, 2018). Information operations are used by the United States intelligence community that describes actions that disrupt information systems and streams of geopolitical adversaries (Arif, Stewart, & Starbird, 2018).

"Disinformation describes the intentional spread of false or inaccurate information meant to mislead others about the state of the world" (Arif, Stewart, & Starbird, 2018).

Social media has become a breeding ground for disinformation because of the way systems and networks are created and nurtured online. For example, algorithms derive newsfeeds related to people's preconceived ideas. People are also typically connected with those they already have similar thoughts with, which cause a homophily filter bubble (Arif, Stewart, & Starbird, 2018). These structural issues can contribute to the effectiveness of disinformation being spread.

Not to mention, the ecosystem of social media is influenced by hot bots, fake news websites, conspiracy theorists, and trolls are influencing mainstream media, influential bloggers and ordinary individuals who are now all amplifying the propaganda in social media platforms (Arif, Stewart, & Starbird, 2018).

Facebook has even acknowledged that their platform had been used for "information operations" to influence the United States Presidential election, in an April 2017 report (Arif, Stewart, & Starbird, 2018). Other platforms such as Twitter and Reddit came forward to say their platform has also been involved in information operations by the Internet Research Agency known to be a Russian "troll farm."

With that in mind, they analyzed how the RU-IRA accounts participated in online discussions about the #BlackLivesMatter movement and shootings in the U.S. during 2016.

There was a list released by Twitter of 2,752 RU-IRA-affiliated troll accounts in November 2017. The researchers begin by analyzing the behavioral network ties to narrow down cases to conduct qualitative research. After they got the 29 accounts integrated into the information network, they conducted a bottom-up open coding on the digital traces left behind by these accounts such as tweets, profiles, linked content, and websites.

The initial dataset included about 58.8 million tweets from December 31st, 2015, and October 5th, 2016. They used the open Twitter API.

Some different terms they searched for were "gunshot, gunman, shooter, and shooting". Then for the #BlackLivesMatters, they searched for "BlackLivesMatter, or "AllLivesMatter and "BlueLivesMatter."

They first conducted a network analysis to understand the 22,020 accounts promoting these issues. They then cross-referenced those accounts with the list released by Twitter and found that 96 RU-IRA accounts from Twitter's list were present in the data, and the subset of the troll accounts tweeted at least once with the various hashtags. These accounts interacted with many accounts surrounding them, which potentially have a more significant impact on influence within the networks (Arif, Stewart, & Starbird, 2018).

After the network analysis was completed, they began the qualitative analysis with the 29 accounts. Those accounts produced 109 tweets, which were retweeted 1,934 times in their data collection (Arif, Stewart, & Starbird, 2018).

They then analyzed three central units, including the profile data, tweets with a focus on original content including memes, and external websites, social platforms, and news articles, these accounts linked to "follow the person" (Arif, Stewart, & Starbird, 2018).

A structural analysis was conducted between the 22,020 twitter accounts and the 58,695 retweets these accounts got from their content. They used a community detection algorithm to identify the clusters systematically. The clusters were best identified as a difference along political lines. One side of the summary bios was involved with the Democratic presidential candidate, and the Black Lives Matter campaign while the other community-supported Trump and the MAGA hashtag claiming Make America Great Again (Arif, Stewart, & Starbird, 2018).

Their results suggested that information operations were occurring and that while some social media does bring us together when these platforms such as twitter are being targeted, there are accounts deliberately trying to pull people apart. It aligns with other literature claiming that the tactics used for disinformation are ideologically fluid and seek to exploit the social divides (Arif, Stewart, & Starbird, 2018).

Many of the profiles analyzed were deceitful in pretending to be a certain kind of person, such as an African American that fit the stereotypical thinking. Another finding was that these accounts tied to Russia were often linked with their websites that undermine traditional media in favor of alternative media websites that are set up for supporting information operations (Arif, Stewart, & Starbird, 2018).

These examples highlight how information operations can invoke content that is not always politically persuading their followers by true or false claims. However, they are affirming and representing personal experiences and shared beliefs that reconfirm what people already believe based on stereotypes that may or may not be accurate. These accounts blend into the communities they target, which help them become more persuasive socially and emotionally.

This article of research opens the doors to understanding the mechanisms these information operations accounts use to manipulate people and what their broader goals are in terms of shaping online political discourse, primarily in the United States (Arif, Stewart, & Starbird, 2018).

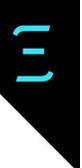

Overall these accounts use fictitious account identities to reflect and shape social divisions and can undermine the trust in information in places such as the "mainstream media" (Arif, Stewart, & Starbird, 2018). Furthermore, because of their tactics used that resonate with the actual persons they are targeting, they are more successful because they understand how and why the people of that targeted community think the way they do and feed information they already believe in, which strengthens their beliefs.

## Challenging Truth and Trust: A Global Inventory of Organized Social Media Manipulation
(Original *Oxford Internet Institute* report by Samantha Bradshaw, Philip N. Howard)

In a short time, social media platforms' roles have expanded from their nascent stages—places for users to connect, share entertainment, discuss popular culture, and stay in touch with each other's day-to-day lives—to fields of operations for large-scale political and ideological warfare. On these online theaters, "cyber troops," carry out missions to manipulate public opinion for political purposes by disseminating and amplifying "computational propaganda" (automation, algorithms and big-data analytics to manipulate public life). While many readers may be familiar with the large, robust cyber operations based in Russia, China, US, or North Korea, this 2018 report by Samantha Bradshaw and Philip N. Howard at the Oxford Internet Institute illuminates formally-organized social media campaigns that are developing across the world, in countries large and small, rich and poor, authoritarian and democratic alike.

Bradshaw and Howard point out that in the past, governments relied on "blunt instruments" to block or filter information, but that modern social media platforms allow for more precise information control, as they have the capabilities of reaching large numbers of people while simultaneously allowing for micro-targeting of people based on location or demographic traits. This versatility is precisely what has made social media platforms suitable tools to shape discourse and nudge public opinion. The value in being able to control discourse online is evidenced by the growth in coordinated attempts to influence public opinion documented in 48 countries in this report, compared to the previous year's 28 countries (but authors do note that their data may not be comprehensive).

While cyber troops function across all sorts of governments, the authors point out that their roles are not one and the same across space. In emerging and Western democracies, political bots are being used to poison the information environment, polarize voting constituencies, promote distrust, and undermine democratic processes. In authoritarian regimes, governing parties use computational propaganda as just one tool in a portfolio of tactics to shape the narrative of the ruling party, stomp out counter-narratives, and subvert elections.

Computational propaganda operations can be targeted at both foreign and domestic audiences, and conducted by government agencies, politicians and parties, private contractors, civil society organizations, or citizens and influencers. They may deploy a number of "valence strategies" (characterizing attractiveness or averseness of content), by spreading pro-government or -party propaganda, attacking the opposition or mounting smear campaigns, or diverting conversations or criticism away from important issues. Often, cyber troops conduct operations through fake accounts, which may be automated accounts, human-controlled, or hybrid/cyborg accounts—in which operators combine automation with elements of human curation, making them particularly difficult to identify and moderate.

Table 2: Social Media Manipulation Strategies: Messaging and Valence

| Country | Fake Account Type | Pro-Government or Party Messages | Attacks on the Opposition | Distracting or Neutral Messages | Trolling or Harassment |
|---|---|---|---|---|---|
| Angola | 🤖 | | ■ | | |
| Argentina | 🤖 | ■ | ■ | | ■ |
| Armenia | 🤖 | | | ■ | |
| Australia | 🤖 | | | | |
| Austria | 👤🤖 | | ■ | | ■ |
| Azerbaijan | 👤🤖⚙ | | ■ | ■ | ■ |
| Bahrain | 👤🤖⚙ | | ■ | | ■ |
| Brazil | 👤🤖⚙ | | ■ | ■ | ■ |
| Cambodia | 🤖⚙ | | | | |
| China | 👤🤖⚙ | ■ | | ■ | ■ |
| Colombia | 👤 | | | | |
| Cuba | 👤🤖⚙ | | | ■ | |
| Czech Republic | | | | ■ | |
| Ecuador | 👤🤖⚙ | ■ | ■ | | ■ |
| Egypt | 👤🤖⚙ | ■ | ■ | | ■ |
| Germany | 👤🤖⚙ | ■ | ■ | | ■ |
| Hungary | 👤 | ■ | ■ | | ■ |
| India | 🤖⚙ | ■ | ■ | | ■ |
| Iran | 👤🤖⚙ | ■ | ■ | ■ | |
| Israel | 👤🤖 | ■ | | | |
| Italy | 🤖⚙ | | ■ | | |
| Kenya | 🤖⚙ | | ■ | | |
| Kyrgyzstan | 👤 | ■ | | | |
| Malaysia | 🤖⚙ | | | | |
| Mexico | 👤🤖⚙ | ■ | | | ■ |
| Myanmar | 🤖 | ■ | | | |
| Netherlands | 🤖 | | | | |
| Nigeria | | ■ | | | |
| North Korea | 👤 | ■ | | | |
| Pakistan | 🤖 | | ■ | | |
| Philippines | 👤🤖 | ■ | ■ | | ■ |
| Poland | 👤 | | ■ | | ■ |
| Russia | 👤🤖⚙ | ■ | ■ | ■ | |
| Saudi Arabia | 🤖⚙ | | | ■ | |
| Serbia | 👤🤖⚙ | ■ | ■ | | ■ |
| South Africa | 👤🤖⚙ | ■ | | | ■ |
| South Korea | 👤🤖⚙ | ■ | | | ■ |
| Syria | 🤖 | | | | ■ |
| Taiwan | 👤🤖⚙ | ■ | | ■ | |
| Thailand | 👤🤖 | | | | |
| Turkey | 👤🤖 | ■ | ■ | | ■ |
| Ukraine | 👤🤖 | | ■ | | ■ |
| UAE | 👤🤖 | | ■ | | ■ |
| United Kingdom | 👤🤖 | | | ■ | ■ |
| United States | 👤🤖⚙ | | | ■ | ■ |
| Venezuela | 👤🤖 | ■ | | ■ | |
| Vietnam | 👤 | ■ | | | |
| Zimbabwe | 👤 | | ■ | | |

Source: Authors' evaluations based on data collected. Note: This table reports on the messaging and valence strategies of cyber troops. A filled box indicates evidence found. For fake account types: 👤 = human accounts; 🤖 = automated accounts; ⚙ = cyborg accounts; = no evidence found.

Cyber troops also employ a suite of communication strategies: they may amplify messages by creating content, posting forum comments, replying to genuine or artificial users, suppress other content, by launching targeted campaigns to falsely mass-report legitimate content or users, so that platforms are temporarily or permanently removed from the site. The authors comment that, logically, automated accounts can be found on platforms that make automation easy, namely, Twitter, but that cyber campaigns take place across all common forms of social media. They also note that 1/5 of the countries had evidence of disinformation campaigns operating over chatting applications (WhatsApp, WeChat, Telegram, etc.), many of which are located in the Global South, where these applications are prevalently used, and large public group chats are widespread.

This report also shares the size, resources, status (permanent or temporary), level of coordination, and capacity of countries' cyber troop capacity. The size of cyber troop teams can range from dozens (in countries such as Argentina or Kyrgyzstan), thousands or tens of thousands (UK and Ukraine, respectively) or even millions (China). Countries differ in the budgets they allocate towards computational propaganda, the degree to which their teams coordinate with other firms and actors, and how often they operate, be it full-time and year-round, or just around critical dates or occasions, like elections.

**Table 4: Cyber Troop Capacity**

| Country | Team Size | Resources | Status | Coordination | Capacity |
|---|---|---|---|---|---|
| Angola | Newly Formed | .. | Temporary | Low | |
| Argentina | 30-40 | Multiple contracts valued at 14 million Pesos and 11 Million Pesos | Previously Active | Low | |
| Armenia | Newly Formed | .. | Temporary | Low | |
| Australia | 900 | .. | Temporary & Permanent | Low | |
| Austria | .. | .. | Temporary | Low | |
| Azerbaijan | 50,000 | .. | Temporary | Low | |
| Bahrain | .. | .. | Permanent | Low | |
| Brazil | 60 | Multiple contracts valued at R10,000,000, R130,000 R24,000 | Permanent | Medium | |
| Cambodia | .. | .. | Temporary & Permanent | Low | |
| China | 300,000-2,000,000 | .. | Permanent | High | |
| Colombia | .. | .. | Temporary | Low | |
| Cuba | .. | .. | Permanent | Medium | |
| Czech Republic | .. | .. | Permanent | Low | |
| Ecuador | .. | Multiple contracts valued at $200,000. | Permanent | Medium | |
| Egypt | .. | .. | Permanent | Low | |
| Germany | <300 | .. | Temporary & Permanent | Low | |
| Hungary | .. | .. | Permanent | Medium | |
| India | .. | .. | Temporary | Low | |
| Iran | 10,000-20,000 | .. | Permanent | Medium | |
| Israel | 400 | Multiple contracts valued at $778,000; $100,000,000 | Permanent | High | |
| Italy | Newly Formed | .. | Temporary | Low | |
| Kenya | Newly Formed | One contract valued at $6,000,000 | Temporary | Low | |
| Kyrgyzstan | 50 | Multiple contracts valued at $2000. $3-4 a day per person | Temporary | Low | |
| Malaysia | .. | .. | Temporary | Low | |
| Mexico | 1,000 | Multiple contracts, one valued at $600,000. €520 per month per person | Temporary | Medium | |
| Myanmar | .. | .. | Temporary & Permanent | Medium | |
| Netherlands | Newly Formed | .. | Temporary | Low | |
| Nigeria | .. | One contract valued at $2,800,000 | Temporary | Low | |
| North Korea | 200 | .. | Permanent | Medium | |
| Pakistan | Newly Formed | .. | Temporary | Low | |
| Philippines | 400-500 | Multiple contracts valued at $200,000+. | Permanent | Medium | |
| Poland | .. | .. | Temporary | Low | |
| Russia | 400-1000 | Annual Budget $10,000,000 | Permanent | High | |
| Saudi Arabia | .. | .. | Permanent | Medium | |
| Serbia | 100 | Average Monthly Salary €370 | Permanent | Medium | |
| South Africa | .. | One contract valued at $2,000,000 | Temporary | Low | |
| South Korea | <20 | .. | Previously Active | Low | |
| Syria | .. | Multiple contracts valued at $4,000 | Permanent | Medium | |
| Taiwan | .. | .. | Temporary | Low | |
| Thailand | Newly formed | .. | Permanent | Low | |
| Turkey | 6,000 | Multiple contracts, one valued at $209,000 | Permanent | Medium | |
| Ukraine | 20,000-40,000 | .. | Permanent | Medium | |
| UAE | .. | Annual Budget $10,000,000+ | Permanent | High | |
| United Kingdom | 1500 | Multiple contracts for elections, total value approximately £3,500,000 | Temporary & Permanent | Medium | |
| United States | .. | Multiple programs valued at $50,000,000; $200,000,000; $202,000,000 | Temporary & Permanent | High | |
| Venezuela | 500 | .. | Permanent | Medium | |
| Vietnam | 10,000 | .. | Permanent | Medium | |
| Zimbabwe | Newly formed | .. | Temporary | Low | |

**Source:** Authors' evaluations based on data collected. **Note:** This table reports on cyber troop size, resources, team permanency, coordination, and capacity. For capacity: = minimal capacity, = low capacity, = medium capacity, = high capacity

The report concludes by calling democracies to take action by formulating guidelines to discourage bad actors from exploiting computational propaganda: "To start to address these challenges [outlined in the report], we need to develop stronger *rules and norms* for the use of social media, big data and new information technologies during elections." Notably, the terms "rules and norms" leave ambiguity with respect to those who should be developing, implementing, and enforcing said reforms: social media platforms or governments? This was likely intentional, as the conversation around who should regulate speech in democracies warrants a paper in its own right.

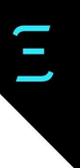

# Can WhatsApp Benefit from Debunked Fact-Checked Stories to Reduce Misinformation?

**([Original paper](#) by Julio C. S. Reis, Philipe de Freitas Melo, Kiran Garimella, Fabrício Benevenuto)**

Facebook may own WhatsApp, but it is different from that of typical social media sites such as Facebook and Twitter. WhatsApp has end-to-end encryption that has made this app unique in communication with others. WhatsApp has over 1.5 billion users and has become a source for sharing news in countries like Brazil and India, where smartphone's use for news access is higher than other devices (Reis et al., 2020). This research study focuses on Brazil and India's two countries and how misinformation has affected the democratic discussion in these countries. There are over 55 billion messages sent a day, with about 4.5 billion messages are images (Reis et al., 2020). Due to the nature of encryption, there is no way that WhatsApp monitors or flags inappropriate or potentially dangerous or fake images as Facebook has the capability of doing. The researchers propose an approach with machine learning, where WhatsApp can automatically detect when a user shares images and videos that have previously been labeled as misinformation with the Facebook database. This would abide by the E2EE and not compromise the encryption or privacy of the user (Reis et al., 2020).

Facebook already has a lot of partnerships with fact-checking agencies around the world, and so the database would not be difficult to obtain. Algorithms would be implemented for hashing and matching similar media content. "A hashing algorithm provides a signature to represent an image or video" (Reis et al., 2020). The researchers were focused on two types of hash functions for this proposal. The first being cryptographic has and the second being perceptual hash. A cryptographic hash is a one way hash function based on techniques like MD5 or SHA and processes a string has given an image. It would be used to identify exact matches only, whereas the perceptual hash could identify similar images and be notified even if the image was altered (Reis et al., 2020).

There are already multiple algorithms, including Facebook PDQ hashing, that allows this to be done.

Another part of this model would be once a user intends to send an image, WhatsApp checks whether it is already in the hashed set. If so, the warning confirmation asks if the user wants to share this information (Reis et al., 2020). When the recipient user gets the message, WhatsApp decrypts the image on the phone, obtains a perceptual hash, and the content is then flagged if it is in the already checked database (Reis et al., 2020). The warning message would also include where the item was already fact-checked.

This new method could also be a benefit for Facebook as they could collect data on how many times a match occurred and establish the prevalence and virality of different types of misinformation and collect information about the users who repeatedly send such content (Reis et al., 2020).

With this idea in mind, the researchers went ahead and tested it in Brazil and India. They had 17,465 users in Brazil, with 34,109 images and 63,500 users in India with 810,000 images. The dataset they used was publicly available.

In the study, the fact-checked images by crawling all images from popular fact-checking websites from Brazil and India. Then, they obtained the date in which they were fact-checked. Next, they used Google reverse image search to check whether one of the main fact-checking domains were returned. If the image passed their test, it was added to the last collection, which has over 100,000 facts checked pictures from Brazil and about 20,000 from India (Reis et al., 2020).

Next, they used the PDQ hashing to implement their algorithm of clustering similar or identical images together.

In their findings, the results showed that 40.7 percent of the misinformation images in Brazil and 82.2 percent of the misinformation image shares in India could have been avoided by flagging the image and preventing it from being forwarded after being fact-checked (Reis et al., 2020).

This study shows just how important it is for technology companies to inform their users of the information they are sending and make an educated decision on what information they want to spread to others.

# Go Wide: Article Summaries

### Facebook Groups Are Destroying America
([Original *Wired* article](#) by Nina Jancowicz, Cindy Otis)

Facebook groups are a great place to meet like-minded people, but in this case, birds of a feather flocking together might be a bad idea, especially as it relates to the spreading of disinformation. We have covered this topic extensively in the past, one of the things that we appreciated about this article was how closed groups that congregate around special topics can become sources for amplifying disinformation. A concern that we've been studying at the Montreal AI Ethics Institute as it relates to disinformation is how these closed groups tend to become a hindrance for researchers who are trying to study the effects that they have on spreading disinformation. Just as is the case with P2P messaging apps where researchers don't have access to the larger network effects on how disinformation is propagating through the network, closed groups also hamper the ability of researchers to study more deeply that these closed groups have on Facebook. One of the suggestions made by the researchers quoted in the article asks for Facebook to change the privacy settings of the group to public once it exceeds a certain size so that researchers are better able to study them.

By working through some of the groups as members, the researchers were able to see how bad actors created Sybil-styled personas that tailored the messaging to the group and helped to propagate disinformation through gullible members of that group who believe the information from the group to be coming from a trusted and safe space because of the closed nature of the group. Another suggestion made by the researchers is to have the platforms provide more transparency on the ownership and management of the groups so that users are able to see patterns when there might be bad actors who are using several groups to coordinate and amplify their disinformation campaigns. While Facebook has taken steps to stop providing other suggested groups related to conspiracy and other topics, they still show up in the Discover tab. Researchers suggest that an approach to mitigate this would be to bar them from appearing at all so that users would have to search from them explicitly which would limit the amplification of these disinformation campaigns.

## Climate Denial Spreads on Facebook as Scientists Face Restrictions
(Original *Scientific American* article by Scott Waldman)

In a time when the Earth has had a brief respite from carbon emissions due to a slowdown in human activity, leveraging disinformation tactics online to further polarize the climate change debate is particularly problematic. Facebook has weakened protections against the spread of climate change related disinformation. The fact-checking mechanisms that are put in place to prevent the abuse of the platform faced some serious threats in terms of what might be a bad precedent for the future of the platform when Facebook overruled the decision provided by an independent fact-checking body. Essentially, it seems that they are now allowing disinformation to spread if they are labeled as opinions which leads the content moderation space into a very slippery slope.

A leading climate scientist's posts were branded as political thus requiring her to provide more private information before she will be allowed to promote and publish posts that she creates which are educational posts debunking some of the climate change related disinformation on the platform. It places an undue burden on scientists who do this sort of work in public interest while organized groups that target the most susceptible populations on the platform with coordinated disinformation campaigns run rampant without any oversight.

In another case where the platform had reversed the label of "false" on a piece of content after it supposedly received pressure from a conservative that moved them to make the change in the label that was provided by a fact-checking organization in their verified network, observers are wondering if these are one-off incidents or this is a signal towards a broad policy change. In revising how Facebook governs content related to climate change under the "environmental politics" category, the platform is placing undue burdens on good samaritans on the platform while allowing for malicious actors to take advantage of the lax policies allowing disinformation to run unchecked harming millions.

## How Fake Accounts Constantly Manipulate What You See on Social Media – and What You Can Do About It
(Original *The Conversation* article by Jeanna Matthews)

In times of discord, issues can be used by malicious actors to polarize and further divide society via social media. Fake accounts that are operated by these actors

The State of AI Ethics, October 2020                                                                          55

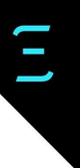

who seek to spread disinformation go by the moniker of "sock puppets" because they are operated by someone else's hand. Sometimes the deception is easy to spot by looking at the account history and the pattern of use.

But, as covered in the learning communities at the Montreal AI Ethics Institute, there are more sophisticated information operations that make this deception much harder to unearth. Even for those who study this phenomenon, often it is hard to separate truth from fiction. When there is a legitimate middle ground, divisiveness in society weakens our democratic institutions. Platform companies that have the most power to correct this imbalance have the fewest incentives to invest in addressing this problem because divisive content is good for business. Sometimes, this failure to act gets attributed to impinging on freedom of speech. But, this freedom of speech issue creates many other problems that are detrimental to social order. The author of this article calls for readers to use social media sparingly, just as we would control our use of addictive substances.

## Deepfakes Are Becoming the Hot New Corporate Training Tool
(Original *Wired* article by Tom Simonite)

Deepfakes are being used to create corporate training videos. In the current pandemic where filming in-person is not possible, the use of such tools offers a unique opportunity to leverage this technology. The agencies providing AI-enabled solutions point to the benefits that such generated content has in terms of providing more diversity in both the voices and appearances of the avatars in the videos.

Typically for small, resource-constrained organizations, they would be limited in their visual story-telling capabilities when they don't have access to large creative agencies. But, the use of such services levels the playing field and allows smaller organizations to compete in the marketing arena with bespoke visuals. Another benefit of utilizing generated digital avatars is that they enable seamless translation into multiple languages allowing locally-tailored content.

The avatars utilize the likeness of real people. These people are compensated based on the amount of footage used in creating these videos. One of the CEOs interviewed also pointed to the benefit of widening beauty standards as the videos will allow for more diversity in the avatars compared to real actors. The one caveat that researchers emphasized is how such efforts might lull companies into a false sense of accomplishment in achieving diversity while creating little, real change on the ground.

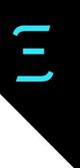

## Memes, the Pandemic and The New Tactics of Information Warfare

**(Original *C4ISRNET* article by Mark Pomerleau)**

The pandemic has brought on new dimensions to this problem by equipping malicious actors with more tools that they can wield in their information operations. Researchers point out that adversaries don't even need to create new content, but merely amplify existing divisive content in a host country to achieve their aims. It has the benefit of leveraging contextual and cultural sensitivity that outsiders would never understand, thus making the material much more effective. Secondly, using what is called memetic warfare, adversaries use compact, visually-appealing content that has a quick, emotional punch to convey their message to their target audiences. Since memes are typically not signed by their creators, they evoke an emotional response, and are susceptible to being shared very widely and quickly.

State-backed adversaries also use this tactic to inflict harm and derail genuine efforts made by a nation towards building a positive society. They often play both sides of the debate, sowing confusion and eroding the trust that the public has in their governments. Finally, sometimes bureaucracy can become an internal enemy by slowing down responses to the spread of disinformation. By the time the disinformation gets debunked, the damage has taken place, and adversaries have succeeded in their goals.

## "Outright Lies": Voting Misinformation Flourishes on Facebook
**(Original *ProPublica* article by Ryan McCarthy)**

Voting is one of the fundamental pillars of democracy, which is why Facebook includes voting-related misinformation to be a category of its own whereby content that misrepresents voting violates their community standards and is liable to be taken down. With a major election coming up this year in the US, researchers have found that the platform is rife with misinformation. Many posts that violate their community standards are left unattended without any action. The platform has kicked the can down the road in terms of harsher clampdowns on misinformation on its platform. Some of the posts blended opinions and factual errors, outright lies about voting, or misleading claims about the reliability of different voting methods.  While they are committed to providing accurate information about voting, they also commit to free speech rights, and both of these might conflict.

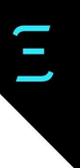

The platform has made the concession that they are considering banning political ads from the platform (which has recently been put into effect leading up to Election Day in the US), but research has found that misinformation spreads more through posts and political ads largely contain factually correct information, thus limiting the impact of the ban on political ads in addressing the underlying issue. When misinformation spreads and even those who are skeptical of it see it proliferating the platform, there is a sense of disenfranchisement among those who are explicitly targeted, often people of color. Sowing distrust in the communities is an effective way to depress voter turnout and skew the results of the election. Slow action by contracted fact-checkers also gives enough spotlight to misinformation that it achieves its goals well before the content gets taken down.

### Hackers Broke Into Real News Sites to Plant Fake Stories
(Original *Wired* article by Andy Greenberg)

The success of disinformation lies in its ability to deceive the intended audiences into believing the content is trustworthy and truthful. While it might be hard to generate content and have it spread via social media to gain enough legitimacy to move the needle, there is tremendous potential for harm if such content gets wrapped in the veneer of respected news outlets. In particular, information operatives have leveraged the technique of hacking into content management systems (CMS) of media houses to plant fake news stories.

Researchers identified the use of this technique leading to divisiveness in eastern Europe, targeting NATO related news items along with relations of those countries with the US. While there are reporting mechanisms, if the content stays up long enough, it has the potential to be copied and impact people nonetheless. Emphasis on cybersecurity becomes critical in combating such attacks. In addition to fighting disinformation along the lines of content, provenance, and other dimensions, all of these efforts get subverted if appropriate guards are not in place.

### AI-Generated Text Is the Scariest Deepfake of All
(Original *Wired* article by Renee DiResta)

This article highlights why AI-generated text might be the deepfake that we must pay attention to in our fight against disinformation. Specifically, the pervasiveness of deepfake text across all channels and relative lightweight analysis done on it so far makes it quite dangerous. There are two kinds of media that needs to be moderated, one that is synthetic and one that is merely modified, that is, altered taking original content slightly to create something that is meant to deceive.

One of the first problems with text deepfakes is that we must now start to tune not only our content moderation tools to detect them, the first line of defense, but also become more discerning consumers of content. There has been press coverage talking about video and audio deepfakes which will hopefully sensitize consumers to watch out for them. But, there are very limited examples covered in popular media that talk about this. With the recent GPT-3 model, there have been some impressive demonstrations that have been put forward by those who were given beta access to the models. While video and audio deepfakes might be deployed at strategic moments to trigger particular outcomes, say compromising footage before an election, text poses larger problems in that it could be used gradually over time on public fora to sway opinions in the form of a "majority illusion" that shifts public opinion on a subject subtly and without checks.

Better media and digital literacy seem to be our best defenses rather than technological solutions that risk becoming outmoded in an arms race.

## Why Wikipedia Decided to Stop Calling Fox a 'Reliable' Source
([Original *Wired* article](#) by Noam Cohen)

For an encyclopedia that is built by community members, volunteers, and run by a non-profit organization, the impact that Wikipedia has had on the knowledge ecosystem of the world is immense despite strong criticism that it isn't thorough enough. Given how many people visit the website to obtain information and how highly it ranks in search engine results, it is essential that information there is as accurate as possible. But this can be hard to achieve for topics related to politics because the volunteer writers/editors don't always reach a consensus.

Entries related to Karen Bass received a lot of attention in recent news cycles and editors were working in overdrive to ensure that the information included in her Wikipedia entry was accurate. In some places where Fox News was cited as a source, the editors took that information down, citing that it wasn't a reliable source on that subject. The community editors have decided that instead of doing post-hoc verification, they are branding sources as credible or not in different subject areas. For example, while Fox News is not considered by them to be a reliable source for politics related items, it continues to be maintained as a potential source in other areas. Such an approach helps to avoid bickering and confusion in the distributed editing workflow for the encyclopedia and helps to keep moving things forward.

# 4. Humans & AI

**Opening Remarks** by **Amba Kak (Director of Global Policy & Programs, New York University's AI Now Institute)**

The crucial question, of course, is: *which humans*? As this chapter explores, the universal, abstract "human" subject of AI ethics discourse is being challenged by a growing body of research and advocacy. Abstraction is replaced by research grounded in 'real life' contexts, the intersecting marginalizations of individuals and communities, and the power dynamics that mediate interactions between people and automated systems.

Recent research on disability and AI directly confronts the ways in which AI systems routinely produce and enforce universalized standards of "normal" and "ability" attributed to being 'human'. The [paper](#) summarized in this chapter identifies the costs of being an "outlier" to these standards. Disability rights movements' have historically fought to identify and dismantle structural bias across physical and digital spaces, and this paper would benefit from being read in conjunction with [research](#) that draws from these lessons to chart a path forward for advocacy.

Universalizing narratives around AI and "the future of humanity" often concede too much power to these systems and those that build and profit from them. The recent Netflix documentary *The Social Dilemma* brought the topic of technology dystopia to dinner table conversations around the world. It left many with the feeling that social media algorithms "knew them better than they knew themselves" — no wonder they were hooked. Two papers in this chapter provide a powerful and necessary antidote to this pessimism. The [paper](#) on the Ubuntu framework argues that "reducing humans to data points" is doomed to be an incomplete process that will fail to capture the wholeness of a person and their myriad individual and group identities. The final [article](#) reminds us that the data capture by firms is optimized for their own profit, structured not to "know" us as much as to categorize and profile us *as consumers* in pursuit of the firms' pre-defined goals of user retention or maximizing time spent on the platform. Narratives that endorse the all-knowing platforms can, therefore, concede too much power. As Donna Haraway argued in her seminal paper on Situated Knowledges in 1988 *"that view of infinite vision is an illusion, a god trick"*.

Another article challenges the now-popular understanding that machine learning algorithms can replicate "human thinking". They argue that anthropomorphizing these systems can ignore the *"context and life-long learning that is done by humans that gives them perceptual ability"*. The paper demonstrates how this reductionist thinking has led to overestimating the efficiency and accuracy of these systems. Recent research on facial recognition testing systems points to similarly poor performance in real-life contexts (as compared to experimental settings), even as the industry continues to rely on these standards to back claims of accuracy.

The "human-in-the-loop" discourse is another example of where an abstract understanding of human intervention by those deploying these systems can lead to tokenistic policy solutions for addressing the harms of automation. The paper in this chapter is part of a growing body of research (see here, for example) which argues that AI governance must contend with the full socio-technical system of human-algorithm collaboration, rather than consider the algorithm or human in isolation.

And what of the humans whose labour sustains the globally distributed supply chain of many modern AI systems? More recent research, like the article in this volume, focuses on studying the material conditions in which workers provide services like data labeling and content moderation, with a concentration in the Global South. This conversation should more proactively forefront questions of the global political economy of AI production, and the complex ways in which it impacts workers' livelihoods and national economies in different parts of the world. As scholar Noopur Raval recently opined on social media, metaphors like "ghost" or "shadow" workers while often used in good faith to illuminate poor working conditions that are hidden from mainstream Western gaze, can inadvertently remove agency from workers who are recognized in their own local contexts.

Amba Kak (**@ambaonadventure**)
Director of Global Policy & Programs, NYU's AI Now Institute

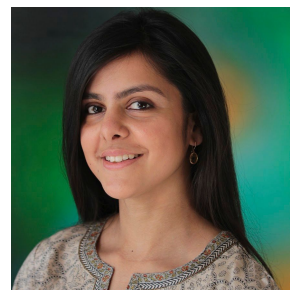

Amba is the Director of Global Policy & Programs at New York University's AI Now Institute where she develops and leads the institute's global policy engagement, programs, and partnerships, and is a fellow at the NYU School of Law. She is also on the Strategy Advisory Committee of the Mozilla Foundation.

# Go Deep: Research Summaries

## Aligning Superhuman AI with Human Behavior: Chess as a Model System
([Original paper](#) by Reid McIlroy-Young, Siddhartha Sen, Jon Kleinberg, Ashton Anderson)

Artificial Intelligence (AI) is becoming smarter and smarter every day. In some cases, AI is achieving or surpassing superhuman performance. AI systems typically approach problems and decision making differently than the way people do (McIlroy-Young, Sen, Kleinberg, Anderson, 2020). The researchers in this study (McIlroy-Young, Sen, Kleinberg, Anderson, 2020) created a new model that explores human chess players' behavior at a move-by-move level and the development of chess algorithms that match the human move-level behavior. In other words, the current system for playing chess online is designed to play the game to win.

However, in this research study, they found a way to align the AI chess player to play in a more human behavioral manner when making decisions on the next move in chess. They found by applying existing chess engines to the data they had did not predict human movements very well. Therefore, their system called "Maia" is a customized version of Alpha-Zero trained on human chess games that predict human moves with a higher accuracy rate than existing engines that play chess. They also achieve maximum accuracy when predicting the decisions made by players at specific skill levels. They take a dual approach when designing this algorithm. Instead of asking, "what move should be played?" They are asking, "What move will a human play?"

The researchers were able to do this by repurposing the Alpha Zero deep neural network framework to predict human actions rather than the most likely winning move. Instead of training the algorithm on self-play games, they taught it on human games that were already recorded in datasets to understand how humans play chess. The next part was creating the policy network that was responsible for the prediction. From this, "Maia" was built and has "natural parametrization under which it can be targeted to predict human moves at a particular skill level" (McIlroy-Young, Sen, Kleinberg, Anderson, 2020).

The second task for "Maia" they developed was to figure out when and whether human players would make a significant mistake on their next move, called "blunder."

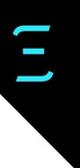

For this study, they designed a custom deep residual neural network and trained it on the same data. They found that the network outperforms competitive baselines at predicting whether humans will make a mistake (McIlroy-Young, Sen, Kleinberg, Anderson, 2020).

By designing AI with human collaboration in mind, one can accurately model granular human decision making. The choices developers make in the design can lead to this type of performance. It can also help understand the prediction of human error (McIlroy-Young, Sen, Kleinberg, Anderson, 2020).

## Toward Fairness in AI for People with Disabilities: A Research Roadmap

([Original paper](#) by Anhong Guo, Ece Kamar, Jennifer Wortman Vaughan, Hanna Wallach, Meredith Ringel Morris)

In this position paper, the authors identify potential areas where Artificial Intelligence (AI) may impact people with disabilities (PWD). Although AI can be extremely beneficial to these populations (the paper provides several examples of such benefits), there is a risk of these systems not working properly for PWD or even discriminating against them. This paper is an effort towards identifying how inclusion issues for PWD may impact AI, which is only a part of the authors' broader research agenda.

The authors note that this systematic analysis of interactions between PWD and AI is not an endorsement of any system, and that there may exist an ethical debate of whether some categories of AI should be built. It's important to note that this analysis is a starting point towards this theme, and may not be exhaustive.

The paper is then divided into considerations across many AI functionalities and possible risks when used by PWD. The authors covered computer vision (identification of patterns in still or video camera inputs), speech systems (systems that recognize the content or properties of speech or generate it from diverse inputs), text processing (understanding text data and its context), integrative AI (complex systems based on multiple models), and other AI techniques.

**Computer Vision** – Face Recognition: The authors hypothesize that such systems may not work well for people "with differences in facial features and expressions if they were not considered when gathering training data and evaluating models". For example, people with Down syndrome, achondroplasia, or cleft/lip palate. Systems may also malfunction for blind people, who may not show their faces at an expected angle or who may use dark glasses. Finally, emotion and expression processing algorithms may malfunction for someone with autism, Williams syndrome, who suffered a stroke, Parkinson's disease or "or other conditions that restrict facial movements".

**Computer Vision** – Body Recognition: "Body recognition systems may not work well for PWD characterized by body shape, posture, or mobility differences". Among some examples, the authors point to people who have amputated limbs or someone who experiences tremors or spastic motion. Regarding people with differences in movement, systems may malfunction for "people with posture differences such as due to cerebral palsy, Parkinson's disease, advanced age, or who use wheelchairs". The paper cites an Uber self-driving car accident, in which the car hit someone walking a bicycle.

**Computer Vision** – Object, Scene, and Text Recognition: Many of these systems are trained in high quality pictures, usually taken by sighted people. It's to expect that these systems may malfunction while trying to detect objects, scenes, and texts from images taken by a blind user, or someone who has tremors or motor disabilities.

**Speech Systems** – Speech Recognition: Automatic Speech Recognition (ASR) may not work well for "people with atypical speech". It's known that such systems work better for men than women, while malfunctioning for people of very advanced ages or with stronger accents. The authors point to speech disabilities, such as dysarthria, that need to be taken into consideration for a fair construction of those systems. Further, ASR locks out people who cannot speak at all.

**Speech Systems** – Speech Generation: Systems may include Text To Speech (TTS) technologies. These systems may be challenging for people with cognitive or intellectual disabilities, who may require slower speech rates.

**Speech Systems** – Speaker Analysis: These systems can identify speakers or make inferences about the speaker's demographic characteristics, potentially being used for biometric authentication. These systems may malfunction for people with disabilities that impact the sound of their speech. Further, Analysis trying to infer sentiments may fall short for austitic people.

**Text Processing** – Text Analysis: Some systems, such as spelling correction and query rewriting tools, may not handle dyslexic spelling. Moreover, since autistic people express emotion differently, systems that infern sentiments from text may also fall short for this population.

**Integrative AI** – Information Retrieval (IR): These are complex systems, such as the ones that power web search engines. It is possible that IR amplifies existing bias against PWD. For example, search results can return stereotypical content for PWD, while targeted-advertising may eventually exclude PWD from products or even employment opportunities.

**Integrative AI** – Conversational Agents: These agents are present in various services, such as healthcare and customer service. These systems may amplify existing bias in their results, if not trained properly. Further, people with cognitive disabilities may encounter poor experience while utilizing these services. It is important that these systems can adapt to the users' needs, such as reduced vocabulary or expression in multiple media.

**Other AI Techniques:** For example, outlier detection. These systems usually flag outlier behaviour as negative, tied to punitive action. For example, input legitimacy (use of CAPCTHAs or other mechanisms to separate humans from bots), may not work well for people with atypical performance timing, such as someone with motor disabilities or visual impairments.

The authors exposed in this opinion paper ways in which AI can negatively affect PWD, which usually reflects in a worse quality of service, underrepresentation, or stereotyping for these populations. Some of the cases mentioned in the paper are hypothesis, while some are backed up by evidence. The authors also propose a broader research roadmap for AI fairness regarding PWD, including testing the hypotheses presented, building representative datasets, and innovative new AI techniques "to address any shortcomings of status quo methods with respect to PWD".

## From Rationality to Relationality: Ubuntu as an Ethical & Human Rights Framework for Artificial Intelligence Governance
([Original *Harvard University* discussion paper](#) by Sabelo Mhlambi)

If one word were to describe this paper, it would be 'Relationality'. This word is pivotal and of paramount importance to the Ubuntu ethical framework. We are no longer to recognise personhood in terms of 'rationality', but in terms of relationality. What this move does is to recognise humans as fundamentally communal, and treat this as a point of departure when considering how to govern AI, no longer considering personhood (for both humans and machines) with the benchmark of rationality, but rather in their relations to other facets of existence. With the Ubuntu framework (found in Sub-Saharan and Southern Africa), the mark of a person is acknowledging that personhood cannot be treated as an individual milestone, but something that is prolonged and realised by how someone represents and interacts with their community. How this came about, how this all fits together, and how this challenges the Western ethical strata which we apply to our governance of AI, I will endeavour to explain now.

Broken down etymologically, 'ubu' "indicates a state of being and becoming" (Mhlambi, 2020, p. 14), and 'Ntu' "evokes the idea of a continuous being or becoming a person" (Mhlambi, 2020, p. 14), an incomplete person then becomes a complete person through serving their community. To clarify, personhood is a state that a person continually possesses through its commitment to relationality, rather than a quality possessed by a human, such as rationality. As a result, personhood can be taken away, as well as regained, determined against the three inter-dependent pillars of Ubuntu: social progress, social harmony, and human dignity. Crimes committed against these (such as algorithmic bias marginalising certain communities) renders a person (or potentially a machine) "Akumu-Ntu lowo", meaning 'not a person'. The entity in question is still a person biologically (or is still a machine in terms of mechanics), but no longer a possessor of the state of personhood due to their crimes against the community that have harmed 1 or more of the three pillars.

With these three pillars, AI's ability to possess personhood would then need to be considered in terms of its service to the community it's involved in. Its personhood state would be lost through its digital surveillance stripping away human dignity by eradicating the idea of privacy, while its analysis reducing humans to data points eliminates its relationality to other humans in the process. The prioritisation of the individual in this way harks back to the relentless reference to the individual in terms of personhood, namely via treating rationality as a quality instead of a state. Instead, the Ubuntu framework would aim to guide AI governance towards conducting AI in terms of its relation to the interconnectedness of society. Not only is AI to serve human dignity, but also social progress and social harmony. Such maxims include the environment and those that reside in it, and so treating the AI as a person through being individually rational will only serve to negate its reach to the environment, and to those who reside in it, including marginalised groups. AI's ability to influence nearly all facets of humanity points towards a need for a framework which can account for this interconnectedness, and that is what Mhlambi sees in Ubuntu.

Nevertheless, this framework acknowledges that although humans rely on one another to be considered a person (whereby AI would need a similar affirmation), humans have liberty in their choices. "Umu-Ntu ngumu-Ntu nga ba-Ntu" (Mhlambi, 2020, p. 19), meaning a person is a person, emphasises how both bad and good acts can be committed, which could then be reflected in the AI models deployed. Examples of this explored by Mhlambi include surveillance capitalism, and data colonialism, both brought about by the human decision to focus on the individual. This will be my last point of focus.

Surveillance capitalism is the practice of observing as many aspects of human behaviour as possible, in order to predict and then manipulate such behaviour. Examples such as predictive policing focus on the individual's data points and how they behave, stripping them of their human dignity, and treating them as separate from the whole (their community). Once this is done, both the individual without the whole, and the whole without the integral part that is the individual, both weaken. Data coloniality then seeks to alter the modes of perception of society, utilising data to define new social relations that are driven by the predatory, extractive and historical processes of colonialism justified in the name of computing. The surplus of data created by the technological revolution opened up the need for new data markets, just like the industrial revolution did for Britain. Like the colonizers of southern-Africa creating the demand for technology on the continent justified in the name of progress, companies are looking for new data to be collected to fuel their algorithms justified in the name of computing. Ubuntu can then offer a defence against such problems. As previously mentioned, data surveillance removes the individual from the whole and strips their privacy away, rendering data surveillance as unsustainable. Data coloniality is then to be seen in the same harmful light as its predecessor, whereby if left unchecked, marginalised communities will suffer. Namely, in the form of the mis-understanding of their experience through disproportionate representation in big-tech companies. Focussing on the individual through these methodologies negates the importance of the interconnectedness they possess with their community and most importantly, relationality with fellow humans. Not only are these methods invasive and presumptuous, they strip the very essence of personhood through individualism. AI is then to be governed accordingly, prioritising the connectedness possessed by humans, and their dependence on one another for their personhood, rather than as 7 billion individuals with data to be mined.

Overall, if our perception of personhood changes, so does AI. AI would no longer be set the benchmark of rationality, but rather perceived as a person in terms of its duties to the community. For AI to best serve the people, it needs to be considered in connection with the people, involving all corners of the people. It's time to embrace the Ntu part of Ubuntu, and immerse AI in the interconnected human ecosystem, rather than allowing it to be fractured by the dangers of surveillance capitalism, and data coloniality.

# A Case for Humans-in-the-Loop: Decisions in the Presence of Erroneous Algorithmic Scores

([Original paper](#) by Maria De-Arteaga, Riccardo Fogliato, Alexandra Chouldechova)

The paper highlights the risks of full automation and the importance of designing decision pipelines that provide humans with autonomy, avoiding the so-called token human problem when it comes to human-in-the-loop systems. For example, when looking at the impact that automated decision aid systems have had on the rates of incarceration and decisions taken by judges, it has been observed that the magnitude of impact is much smaller than expected. This has been attributed to the heterogeneity of adherence to these decision aid system outputs by the judges.

There are two phenomena that are identified: algorithmic aversion and automation bias. In algorithmic aversion, users don't trust the system enough because of prior erroneous results and in automation bias, users trust the system more than they should ignoring erroneous cases.

There are also other errors that arise in the use of automated systems: omission errors and commission errors. Omission errors occur when humans fail to detect errors made by the system because they are not flagged as such by the system. Commission errors are the case when humans act on erroneous recommendations by the system, failing to incorporate contradictory or external information.

One of the case studies that the paper considers is to look at child welfare screening systems where the aim is to help streamline the incoming case loads and to determine whether they warrant a deeper look. What they observed that was noticeable was that the humans that were being assisted by the system were better calibrated with the assessed score rather than the score that they were shown by the system. In screening-in cases, especially even when the scores shown by the system were low, the call workers were incorporating their experience and external information to include these cases rather than ignoring them as recommended by the system. Essentially, they were able to overcome omission errors by the system which showcases the power of empowering users of the system with autonomy leading to better results rather than relying on complete automation. The study conducted by the authors of the paper showed higher precision in post-deployment periods: meaning that more of the screened-in referrals were being provided with services which demonstrated that this combination of humans and automated systems where humans have autonomy led to better results than just using humans alone or relying fully on automated systems.

One of the important things highlighted in the paper is that when inputs related to previous child welfare history were being miscalculated, because of the degree of autonomy granted to the workers allowed them access to the correct information in the data systems, it allowed them to take that into consideration, enabling them to take better informed decisions. But, this was only possible because the workers prior to the conduction of this study had been trained extensively in handling these screen-ins and thus had experience that they could draw on to make these decisions. They had the essential skills of being able to parse through and interpret the raw data. On the other hand, cases like the catastrophic automation failures like with the Air France flight a few years ago when the autopilot disengaged and handed back control to pilots, the decisions that were made were poor because the human pilots never had training without the assistance of the automated system which limited not only their ability to take decisions independent of the automated system but also their wherewithal to judge when the system might be making mistakes and avoid the omission and commission errors.

The authors conclude by mentioning that designing such automated systems in a manner such that humans are trained to not only acknowledge that the system can make errors but also know how to fall back to "manual" methods so that they are not paralyzed into inaction.

## Learning to Complement Humans
([Original paper](#) by Bryan Wilder, Eric Horvitz, Ece Kamar)

The domain of humans and machines working together has been studied in various different settings but this paper takes a particular look at how this manifests itself in the context of machines having to make a decision on when to defer to human decision-making and then combining human and machine inputs.

They look at both discriminative and decision-theoretic approaches to understanding the impacts of this approach and take into account as a baseline first constructing a predictive model for the task and then a policy for choosing when to query a human treating the predictive model as fixed.

There is a focus on complementarity because of the asymmetric nature of errors made by humans and machines. This becomes even more important when the ML model has limited capacity and we need to make the choice to focus its efforts on particular parts of the task. The way that this work differs from others in the domain is that they emphasize joint training that explicitly considers the relative strengths and weaknesses of humans and machines.

A discriminative approach is one where there is a direct mapping between the features and the outputs without building intermediate probabilistic representations for the components of the system. Under relaxed assumptions as formulated in the paper, the authors take the approach of taking the decision from the human when the human is chosen to be queried before making a prediction. In a decision-theoretic approach, there is the possibility of following up on the first step with the calculation of the expected value of information from querying the human. Taking an inductive approach from the fixed value of information (VOI) system allows for the model to start from well-founded probabilistic reasoning and then fine-tune for complementarity with the human.

From the experiments run by the authors, it is clear that a joint approach leads to better results compared to the fixed counterparts when these are optimized for complementarity. Though, in deeper models, this approach is tied to the counterparts or makes modest improvements but never performs worse. An insight that the authors present is that in a lower capacity model, there is high bias and which makes aligning the training procedure with humans all the more important given that some errors will be inevitable. Another experiment that they ran on the CAMELYON16 dataset showed that the gaps increased significantly in an asymmetric costs regime, especially boding well for cases where there are particular kinds of errors that we would like to avoid in practice – for example missing diagnosis in medicine.

Finally, the authors conclude that the way the distribution of errors shift is such that the model is better able to complement the strengths and weaknesses of both humans and machines, something that we should try and include in the design of our machine learning systems if we are to make safer, and more robust systems.

# Go Wide: Article Summaries

## The Humans Working Behind the AI Curtain
(Original *Harvard Business Review* article by Mary L. Gray, Siddharth Suri)

Just as in the Wizard of Oz and Dorothy peeking behind the curtain to discover strings being pulled and shattering an illusion, a lot of AI companies who are trying to latch onto the hype surrounding AI, as covered before in our newsletters, utilize low-cost human workers who are asked to step into smoothen out the rough edges of AI systems. This labor is often in unregulated environments where there is limited training and guidance provided to the workers. Especially when they are called upon to moderate disturbing online content, review posts that have been flagged for violating community guidelines, and many other tasks that help to keep the internet from being overwhelmed with spam and undesirable content.

One of the discouraging things about the labor that backs this faux automation is that they are rarely compensated in a fair manner and they are hidden in the shadows so that companies can maintain the aura that they have well-functioning automation in place, when it is more akin to a human-backed fly-by-the-wire system that helps to smoothen out the problems that occur when the AI system doesn't function as intended or is non-existent to begin with. As AI is rolled out more widely, the author points out that this is the paradox of automation's last mile, giving a nod to the last-mile problem faced by delivery companies where the last-mile to deliver the product from the source to destination is the most challenging and often requires resorting to highly labor-intensive and manual processes.

Given the severe lack in training, work environment regulations, protocols for regulating content, there is a high risk of uneven enforcement of the policies of different platforms which harms the experience of users. This harm often tends to fall disproportionately on those who are already marginalized further stripping them of their ability to express themselves freely online. We need to make a strong call for the companies to be transparent about their labor practices so that consumers are able to effectively evaluate them for unfair practices and use their attention and dollars to steer the market towards better social outcomes.

## Challenges of Comparing Human and Machine Perception
([Original *The Gradient* article](#) by Judy Borowski, Christina Funke)

The authors of this research highlight some common pitfalls to help create more realistic expectations by identifying some of the challenges that arise when we try to compare machine perception performance against that of humans.

The first pitfall that they focus on is how humans are too quick to conclude that machines have learned human-like concepts. As an experimental setup, borrowing from the Gestalt theory of contour closeness, they design an experiment to analyze if indeed visual processing is done similar to how humans perceive visuals. What they find is that the machine models actually pick up cues in the data in a very different way compared to how humans process images and one of the measures to counter against coming to conclusions that humans and machine analogue systems works in the same way is to thoroughly examine the model, the decision-making process, and the dataset to arrive at conclusions that are rooted in empirical evidence rather than unnecessarily anthropomorphizing these systems.

The second pitfall that they identify is that it is hard to draw conclusions about the generalizing that the model can do beyond the tested architectures and training procedures. Analyzing same-different and spatial analysis tasks, they show that there is a problem in making unfair comparisons between human and machine perception because the baselines might be very different. They find that the low data regime isn't all that meaningful in making inferences about what the system performance is going to be like in the wild. The learning speed also greatly depends on the starting conditions of the system. What they mean by that is that there is a lot of context and life-long learning that is done by humans that gives them their perceptual ability and doing a baseline comparison with machines in that setting becomes really hard unless an analogous starting condition can be created for the machine as well.

Building on this, the final pitfall that the researchers explore is finding strong parallels in the experimental setups to compare humans and machines. When doing comparisons on drops in performance on recognizability when images are progressively cropped, aligning the testing conditions so that they are tailored to humans and to machines actually leads to conclusions that are quite different from existing studies.

Keeping these ideas in mind would lend a higher degree of realism to the expectations that we have from the systems and more pragmatically evaluate the capability of these systems.

## How Not to Know Ourselves
(**[Original *Data & Society: Points* article](#)** by Angela Xiao Wu)

Data from large, online platforms provide convenient data fodder for algorithmic analysis, especially given the penchant for deep learning to lean on big data for high predictive performance. Yet, this is plagued by what researchers in this article called "measurement bias." It refers to the collection of metrics that are fueled by the platforms' desire to run behavioral experiments in a bid to improve stickiness in user retention. Non-public actors increasingly control a large portion of this data with the sole intention of making profits and building their image.

The ascendance of platforms' power has left researchers asking critical questions on the representativeness of the data and privacy concerns. The captured data is "administrative data"; helping platforms meet goals of harvesting intermediation fees, advertiser revenue, and venture capital funding. The social science research implications are incidental to this process.

The subsequent data generation through user interactions is thus highly biased: it gets shaped by iterative experimentation carried out by the platforms to meet the goals stated above. Whether this data is a representation of user behavior or a result of the successful manipulation of users' behavior by the platform is hard to tell. Treating it as a representation of human society is reductionist. It risks putting platforms in a place where they are the entirety of the world, handing over even more power to these large, opaque organizations.

# 5. Labor Impacts

**Opening Remarks** by Thomas A. Kochan (George M. Bunker Professor, MIT Sloan School of Management)

AI represents one of the greatest challenges *and* opportunities facing workers and society today—and tomorrow. This chapter does an excellent job of outlining the risks of managing AI poorly and the opportunities for using it constructively. The key is to involve workers themselves in shaping how AI will affect their jobs and their privacy.

A proactive, proworker strategy starts by educating workers about how AI can be used to augment and improve work. Armed with this knowledge, workers should insist on having a voice in defining the problems AI is asked to address, not leaving problem definition to AI inventors or vendors. Then workers need to be involved in shaping the algorithms/rules that are built into AI models so they don't discriminate. Finally, those workers whose jobs are adversely affected need to be treated fairly. By touching all these key points this chapter provides a useful roadmap for using AI for the common good.

Thomas A. Kochan (**@TomKochan**)
George M. Bunker Professor, MIT Sloan School of Management

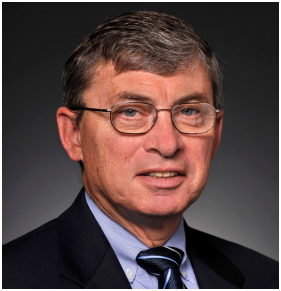

Thomas A. Kochan is the George Maverick Bunker Professor at the MIT Sloan School of Management and a faculty member in the MIT Institute for Work and Employment Research. Kochan focuses on the need to update America's work and employment policies, institutions, and practices to catch up with a changing workforce and economy. His most recent book is *Shaping the Future of Work: A Handbook for Action and a New Social Contract* (Routledge, 2020).

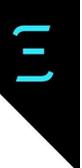

# Go Deep: Research Summaries

## On the Edge of Tomorrow: Canada's AI Augmented Workforce
([Original *ICTC* report](#) by Ryan McLaughlin, Trevor Quan)

Following the 2008 financial crisis, the pursuit of economic growth and prosperity led many companies to pivot from labor intensive to capital intensive business models with the espousal of AI technology. The capitalist's case for AI centered on potential gains in labor productivity and labor supply. The demand for AI grew with increased affordability of sensors, accessibility of big data and a growth of computational power. Although the technology has already augmented various industries, it has also adversely impacted the workforce. Not to mention, the data, which powers AI, can be collected and used in ways that put fundamental civil liberties at risk. Due to Canada's global reputation and extensive AI talent, the ICTC recommends Canada take a leadership role in the ethical deployment of this technology.

It is anticipated that by 2023, over 305,000 people will be working in the Canadian digital economy. However, it is also expected that many jobs that exist today will become obsolete at the hands of AI technology. The ICTC developed an AI Labor Augmentation Model to showcase those occupations that are most at risk of AI replacement. The Model found that jobs such as bookkeepers, payroll administrators and administrative assistants, which entail routine administrative tasks, are most at risk. According to the ICTC, 70% of people with occupations that are most likely to be replaced by AI are held by women. The Model also found that jobs such as psychologists, dentists and legislators, which require complex problem solving, judgment calls and qualitative analysis will remain the exclusive domain of humans.

In addition to concerns of labor replacement, the technology is also at risk of perpetuating bias and discrimination. This is because an algorithm is only as fair and just as the data from which it learns. For example, data on bond risk scores in the judicial system led AI software, trained on biased data, to be twice as likely to falsely label black defendants as future criminals rather than white defendants.

This risk becomes acutely problematic when the technology is not explainable or transparent and thereby, unaccountable. Canada is uniquely positioned for leadership in this field as it is one of the first countries with a national AI strategy, Canada ranks 3rd in the world for AI research, and has the largest proportion of AI ethics committees in the world. Canada also houses leading AI organizations in cities including Toronto, Montreal and Edmonton. As such, Canada should step into its role as a global leader on AI, since leadership over the technology's social implications is critical for AI to have a net positive impact.

## AI in Context: The Labor of Integrating New Technologies
([Original *Data & Society* report](#) by Alexandra Mateescu, Madeleine Clare Elish)

With the rise of automation and AI-powered systems, we appear to be ushering into a new technological revolution, where the future of human labour seems to be increasingly insecure. Economists have predicted the technologies will soon replace and take over millions of jobs in the US. In reality, the growth and spread of technological innovation have led to a tendency to oversell the merits and capacities of automated and AI-powered technologies, whose smooth operation continues to depend on human labour. Rather than being simply deployed, these technologies require substantial amounts of human labour to be successfully integrated and require new skills from those who work alongside them. The authors chose to study family-owned and retail grocery employees' daily experiences with automated tools, finding that AI technologies reconfigure work practices rather than replace workers. Their observations can help improve the development, assessment and regulation of AI technologies and monitor the unevenly felt risks of experimenting with these new technologies.

Agrotech, being agricultural technologies, is a booming multi-billion dollar industry, which promises to increase farming efficiency and crop yields while diminishing costs by relying on digital data. The pressures of climate change and global food insecurity have added to the commercial drive to invest in agrotech and precision agriculture. Through studying family-owned farms, the authors documented how the transition to data-intensive technologies, such as crop management tools and smart tractors, is neither as smooth nor as automatic as promised. For farmers in the US, adopting these technologies meant changing work routines and reconfiguring their barns and silos' physical infrastructure to be more apt to sensor readings and data collection. The field was reconceptualized as a complex dataset to be managed through digital tools by agrotech developers, who farmers felt had no first-hand knowledge of agricultural practices. Seemingly mundane issues like securing rural broadband internet and learning to use and interpret the data behind these technologies was a big barrier to farmers.

Due to the cyclical and weather-dependent nature of farming, it was difficult for farmers to correlate investments in these new technologies with higher crop yield. The integration and success of these agrotech tools were dependent on the social conditions, financial resources and labour-power of farmers, underscoring the number of elements that must be in place to secure the promised benefits of AI. For smaller family farms, their inability to undertake the financial risks of adopting these systems left them at an even greater disadvantage against large finance-backed conglomerates who could easily absorb the costs of experimentation. Larger agrotech companies, financial institutions and agricultural conglomerates are the ones who will profit from these technologies, leaving small farmers behind and vulnerable to exploitation by these new tech intermediaries and vendors.

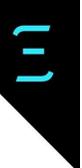

The opening of Amazon Go in 2018, with its automated checkouts and payment transactions, appeared to announce the beginning of a new phase in the retail industry. Retailers face mounting pressures to deploy artificial intelligence technologies to reduce costs and keep competing with chain killers such as Amazon and Walmart and e-commerce. For the competitive grocery industry, automated self-checkouts are seen as a necessary part of cost-cutting experiments. In their examination of grocery stores across the US, the authors detailed the considerable labour required to ensure automated self-service technologies' smooth operation. Frontline workers were needed to help keep self-checkouts and customer-operation scanners running. Rather than being rendered obsolete, frontline workers are charged with new responsibilities such as managing lines, checking for theft, assisting confused shoppers and so on. To retail owners, these new skills and tasks are often invisible and undervalued.

For workers themselves, these shifts in work roles and expectations were often unwelcomed. Grocery store workers are already experiencing a trend of casualization, which has led to decreased benefits and wages. Employees repeatedly complained of their store's pressure to encourage customers to use self-checkouts. Workers, whose performance was measured by speed and efficiency, had little time to provide a "human touch" to customers and expressed the feeling that they were becoming extensions of the machines they had to supervise. Often, they needed to teach themselves basic mechanical and software repairs to keep the flow of shoppers and machines operating. Although automated technologies did not erase frontline workers, it did lead to negative changes in their working conditions and responsibilities.

While the rise of automated technologies will not mark the end of human labour, it may usher in detrimental changes to America's working conditions. As demonstrated by the grocery retail industry, the emergence of new AI systems or automated technologies can create and reinforce forms of uncompensated, invisible and underpaid labour. The integration of these technologies, and their costs of experimentation, will lead to new economic divides in the farming industry, placing smaller-scale farmers at greater financial risks. As has been hinted by the authors, the rise of automation may widen the gaps between frontline workers and small-scale entrepreneurs and those in the financial industry and managerial positions. To avoid the negative impacts of automation, policy-makers will need to pay greater attention to those who will bear the costs of technological integration and experimentation and ensure that workers are properly compensated.

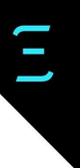

## Artificial Intelligence: The Ambiguous Labor Market Impact of Automating Prediction
([Original paper](#) by Ajay Agrawal, Joshua S. Gans, Avi Goldfarb)

The automation impacts of artificial intelligence have been the subject of much discussion and debate, often hampered by a poor demarcation of the limits of AI. In this paper, Agrawal, Gans, and Goldfarb build upon the framework that they introduced in their book Prediction Machines, that AI is fundamentally a prediction technology, and as it improves it will decrease the cost of prediction resulting in wider use of the technology in prediction tasks. The effect this has on labor then depends on the relative importance of prediction in a given job. They identify four possibilities.

The first is that many job tasks are pure prediction, for example forecasting work within operations departments, legal summary work done by paralegals, and email response work done by executive assistants, are all tasks that can be substituted by AIs as is, and possibly with greater efficiency, threatening these jobs if they do not have other high value-add tasks to do. A second possibility is that while a task may have a decision-component beyond prediction, this would no longer be important if predictions were better and cheaper. They give the example of autonomous vehicles, as driving is a common task that involves both prediction (what is happening in the environment around you and the potential payoffs of each decision) and judgement (what is the right action to take given this information). This judgement component may only be important because humans cannot make as fast and accurate a prediction as an autonomous vehicle. Should understanding the environment and the outcome of each action be nearly instantaneous and highly accurate, picking the one with the highest expected reward given a certain set of rules may result in the human judgement having reduced importance. In these contexts, even decision tasks could be substituted by AIs when the predictions are fast, cheap, and high quality enough.

The third possibility is that AI results in greater labor need as expert judgement becomes an important complement to better prediction. For example, in emergency medicine should diagnostics become better, faster, and cheaper through better AI, medical staff are now able to have a more accurate understanding of their patient needs, being able to prioritize workloads better and make targeted interventions. This increased productivity could in turn make hospitals more efficient, requiring more staff to provide even more care. Finally, the possibility exists that new types of tasks are created by the advent of AI. We can already see this appear somewhat in the data labelling industry that has arisen to support the models being deployed. As prediction becomes better and cheaper there may be more tasks that were simply unfeasible when prediction was poor and costly that are now able to be done.

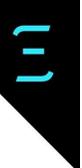

This paper provides a valuable framework for decomposing the impacts of AIs on jobs and provides a language for understanding the different effects it can have on workers depending on their specific circumstances. This theoretical understanding can provide the basis for more nuanced quantifications of the employment impact of new AIs, help policy makers and educators understand the skills that will have stable demand, and help businesses prepare for worker transition by identifying which employees require skills support.

## Working Algorithms: Software Automation and the Future of Work
([Original PhD dissertation](#) by Benjamin Shestakofsky)

Predictions on automation, job loss and the future of labour are often grounded in macro-level phenomena, such as labour markets, job categories and work tasks. These large-scale observations often fail to consider the lags and gaps in AI systems, who cannot fully operate without human assistance. Shestakofsky, using the example of the startup AllDone, demonstrates how human workers will not be replaced but will continue to work alongside machines. His close examination of the startup's development reveals how AI systems continue to rely on human skills and complementary forms of emotional labour. The author also delves into instances where human labour is preferred over computational labour for cost-saving and strategic reasons.

AllDone was a tech startup that sought to transform local service markers by building a website that would connect buyers and sellers of small services, from drape-making to cleaning and construction jobs. At the first stage of its development, when AllDone was looking to attract users to its system, it relied upon a Filipino contract team's efforts to collect information and target potential users to conduct a digital marketing campaign. This team of workers provided what the author termed "computational labour" when software engineers at AllDone did not have the resources to design and develop an automated system. After acquiring a customer base, AllDone turned its attention to securing sellers on the website. Many sellers were small entrepreneurs or individual workers who did not understand the system's design and rules and often voiced frustration over a lack of responses to quotes. To help repair this knowledge gap, AllDone hired a team of customer service agents who would patiently explain the system to new sellers and advise them on how to improve their profile. This team provided the emotional labour needed to help users adapt to the system. When AllDone sought to extract greater profits from its users and sellers in its third stage, it relied on both emotional labour to convince users to keep their subscription and computational labour to prevent sellers from circumventing and gaming the new rules.

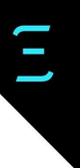

The author credits AllDone's internal dynamism by responding to new startups' challenges and creating new complementarities between humans and machines. The rough edges of machine systems will almost always require complementary human workers to smooth it out. The company's limited resources meant that it could not automate every task. Even with its AI systems, software engineers often called on workers' assistance to quickly collect information or test out features. Human touch was also necessary to create a loyal customer base. While more forms of economic exchange will likely become technologically mediated, key emotional skills of persuasion, support and empathy are still difficult to automate. The context here is also important, as AllDone's development was also dictated by venture capitalists' expectations that funded its growth. The author does not touch upon how this dynamism might work in a larger and more established corporation with bigger resources at its disposition.

While some new complementarities may appear between humans and machines, it also appears that old habits die hard. Shestakofsky's study reveals how, even with new technologies and expectations of changes, longstanding trends and divides in labour continue to persist. The emotional labour of counselling potential sellers and reassuring customers was accomplished by the women who led the phone line support staff. Women have long been characterized as support and care workers. These perceptions are replicated in technologies and within the tech industry, as seen by female voices' ubiquity in virtual home and phone assistants. The cheap and disposable labour is extracted from the global south, where underpaid Filipino workers are given short-term contracts to fill in when computational resources are missing. Their efforts, in comparison with North American working and wage standards, are largely uncompensated. Outsourcing, offshoring and contracting labour to the global south will continue to persist even in a world of software automation, with underpaid workers providing human assistance needed to create and keep running these systems and the managerial team accruing wealth. Rather than pushing humans out of production, tech automation may be replicating and reinforcing inequities at the domestic and global scale as well as furthering the divide between the managerial and working class.

A few words of caution, the startup appears to no longer be in existence, most of the observations took place in 2011, and there is a limit to how much we can extrapolate from a single observational case study. Shestakosky's article nonetheless provides interesting insights in how we think about human-machine interactions and the future of labour. Instead of fearing the replacement of human workers, we can turn our attention to ensuring the positive development of human-machine complementarities and preventing vulnerable people from bearing the brunt of emotional and underpaid labour.

# **Go Wide: Article Summaries**

## **Protecting the Essential Workers of the Data Supply Chain**
**([Original *IDinsight* article](#) by Ruth Levine)**

In-person data collection is an essential tool for understanding our world. With 2020 being the year for the census in the US, this article makes concrete arguments in asking the WHO to come forth with guidelines on safe in-person survey methodologies. Not only is this data critical for administrative purposes, but the SDGs from the UN are also highly dependent on household survey information.

Some might argue for the use of phones and other technologies to capture equivalent data. But this is fraught with problems since technology penetration is uneven and tends to exclude the already marginalized from adequate representation. For example, studies have demonstrated that women get systematically less representation when relying solely on data collected via phone surveys. Such second-best alternatives are very risky when making consequential decisions.

Evidence-based guidelines from the WHO can help bridge this gap allowing surveyors to return to in-person data collection, protecting their health and of the interviewees. Some suggestions made in the article include shortening the surveys to focus on questions that require in-person conversations, providing workers with private transport, regular testing, carrying additional supplies of face masks and sanitizers that could be handed out as tokens of appreciation, and more. Bolstering these recommendations with the advice from medical experts will help us get back to high-quality data collection that is critical to the proper functioning of our society.

## **Data, Compute, Labour**
**([Original *Ada Lovelace Institute* article](#) by Nick Srnicek)**

This article moves the conversation beyond the familiar trope of data powering the flywheel of building monopolies in the technology domain. Zooming out a bit, we see that there are other factors that play an equally important role in spinning that flywheel - namely, compute capacity and labor. In particular, as highlighted in research work from the Montreal AI Ethics Institute [here](#), compute capacity plays a polarizing role in how research and development is done in the domain of AI. As we move towards using larger models that are even more computationally expensive, this has a strongly prohibiting effect on those that don't have access to large clusters of computing power (i.e. typically people outside of universities and large corporations) which chills the scientific discovery process. By no means are we advocating that large models shouldn't be explored, instead we ask that people investigate a bit more on how we can move towards more compute-efficient architectures and approaches. In the same vein, having more public goods-styled compute facilities and data stores will also allow more people outside of these traditional venues to participate in AI research.

The other factor which is labor, especially highly-skilled workers who are able to work with this hardware are often scooped up from various places and concentrated in corporate hubs where they are paid exorbitant amounts of money. This has notably had effects in how academia has been losing star researchers to corporations (even when they have dual roles, they still are able to dedicate less time to the nurturing of the next generation of talent. Some might argue that there is benefit in having these dual appointments because of the mix of industry and research experience that they can bring in, yet only time will tell if that effect actually gets realized in practice).

## An Ethics Guide for Tech Gets Rewritten With Workers in Mind
([Original *Wired* article](#) by Arielle Pardes)

Ethics needs to be discussed as early in the lifecycle of product development as possible. Akin to cybersecurity, the cost of doing so is orders of magnitudes less, and the efficacy much higher when following this pattern. So why is it that this doesn't happen often? This article talks about a new toolkit from the Omidyar Network that builds on their prior work. Specifically, this iteration focuses on mechanisms that empower developers and those who are on the ground, creating products and services to feel confident in bringing up these issues.

"The kit includes a field guide for navigating eight risk zones: surveillance, disinformation, exclusion, algorithmic bias, addiction, data control, bad actors, and outsize power." Absolutely in line with the work that we do at the Montreal AI Ethics Institute, we believe that there is a dire need for handy formulations that we can frictionlessly integrate into the design, development, and deployment phases of AI. While the unofficial mantra of Silicon Valley is to "move fast and break things," such tools might help to slow down this process and offer opportunities for reflection that can help incorporate ethics, safety, and inclusion in the systems that we build.

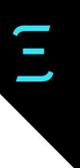

# Closing Remarks by Katya Klinova (AI, Labor & the Economy Research Program Lead, Partnership on AI)

As we enter an era of rapid advancements in AI, the literature assessing the impact of these advancements on the labor market has similarly seen dynamic developments. In the last few years, the field has moved from mainly focusing on projecting the magnitude of future job displacements to developing theoretical frameworks that illuminate generalizable patterns about the consequences of automation on the economies of today and tomorrow.  Works summarized in this chapter elevate the Future of Work discourse to the point of examining its own shortcomings. Together, they pose a necessary question: Have we become so mesmerized by technological change that our analysis has largely neglected the work of the humans that enable it?

Hidden behind both the "artificiality" and "intelligence" of AI is an often underpaid and precarious workforce of uncelebrated human beings. Alexandra Mateescu and Madeleine Clare Elish emphasize that technology almost never gets simply "deployed," usually requiring patient and attentive labor to integrate it into existing processes and contexts. It is fundamentally the care of people watching over "smart" machines that power the self-checkout line in a grocery store — and prevents customers from constantly throwing up their hands in frustration. The importance of this crucial human labor remains unrecognized because acknowledging it would be to admit that these technical systems are not so autonomous and often not so intelligent, either.

It is easy to identify similar examples of human workers playing a critical role in enabling technological innovation. The flexibility of "computational labor" described by Shostakofsky in "Working Algorithms," for instance, enables dynamic experimentation and testing of new products and features, while the emotional labor of people supporting newly developed systems smooths out the wrinkles and rough edges that such systems inevitably possess. It is humans who "teach" algorithms to differentiate cats from dogs, pedestrians from street signs, and verbs from nouns. And it is humans who come to the rescue and offer their imperfect human judgment when algorithms get stuck, unable to move past a low-confidence prediction. We see humans playing central roles in all technical systems, but instead of appreciating and honoring their work, innovators often appear more interested in trying to automate humans out of relevance. As if in pursuit of an implicit goal of "pure artificiality," their AI systems are tasked, above all, with "beating" human performance.

But the purpose of technological progress should be in supporting human and planetary flourishing, not in attaining "pure artificiality" or pure autonomy for their own sake. Thus it is fitting to broaden the scope of the Future of Work research and discourse beyond just thinking about what jobs the march of AI will displace or how displaced workers can adjust or "upskill." We must also ask how the hidden work of ensuring machines' productivity and their smooth integration into our lives can start getting the recognition, benefits, and pay that it deserves. It is very encouraging to see the growth in scholarship spotlighting the role of human workers who are absolutely essential to the artificially intelligent performance of technical systems.

Katya Klinova (**@klinovakatya**)
AI, Labor & the Economy Research Program Lead, Partnership on AI

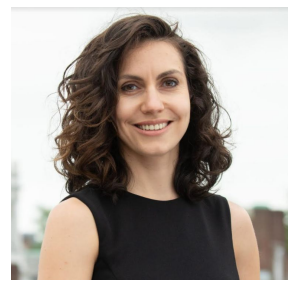

Katya Klinova directs AI, Labor, and the Economy Research Programs at the Partnership on AI. In this role, she oversees multiple flagship projects, including the AI and Shared Prosperity Initiative.

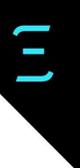

# 6. Privacy

**Opening Remarks** by Abhishek Gupta (Founder, Montreal AI Ethics Institute)

A notion that gains more and more prominence every quarter that we publish this report, it goes without saying that the importance of privacy is paramount. Whether it is talking about contact-tracing applications, voice assistants, your emails, your selfies, or any other digital traces (and we leave a lot of those!) when it comes to how we socialize, work, and go about our lives.

Facial recognition technology certainly has been on everyone's mind this past quarter; while some lauded that with the mandatory use of face-masks, the efficacy of facial recognition technology might go down, this doesn't seem to be the case. Data collection of masked selfies as mentioned in this chapter is helping to train systems that perform well even with face masks being present. Particularly systems from regimes that have more relaxed privacy standards, this is a rising problem. Using data scraped from Instagram has been the unfortunate fuel in this fire. Pressing the need for better research ethics in the space when it comes to the use of public data is critical if we're to make some progress in this space.

Have you ever taken the time to read through an entire privacy policy? I certainly haven't! (much to my own chagrin) But you're not alone. With the complexities of these policies, it is hardly a surprise that we choose not to invest effort into reading them. Senators in the US want to shift the burden for making these policies intelligible onto the companies, akin to the requirements set out in the GDPR. In the cases of IoT systems, having things like privacy nutrition labels can also aid with this process. Losing personal data due to a lack of robust privacy controls can also become a national security issue as we highlight in this chapter. Items like digital voice assistants should be evaluated more thoroughly to ensure that our data is indeed being kept private. An example of where this has failed is highlighted in this chapter as well where researchers were able to use intentional, masked wake words to trigger voice assistants into capturing our data without our knowledge and consent.

Relying solely on anonymization is insufficient, and taking approaches like differential privacy that can provide mathematical guarantees on the level of privacy controls is essential. I also believe that until we get rid of the centralization of data and compute monopolies, privacy concerns will continue to abound as highlighted in this chapter. Eliminating unethical actors like data brokers who are able to collate large amounts of data without scrutiny and relegate people to a digital purgatory without recourse need to be held accountable to higher standards and ultimately banned.

Finally, relying on ideas like geo-indistinguishability to provide privacy for the location-based services that we use can act as an additional layer of protection for something that has become ubiquitous in our lives. There is also much to be borrowed from the field of cybersecurity when it comes to privacy and security in machine learning systems as highlighted in this chapter.

My call to the readers of this chapter is to move beyond the familiar discussions of the abstract principles surrounding privacy in the domain and focus more on the operationalization of these ideas. Surfacing the places where they fail and how we can improve is going to be critical if we're to achieve the (currently) pipe dream of having more robust privacy controls in our use of technology today. I hope to have better news for us next quarter when we meet on this topic, for now, I hope that this chapter prompts some new ideas and nudges you towards operationalizing privacy within your own work.

Abhishek Gupta (**@atg_abhishek**)
Founder & Principal Researcher, Montreal AI Ethics Institute

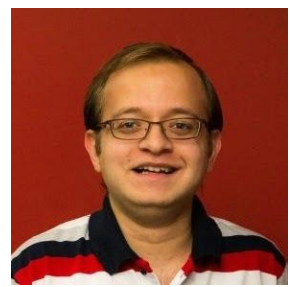

Abhishek Gupta is the founder and principal researcher at the Montreal AI Ethics Institute, seeking to define humanity's place in a world increasingly characterized and driven by algorithms. He is also a machine learning engineer at Microsoft, where he serves on the CSE Responsible AI Board. His book 'Actionable AI Ethics' will be published by Manning in 2021.

# Go Deep: Research Summaries

## Geo-Indistinguishability: Differential Privacy for Location-Based Systems

([Original paper](#) by Miguel Andrés, Nicolás Bordenabe, Konstantinos Chatzikokolakis, Catuscia Palamidessi)

The authors discuss how the onslaught of location-based systems (LBS) has resulted in considerable challenges to locational privacy. Add to this the fact that most of such individual data (about locations) is stored in unknown and arguably unsecure servers, there is a need to safeguard an individual's exact location whilst she uses a LBS. Geo-indistinguishability is the novel mechanism this paper proposes to ensure the balance where a user of a LBS discloses just enough of her approximate location to efficiently benefit from these services, while not divulging her precise location.

**Existing notions of privacy**

While the authors intend to provide a formal notion of privacy (i.e. geo-indistinguishability), they initiate the conversation by covering some existing ideas on privacy. These include:

- Expected distance error, which is a location-obfuscation mechanism resulting in an adversary to inaccurately determine an individual's location. The obfuscation can occur in different ways – for instance, to throw off the tracking of an individual's path/location, multiple paths of different users are intertwined, thus, perturbing the adversary.

- k-anonymity, which includes concealing the true identity of the user of a LBS by placing her in the midst of a set of users (k). Unlike some other notions, this focuses on protecting an individual's identity, and consequently, her location.

- Differential privacy, which emerges from the field of statistical databases. The notion requires the publication of aggregate data emerging from a dataset, in lieu of individual data. The difference by altering some individual data points, should be negligible and still yield the same results to a query. Given that the notion relies on aggregated information, it is inapt for situations involving a single individual.

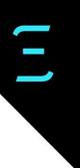

- Location cloaking mechanism, which as the name suggests, aims at concealing the location of a user through location-ranged queries. Essentially, the objective is to cover a range of areas, and conceal locations/regions within this range that the user may consider sensitive.

- Transformation based approaches make the location of a user completely invisible, rather than cloaking it. Through the use of cryptography, the data (including the query sought, as well as the location of a user), are encrypted. Using this encrypted information, the service provider can respond to a query without actually detecting the location of the user.

**Geo-Indistinguishability**

The probabilistic model includes multiple possible locations of a user (denoted by X). Additionally, to obfuscate the precise location, the adversary/attacker is fed variable locations (termed as reported values) to create enough disturbance to insulate the true location of the user. However, the element of probability comes into play contingent on the nature of additional (side) information that the adversary/attacker may possess, which can allow to overcome some of these disturbances, and get a relatively more accurate lock on the location of the user.

Defining geo-indistinguishability – Unlike the standard form of differential privacy which aims at completely protecting the location of a user, geo-indistinguishability is about disclosing just enough elements of such location as would allow the user to access and use the requisite LBS. Hence, while it has some commonalities with differential privacy, it uses different metrics.

**Characterizations of geo-indistinguishability**

The paper also elicits two key characterizations of geo-indistinguishability:

- First, it discusses the hidden functionality which allows the actual location of a user to be concealed from an attacker. Instead of disclosing the actual location, the mechanism introduces a hidden version, which can impact the conclusion(s) of the attacker in discerning the real location of a user. The extent of impact on the conclusions is affected by the distance between the actual and the hidden location. For instance, if an individual is located in Paris and using a restaurant searching app, but a hidden functionality discloses her location as London, then the attacker is likely to be completely thrown off.

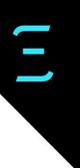

- Second, the authors emphasize how geo-indistinguishability abstracts the side information. Side information essentially can be any ancillary information that may be in the possession of an attacker prior to her using a LBS. For instance, knowledge that an individual is located at an airport, yet not knowing which city's airport. However, as the authors argue, any minimal service request will at least disclose a city, which can then be used to infer the actual location at the city's airport. Therefore, it is necessary to abstract such side information which can be accomplished through geo-indistinguishability.

**Attaining geo-indistinguishability and sample example**

While concealing a singular location is one stage, it is possible for an individual to have multiple locations of interest which she may not want to divulge. For preserving the locational secrecy of these multiple points, the paper suggests two ways. First, to report on the whole set of locations by applying a common obfuscation mechanism to every single location; and second, by reporting an aggregated location, which can be the centroid of the tuple of locations that a user wants to preserve.

Given that creation of controlled noise is a prerequisite for attaining geo-indistinguishability, the authors explore different mechanisms for this with greater nuance. For this, the authors set out the mechanism for creating a continuous plane, which allows them to remap each point on such a plane to the closest point in the discrete domain, and finally, only disclose points close to the actual location of a user (area of interest).

To gauge the viability of geo-indistinguishability as a privacy guarding intervention, the authors test it on different LBS systems. These could include mildly location-sensitive and highly location-sensitive LBS applications. It is the latter where guaranteeing privacy, while delivering adequate and accurate service, is challenging. The authors also state that for a user performing multiple query requests, locational privacy can be guaranteed by performing geo-indistinguishability and obtaining approximate locations to every one of the user's locations for each query request.

The authors also contrast their method of geo-indistinguishability with other methods. For instance, they compare it to the obfuscation mechanism which also involves the creation of randomly selected locations (different from the actual location of the user). However, the obfuscation method cannot abstract prior or side knowledge and is therefore susceptible to breaches. The key difference that the authors bring forth in advocacy of geo-indistinguishability is the balance it affords in safeguarding privacy while suffering from minimal Service Quality Loss (or inaccuracies in responding to the query requests of users).

To conclude, the authors list out some possible expansions to their ongoing work on developing geo-indistinguishability.

## SoK: Security and Privacy in Machine Learning
([Original paper](#) by Nicolas Papernot, Patrick McDaniel, Arunesh Sinha, Michael P. Wellman)

**1. Introduction**

Despite the growing deployment of machine learning (ML) systems, there is a profound lack of understanding regarding their inherent vulnerabilities and how to defend against attacks. In particular, there needs to be more research done on the "sensitivity" of ML algorithms to their input data. In this paper, Papernot et. al. "systematize findings on ML security and privacy," "articulate a comprehensive threat model" and "categorize attacks and defenses within an adversarial framework."

**2. About Machine Learning**

*Overview of Machine Learning Tasks*

Through machine learning, we're able to automate data analysis and create relevant models and/or decision procedures that reflect the relationships identified in that analysis. There are three common classes of ML techniques: *supervised learning* (training with inputs labeled with corresponding outputs), *unsupervised learning* (training with unlabeled inputs), and *reinforcement learning* (training with data in the form of sequences of actions, observations, and rewards).

*ML Stages: Training and Inference*

There are two general stages (or phases) of ML: the *training* stage includes when a model is learned from input data, generally described as "functions $h_\theta(x)$ taking an input $x$ and parametrized by a vector $\theta \in \Theta$.1" Once trained, the model's performance is measured against a test dataset to measure the model's *generalization* (performance on data not included in its training). The *inference* *stage* includes when the trained model is deployed to *infer* predictions on inputs not included during training.

**3. Threat Model**

It's important to note that an information security model should consist of two primary components: the *threat model* and the *trust model*. In the case of ML systems, these components are examined below.

*Threat Model*

A system's *threat model* is based on the adversarial goals and capabilities it's designed to protect itself from. To create a comprehensive model, it's important to first examine the system's "attack surface"—where and how an adversary will attack. Although attack surfaces can vary, the authors view all ML systems within a "generalized data processing pipeline" in which adversaries have the opportunity to "manipulate the collection of data, corrupt the model, or tamper with the outputs" at different points within the pipeline."

*Trust Model*

A system's *trust model* relates to the classes of actors involved in an ML-based system's deployment—this includes *data-owners, system providers, consumers,* and *outsiders*. A level of trust is assigned to each actor and the sum forms the model—this helps identify ways in which bad actors may attack the system both internally and externally.

*Adversarial Capabilities*

The capabilities of an adversary (or a bad actor) refers to the "whats and hows of the available attacks." These capabilities can be further understood by examining them separately at the inference phase and the training phase. At the inference phase, attacks are referred to as *exploratory attacks* because they either cause the model to produce selected outputs or "collect evidence about the model's characteristics." These attacks are classified as either *white box* (the adversary has some information about the model or its training data) or *black box* (the adversary has no knowledge about the model, and instead uses information about the mode's setting or past inputs to identify vulnerabilities).

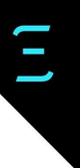

At the training phase, attacks attempt to learn, influence, or corrupt the model through *injection* (inserting inputs into the existing training data as a user) or *modification* (directly altering the training data through the data collection component). A powerful attack against learning algorithms is logic corruption, which essentially modifies the model's learning environment and gives the adversary control over the model itself.

*Adversarial Goals*

Using the CIA triad, with the addition of *privacy*, the authors model adversarial goals both in terms of the ML model itself, but also in terms of the environment in which the model is deployed. The CIA model includes *confidentiality* (concerns the model's structure, parameters, or training/testing data), *integrity* (concerns the model's output or behaviors), and *availability* (concerns access to the model's meaningful outputs or features). In regards to confidentiality and privacy, attacks tend to be targeted at the model and data with the general goal of exposing either one. Specifically, due to the nature of ML models, which have the ability to "capture and memorize elements of their training data," it is difficult to guarantee privacy to individuals included in that dataset. In regards to integrity and availability, attacks tend to target the model's outputs with the goal of inducing "model behavior as chosen by the adversary" and thus undermining the integrity of the inferences. Researchers have also found that the integrity of an ML model can be compromised by attacks on its inputs or training data. Availability is slightly different than integrity, the goal of these attacks is to make the model "inconsistent or unreliable in the target environment." Finally, if access to a system in which an ML model is deployed depends on the model's outputs, then it can be subject to denial of services attacks.

**4. Training in Adversarial Settings**

Training data for ML models are particularly vulnerable to manipulations by adversaries or bad actors. Otherwise known as a *poisoning attack*, the training dataset is inserted, edited, or points are removed "with the intent of modifying the decision boundaries" of the model. These attacks can cause the system to be completely unavailable.

*Targeting Integrity*

In a study by Kearns et. al., researchers looked at the accuracy of learning classifiers when training samples are modified. They found that to achieve 90% accuracy in a model, the "manipulation rate" needs to be less than 10%. Adversaries can attack the integrity of classifiers through *label manipulation*, which includes perturbing the labels for just a fraction of the training dataset. In order to do this, the adversary needs to have either partial or full knowledge of the learning algorithm. Unfortunately, this type of attack not only has immediate ramifications on the model during training but "further degrades the model's performance during inference," thus making it difficult to quantify its impact. Adversaries can also attack the integrity of classifiers through *input manipulation,* which includes corrupting the training dataset's labels and input features at different training points. In order to do this, the adversary needs to have a great deal of knowledge regarding the learning algorithm and its training data. This poisoning attack can be done both directly and indirectly. Direct poisoning primarily focuses on "clustering models," where the adversary slowly moves the center of the cluster so that there are misclassifications during the inference stage.

*Targeting Privacy and Confidentiality*

The confidentiality and privacy of a model during training may only be impacted by the extent to which the adversary has access to the system hosting the ML model—which the authors note, is a "traditional access control problem" and falls outside of the scope of this paper.

**5. Inferring in Adversarial Settings**

During the inference stage, the adversary must mount an attack that will evade detection during deployment since the model's parameters are fixed. There are two types of attackers, *white-box* (they have access to the model's internals, such as the parameters, etc.) and *black-box* (they do not have access, and are thus limited to "interacting with the model as an oracle").

*White-box Adversaries*

These attackers have access to both the model and its parameters, making them particularly dangerous. To attack a model's integrity, these adversaries perturb the model's inputs through *direct manipulation* (altering the feature values processed by the model), *indirect manipulation* (locating perturbations in the data pipeline before the classifier), or other means that impact ML models other than classification, such as autoregressive models or reinforcement learning.

*Black-box Adversaries*

These attackers do not know the model's parameters but do have access to the model's outputs which allows them to observe its environment, including its detection and response policies. The common threat model for these types of adversaries is *oracle*, in which they "issue queries to the ML model and observe its output for any chosen input" in order to reconstruct the model or identify its training data. These adversaries can attack a model's integrity through *direct manipulation of model inputs* or *data pipeline manipulation*. To attack a model's privacy and confidentiality, adversaries may mount *membership attacks* (testing whether a point was part of the training dataset to learn the model's parameters), *model inversion attacks* (extracting training data from a model's predictions), *model extraction* (extracting parameters of a model by observing its predictions).

**6. Towards Robust, Private, and Accountable Machine Learning Models**

In this section, the authors identify parallel defense efforts against attacks to reach the following goals: "(a) robustness to distribution drifts, (b) learning privacy-preserving models, and (c) fairness and accountability."

*Robustness of Models to Distribution Drifts*

In regards to maintaining integrity, ML models need to be robust to *distribution drifts*: instances when there are differences between the training and test distribution. Drifts can occur, for example, as a result of adversarial manipulations. To defend against attacks during training time, proposals for defenders include building a PCA-based detection module, adding a regularization term to the loss function, using obfuscation or disinformation to keep details of the model's internals secret, or creating a detection model that removes data point outliers before the model is learned. Defending against attacks during inference time is difficult due to the "inherent complexity of ML models' output surface," therefore it remains an "open problem." One way to defend against integrity attacks is through *gradient masking*: reducing the model's sensitivity to small changes in their inputs. However, this strategy has limited success because the adversary may be able to use a substitute model to craft adversarial examples because it falls outside of the target model's defense—thus, these examples will also be misclassified. To defend against *larger perturbations*, defenders can inject correctly labeled adversarial samples into the training dataset, making it more "regularized" and robust. However, the authors note, this method is relatively weak "in the face of adaptive adversaries."

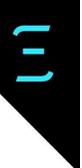

*Learning and inferring with privacy*

*Differential privacy* is a framework used by defenders to analyze if an algorithm provides a certain level of privacy. Essentially, this framework views privacy as the "property that an algorithm's output does not differ significantly statistically for two versions of the data differing by only one record." In order to achieve differential privacy or any other form of privacy, the ML system's pipeline should be randomized. In order to achieve privacy during the training stage, defenders may inject "random noise" into the training data to create *randomized response* or *objective perturbation*. To achieve differential privacy during the inference stage, defenders may also inject noise into the model's predictions. This can impact the accuracy of those predictions, however. A method of protecting the confidentiality of individual inputs to a model is *homomorphic encryption*, which encrypts the data in a way that doesn't require decrypting for it to be processed by the model.

*Fairness and accountability in ML*

Due to the "opaque nature of ML," there are concerns regarding the fairness and accountability of model predictions, as well as growing legal and policy requirements for companies to explain the predictions made by deployed models to users, officials, etc. *Fairness* largely refers to the "action taken in the physical domain" based on the prediction made by a model, ensuring that the prediction and thus the action does not discriminate against certain individuals or peoples. Often, concerns arise regarding the ML's training data, which can be the source of bias that leads to a lack of fairness. The learning algorithm can also be a source of bias if it's adapted to benefit a specific subset of the training data. One method to achieve fairness, illustrated by Edwards et al., is having a model learn in "competition with an adversary trying to predict the sensitive variable from the fair model's prediction." *Accountability* refers to the ability to explain the model's predictions based on its internals. One way of achieving accountability is to measure the "influence of specific inputs on the model output," called *quantitative input influence* by Datta et al. Another way is by identifying the specific inputs a model is most sensitive to,. For neural networks, *activation maximization* can synthesize inputs that activate specific neurons. Unfortunately, the authors note, methods for achieving accountability and fairness may open the ML model to more sophisticated attacks since they provide an adversary with information on the model's internals. However, these methods could also increase privacy.

**7. Conclusions**

The primary takeaway of this paper is that it's essential for the classes of actors involved in an ML-based system's deployment to characterize the "sensitivity of learning algorithms to their training data" in order to achieve privacy-preserving ML. Further, controlling this sensitivity once models are deployed is essential for securing the model. In particular, the authors note, there must be more research into the "sensitivity of the generalization error" of ML models.

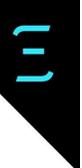

# Go Wide: Article Summaries

**Your Face Mask Selfies Could Be Training the Next Facial Recognition Tool**
([Original *CNET* article](#) by Alfred Ng)

Instagram and other social media platforms that allow us to share pictures with each other are a great way to keep in touch with friends and others who are far away. In the times of a pandemic with a forced social distancing, they serve as an even more potent tool to maintain social connection (albeit virtually). But, when such photos are posted publicly, there is always a strong risk for them being scraped and picked up for purposes that you never intended your pictures to be used for. This came to light with the recent compilation of a dataset containing selfies of people wearing face masks that were scraped from public Instagram accounts. The purpose of this data collection exercise is to allow researchers to build an AI system that can recognize faces from limited information, such as when a significant part of the face is obscured by a mask.

For facial recognition technology which requires access to multiple facial features to make an accurate match, masks serve to disrupt that recognition process which protects protestors from being identified in an automated manner and be matched up with vast databases of face data. Now, when governments are mandating the use of masks to prevent the spread of the disease, this issue has come back in another form, in cases of law enforcement and other activities where authorities rely on face recognition, their efforts are now hampered by the use of masks and they are searching for alternatives to be able to still use the tools and surveillance infrastructure that has been put in place. While there are attempts to digitally add masks to existing photos as a means of training the system, preliminary analysis shows that it isn't quite as effective as getting real pictures that captures the diversity of skin tones, lighting conditions, angles, and other factors that affect the photos.

The researcher who was behind the collection of this data shirked their responsibility by asserting that if people didn't want their photos to be used in this manner, they should make their pages private. This presents a problematic perspective whereby tools enabling greater surveillance and privacy intrusions are built while relegating responsibility to the users of digital services where they are not even aware that such a thing might be happening. Public awareness of such efforts is paramount to combat this kind of blind tech-solutionism that ignores fundamental rights and freedoms of people.

## How Well Can Algorithms Recognize Your Masked Face?
(Original *Wired* article by Tom Simonite)

There is a tremendous rush to collect data to train facial-recognition systems when faces are obscured by people wearing masks. Researchers who work on developing these systems point to the problems that arise when faces are obscured by any element, including face masks. Lower rates of accuracy can lead to false positives which will exacerbate the myriad issues with facial-recognition technologies, including the lower rates of recognition of minorities.

A lot of companies and government entities mentioned in the article claim that they now have the ability to recognize faces even in the presence of face masks, relying on features that are exposed like the eyes, eyebrows, and nose bridges. But, without external validation benchmarks, these claims don't have verifiable backings which makes it hard to judge whether problems with false positives are aggravated with the use of this technology. NIST, which has benchmarks for this, is looking to add face masks digitally to existing photos to create a new benchmark for which it is inviting companies to submit their systems to check their levels of accuracy in a publicly ranked leaderboard.

Chinese and Russian systems tend to perform well because of lighter privacy regulations which makes it easier to work with larger datasets. In China for example, people use this technology ubiquitously as a means of payment using the app AliPay and there is a case to be made how this can be a supplement to other contactless payment technologies, especially when trying to curb the spread of the pandemic. A user of this technology pointed out how he appreciated the usefulness of this working with masks on, decreasing the chances of risk by not having to take the mask off in public places. The important thing will be to continue to respect people's fundamental rights of freedom and leveraging informed consent before getting them to opt-in to such systems.

## Nobody Reads Privacy Policies. This Senator Wants Lawmakers to Stop Pretending We Do.
(Original *Washington Post* article by Geoffrey A. Fowler)

Given that there isn't a consumer privacy law that applies across the board, the proposal highlighted in this article mentions how there is a strong need for recognizing that consent based models which try to imbue agency onto users actually place undue burden on them. Specifically, it is well documented that the average person would require many days worth of time to parse through and comprehend meaningfully the various privacy policies as they relate to the different products and services that they use in their everyday life. In this case, people tend to just click through the "I accept" button without giving it much thought as to how their data is used and who has access to it. Informed consent doesn't mean much, even when thinking about progressive disclosure models that purport to build up an understanding on the part of the user. The key problem lies in the allocation of burden and this proposed bill rightly shifts the burden away from the consumer onto the platform companies that tend to benefit enormously from the users' private data, compared to the miniscule benefits that are offered in return to the users in terms of free service.

This shift also creates an impetus for companies to explore novel business models that don't solely rely on extractive practices of selling users' personal data for targeted advertising. Sen. Brown points out that the inspiration for this came from the Equifax data breach where the American public was caught off guard in that they didn't even know what Equifax was, yet their personal information as held by Equifax had been compromised.

While the bill proposes to prohibit the use of data to generate targeted ads, contextual ads, such as the ones from Google in their search products would pass muster under the proposed requirements. Ultimately, as a researcher opines in the article, we must take incremental steps in privacy law in this context rather than aiming for something perfect which might not be something that will get passed due to differing views in the law-making bodies on this subject.

## IoT Security Is a Mess. Privacy 'Nutrition' Labels Could Help
([Original *Wired* article](#) by Lily Hay Newman)

At DEFCON 2017, someone mentioned that you can't spell "idiot" without "iot". What the panelists and speakers were referring to were the rampant security and privacy challenges that plague IoT systems, especially due to their connected nature and their limited processing capabilities which hamper their ability to have extensive security mechanisms in place. Analogous to other work in the machine learning domain, for example, the dataset nutrition labels, the researchers quoted in this article created a new set of privacy nutrition labels that elucidate the security posture of the IoT system, detailing things like the length of support from the manufacturer, the updates schedule, what data is collected, how is it used, etc.

One of the things that immediately caught our attention was the focus on making the labels accessible to the everyday consumer and supplementing it with functionality that makes these labels machine-readable and interoperable. This is very strongly aligned with the [work at the Montreal AI Ethics Institute](#) in the domain of machine learning security being done by the research staff Erick Galinkin and Abhishek Gupta. As pointed out in the article, one of the benefits of this approach is that consumers will be able to search for this machine-readable information in a manner that allows them to make informed purchasing decisions along the privacy and security features of the product rather than just other technical considerations.

One of the other considerations is to standardize the bill of materials used in a software product so that there is transparency in terms of the open-source and other libraries that underpin the system so that consumers and external researchers can have clarity on potential security and privacy threats as it relates to the system by looking at these labels. While we are far away from widespread adoption, something that is evident is the need to create something that is intelligible to users that empowers them to make informed decisions.

## Privacy Is Not the Problem with the Apple-Google Contact-Tracing Toolkit
**([Original *The Guardian* article](#) by Michael Veale)**

From the creator of the DP3T framework for doing decentralized privacy-preserving proximity tracing, this article offers a fresh take on some of the problems that we face when debating contact- and proximity-tracing apps that have the potential to reshape our social fabric if they are deployed widely. It starts by pointing out the history of passports that were introduced temporarily during WW1 but were retained during the time of the Spanish flu as a means of curbing the spread of that pandemic. Measures introduced during times of emergencies have the potential to persist beyond their original purpose and they can morph into mechanisms that have the power to rethread societal fabric as we know it.

In the push for having decentralized contact- and proximity-tracing, the Apple-Google protocol offers an apparent win that is hard to replace, especially given its ubiquity in terms of covering almost everyone with a smartphone. The UK and France have advocated for centralized protocols citing reasons of reducing fraud and lowering risks of snooping behaviour. But, in the midst of all this, an inherently adversarial framing has emerged which pits large corporations against nation states, each viewing the other as a sovereign entity that doesn't have any other recourse towards arriving at meaningful solutions.

While data privacy can be a strong reason for the push for having decentralized apps, this doesn't detract from the problem of having centralized control over the compute infrastructure who will still be able to assert significant control over society. The article concludes by pointing out that deflating digital power isn't just about the governance surrounding data but more so understanding the systemic forces at play in the underlying infrastructure.

### Researchers Identify Dozens of Words that Accidentally Trigger Amazon Echo Speakers
**(**[Original *VentureBeat* article](#) **by Kyle Wiggers)**

If privacy concerns were the only thing that you worried about when bringing home a smart voice assistant, think again! Our researchers have been advocating for thinking more critically about adversarial machine learning and machine learning security as important considerations in the design, development, and deployment of ML systems, something that has been covered quite frequently in our past newsletters. The research work covered in this article talks about LeakyPick, an approach that was used to check accidental triggering of smart voice assistants to start capturing audio snippets even when they weren't invoked. This has severe privacy implications given that a lot of personal data can be captured inadvertently when the devices are turned on and they send over their information to the central servers for processing.

Using phonemes similar to the wakewords that are used to activate these devices, the researchers tested a wide range of devices and found that they were demonstrable and repeatable instances where the devices woke up and recorded conversations when they weren't invoked using the actual wakeword. Another thing that was highlighted by the researchers was that even secondary checks on the end of the devices and servers let slip some of these accidental trigger words pointing to glaring problems in these devices as they are implemented at present.

### Can We Trust Digital Assistants to Keep Our Data Private?
**(**[Original *AIGA Eye on Design* article](#) **by Liz Stinson)**

Privacy is a topic that is covered ad nauseam in popular media. But, everyday citizens still have minimal awareness about the actual functioning of apps that they download and what happens to their data once they click "I accept." This article explores two conceptual ideas called Personal Privacy Assistant and Kagi that intend to guide the user in understanding various privacy policies.

The Personal Privacy Assistant serves up insights from the privacy policies and presents them to the users so that they can make more informed choices. The current privacy policy regime is notorious for burying information in legalese. Privacy settings are also hard to navigate; people never move away from the default options that benefit the platforms. But such assistants need to be helpful without being annoying. Many design moments need to be configured in the right manner so that they don't fatigue the user. Through a transitory process, they can earn the trust of the user by proving their efficacy and value. Kagi functions similarly, acting as an intermediary between the interests of the user and the platform. It depends on a potential future where there might be greater alignment between the welfare of the users and the business motives of the platform. Till then, we must do the best we can to help people navigate the existing privacy and data regulations.

## The Purgatory of Digital Punishment
([Original *Slate* article](#) by Sarah Esther Lageson)

Crime and punishment, an inexorable part of society, have taken on a new dimension. This article goes into detail on how punishment continues in the digital realm far beyond the asks made of the guilty. Specifically, digital purgatory is far worse in its outcomes for both those who are guilty and especially for those who are not. Data brokers are notorious for trading private data in hidden markets for financial gain. But, the extent to which this happens became apparent to people interviewed in this article when they found that mistaken identities, sealed records, past crimes, and other issues continued to create problems that became impossible to address completely.

While some of these items are supposed to be protected by legal mechanisms, the internet is indiscriminate in keeping these alive, making the "right to be forgotten" something impossible to achieve in practice. One of the problems is that as records are propagated, downloaded, reshared, they become decontextualized and stale making them problematic. The original crime bookkeeping operations are superseded by digitally-powered operations, which churn out millions of public records that are consumable by machines. Uneven rollouts and competing legal and political mandates exacerbate the problems.

Privacy inequities are the unfortunate consequence of this mayhem. Specifically, those who are marginalized bear a disproportionate burden since they have the fewest means to address and fix these problems. The victims are the ones who have the onus to correct their record, while those in power can abdicate their responsibility for maintaining accurate records. The article concludes by making the call for implementing regulations in the same manner as they apply to medical records and credit reports. The companies doing background checks should be held legally accountable for errors. We have a choice to make: resigning to technological determinism is not the answer.

## Differential Privacy for Privacy-Preserving Data Analysis: An Introduction to our Blog Series

([Original *NIST* article](#) by Joseph Near, David Darais, Kaitlin Boeckl)

Doing data analysis on personally identifiable information (PII) is rife with privacy challenges. One promising technique that seeks to overcome many of the shortcomings of other popular methods is called differential privacy. In particular, we would like to do our data analysis such that we can unearth trends without learning anything new about specific individuals. De-identification methods are vulnerable to database linkage attacks. Restriction to aggregated queries is also only feasible when the groups are large enough, and even then, they might be subject to privacy-compromising attacks.

Differential privacy is the mathematical formulation of what it means to have privacy. It is a property of a process rather than a method itself. It guarantees that the output of a differentially private analysis will be roughly the same whether or not your data gets included in the dataset. The strength of privacy gets controlled by a parameter called epsilon that is the privacy loss or the privacy budget. The lower the value of epsilon, the higher the degree of protection of the individual's data. But, queries with higher sensitivity require the addition of more noise, thus potentially diminishing the quality of the results.

The advantages of differentially private methods are as follows: the assumption that all information is private, resistance to privacy attacks based on auxiliary information, and compositionality of different differentially private methods. A minor drawback at the moment is the lack of many well-tested frameworks for implementing differentially private methods in practice but that is changing rapidly.

**Why Personal Data Is a National Security Issue**
([Original *Barron's* article](#) by Susan Ariel Aaronson)

The internet is abuzz with TikTok and what the implications of its potential acquisition sets in terms of a precedent for how software products and services work on a global scale, especially when they come into conflict with regulations and legislation, and perhaps even with political agendas. One of the highlights in the article is stressing the importance of interoperable data legislations since data from consumers in one country can be stored in another and if legislations are widely different, that might lead to all sorts of concerns.

From a national security perspective, in places where the government through instruments like lawful intercepts require that a company disclose records, the governments can combine lawfully obtained data with other large, public datasets, even when anonymized, to glean insights that would be beyond the capabilities of other actors in the ecosystem. This can often lead to knowing details about an individual but also at times about the nation as a whole and what its strategies might be in areas of strategic importance. Trust in data ecosystems requires that homegrown legislation is strengthened to protect the rights of people, at the same time ensuring that it is in sync with the operations in the rest of the world.

# 7. Risk

**Opening Remarks** by **Abhishek Gupta (Founder, Montreal AI Ethics Institute)**

I've been a proponent now for the incorporation of a risk-aware methodology in evaluating machine learning systems, especially from the perspective of machine learning security: something that gets very little attention at the moment in the discussions on AI ethics. While risk is talked about from the perspective of organizational and reputation management, there is a lot more to be unpacked as this chapter showcases.

According to Turing and Church, the complexities of some programs are such that there isn't a way to ascertain their outcomes prior to the actual execution of those programs; building on that, we can see that AI shares some of those similarities: especially when it comes to tackling the unpredictability of AI systems as highlighted in a segment of this chapter. This ties in with the notion of how these systems are often seen as black boxes and hence the demand for explanations of their behaviour. Yet, this overlooks the "token human" problem where humans are essentially powerless when placed in the loop of these autonomous systems and asked to sign-off on the outputs from the systems. This is exacerbated by the fact that humans can lean towards becoming over-trusting of the outcomes from the system.

While we had all been promised flying cars time and again, for example at various World Fairs of yesteryears, when it comes to things like autonomous vehicles (AVs), we see that we are still a long ways away from them becoming a widespread reality on our roads. This is perhaps due to a lack of uniform regulations, doubt in the robustness of the ML systems onboard, and a lack of social acceptance which is alluded to in the calls for explanations on how systems make decisions, often exemplified in scenarios like the "Trolley Problem" (which for the most part is an unrealistic depiction of what actually happens in AVs).

So, when talking about machine learning security, we have to be considerate of the existing brittleness of the systems, as exhibited by flaws whereby we can mount evasion attacks against the processes of the systems through the use of mechanisms like data poisoning, model inversion, model dumping, etc. Being cognizant of the capabilities of the capabilities of the adversaries who try to compromise the systems can help us develop better defenses against such attacks. As an example in this chapter, Facebook adopts a "red team" approach to address some of these challenges.

A couple more interesting examples in this chapter like the ability to pick locks by capturing the audio when you are attempting to insert a key and unlock your front door present other vectors of attacks on physical systems which we can't even fathom; hence also the need to support the research into evaluating the risk of deploying sensors widely and the potential risks that it poses to other security systems that we have in the world. Another example that was really interesting was how we can compromise the safety of machine learning systems by triggering high-energy consumption in them by sending in crafted examples that can reduce the availability of the systems at inference time.

I envision a time where we become more cognizant of the risks that the creation and deployment of our machine learning systems pose, not just from the currently popular ethics frameworks but also from a cybersecurity and traditional physical security perspective. I encourage you to dive deep into the content in this chapter and share it with your colleagues to broaden the discussion on these issues.

Abhishek Gupta (**@atg_abhishek**)
Founder & Principal Researcher, Montreal AI Ethics Institute

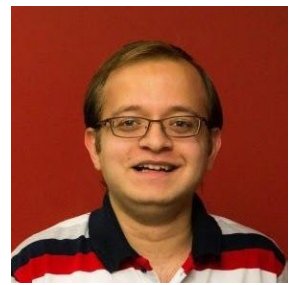

Abhishek Gupta is the founder and principal researcher at the Montreal AI Ethics Institute, seeking to define humanity's place in a world increasingly characterized and driven by algorithms. He is also a machine learning engineer at Microsoft, where he serves on the CSE Responsible AI Board. His book '[Actionable AI Ethics](#)' will be published by Manning in 2021.

# Go Deep: Research Summaries

### Evasion Attacks Against Machine Learning at Test Time
(Original paper by Battista Biggio, Igino Corona, Davide Maiorca, Blaine Nelson, Nedim Srndic, Pavel Laskov, Giorgio Giacinto, Fabio Roli)

Machine learning adoption is widespread and in the field of security, applications such as spam filtering, malware detection, and intrusion detection are becoming increasingly reliant on machine learning techniques. Since these environments are naturally adversarial, defenders cannot rely on the assumption that underlying data distributions are stationary. Instead, machine learning practitioners in the security domain must adopt paradigms from cryptography and security engineering to deal with these systems in adversarial settings.

Previously, approaches such as min-max and Nash equilibrium have been used to consider attack scenarios. However, realistic constraints are far more complex than these frameworks allow, and so we instead see how practitioners can understand how classification performance is degraded under attack. This allows us to better design algorithms to detect what we want in an environment where attackers seek to reduce our ability to correctly classify examples. Specifically, this work considers attacks on classifiers which are not necessarily linear or convex.

To simulate attacks, two strategies are undertaken:

- "Perfect Knowledge" – this is a conventional "white box" attack where attackers have perfect knowledge of the feature space, the trained model itself, the classifier, the training data, and can transform attack points in the test data within a distance of $d_{max}$.

- "Limited Knowledge" – In this "grey box" attack, the adversary still has knowledge of the classifier type and feature space but cannot directly compute the discriminant function $g(x)$. Instead, they must compute a surrogate function from data not in the training set, but from the same underlying distribution.

The attacker's strategy is to minimize the discriminant function $g(x)$ or the corresponding surrogate function in the limited knowledge case. In order to overcome failure cases for gradient descent-based approaches, a density estimator is introduced which penalizes the model in low-density regions. This component is known as "mimicry" and is parametrized by $\lambda$, a trade-off parameter. When $\lambda$ is 0, no mimicry is used, and as $\lambda$ increases, the attack sample becomes more similar to the target class. In the case of images, this can make the attack sample unrecognizable to humans.

The first "toy" example used is MNIST, where an image which is obviously a "3" to human observers is reliably misclassified as the target class "7" against a support vector machine.

The task of discriminating between malicious and benign PDF files was also addressed, relying on the ease of inserting new objects to a PDF file as a method of controlling d☐$_{ax}$. For the limited knowledge case, a surrogate dataset 20% of the size of the training data was used. For SVMs with both linear and RBF kernels, both perfect knowledge and limited knowledge attacks were highly successful both with and without mimicry, in as few as 5 modifications. For the neural network classifiers, the attacks without mimicry were not very successful, though the perfect knowledge attacks with mimicry were highly successful.

The authors suggest many avenues for further research, including using the mimicry term as a search heuristic; building small but representative sets of surrogate data; and using ensemble techniques such as bagging or random subspace methods to train several classifiers.

## Sponge Examples: Energy-Latency Attacks on Neural Networks
([Original paper](#) by Ilia Shumailov, Yiren Zhao, Daniel Bates, Nicolas Papernot, Robert Mullins, Ross Anderson)

Energy use is an important and yet understudied aspect of Machine Learning (ML). Energy consumption can help us gauge the environmental impacts of ML, for one. In this paper, Shumailov et al. show how energy use can also be used for nefarious purposes through sponge examples: attacks made on an ML model to drastically increase its energy consumption during inference. Sponge examples, of course, can make an ML model's carbon emissions skyrocket, but they can also cause more immediate harm. Indeed, increased energy consumption can significantly decrease the availability of the model, increase latency and ultimately delay operations. More concretely, autonomous vehicles undergoing a sponge attack may be unable to perform operations fast enough due to this delay, causing a vehicle to fail to break in time, leading to a collision.

Shumailov et al. propose two hypotheses as to how sponge examples can be generated. For one, a sponge example can exploit how sparsely activated some

The State of AI Ethics, October 2020                                                                 109

hidden layers of the neural network are when the sum of inputs to a neuron is negative. By adding inputs that lead to more neuron activations, this increases the model's energy used because of the larger amount of operations performed.

Secondly, sponge examples can soak up large amounts of energy by exploiting the energy-latency gap: "different inputs of the same size can cause a deep neural network (DNN) to draw very different amounts of time and energy" (Shumailov et al., 2020). The authors use Transformer as their example — an ML model that takes words as its data. The token input size and the token output size (the number of individual words), as well as the input and output embedding spaces' size, can be increased by a remote attacker with no access to the model's configuration or hardware. These increases can yield non-linear increases in energy use; in other words, energy consumption goes up exponentially as token input, token output, or embedding space size increase linearly.

The paper explores three threat models. First, a white box setup, where the attacker knows the model's parameters and architecture. Second, an interactive black box setup, where the attacker does not know the parameters and architecture of the model, but can measure the energy consumption remotely as well as the time necessary for an operation to run. Third is the clueless[1] adversary setup, where the attacker has none of the information of the two prior setups, and can only transfer sponge examples to this new model without previously having interacted with it.

In the cases of the white box and interactive black box setups, an attacker can create a sponge example attack through genetic algorithms. In this context, a genetic algorithm would continually select for the top 10% of inputs with the highest energy consumption, these becoming the "parents" for the next "generation" of inputs. Genetic algorithms can thus help maximize the damage a sponge example attack can have by providing inputs that consume extremely high amounts of energy.

In a white box setting, an attacker can likewise launch a sponge example attack by using an L-BFGS-B algorithm to generate inputs that increase all the activation values throughout the model, forcing more operations to be undertaken and causing energy consumption to surge.

As for the clueless adversary setup, the energy consumption of hardware (CPUs, GPUs, and ASICs) can be determined without an attacker having access to the model (through calculations or through the NVIDIA Management Library, for instance). The authors perform experiments on NLP (Natural Language Processing) tasks and Computer Vision tasks to evaluate the performance of sponge examples across models, hardware, and tasks. Shumailov et al. find that sponge examples are transferable across both hardware and models in the white box setup, in the interactive black box setup, and even in the clueless adversary setup, where performing an attack is most difficult.

To defend against an adversary exploiting a sponge example, the authors suggest two methods. First, a cut-off threshold, where the total amount of energy consumed for one inference cannot be higher than a predetermined threshold. This could prevent sponge examples from impacting the availability of the machine learning model. This, however, applies to scenarios where battery drainage is the main concern.

To address delays in real-time performance, which could have deadly consequences in autonomous vehicles or missile targeting systems, the authors believe these systems must be designed to function properly even in worst-case performance scenarios, and perhaps be equipped with a fallback mechanism for instances where these systems fail completely.
The paper ends on a call for more research to be done regarding the carbon emissions of machine learning at the stage of inference. Most of the research done on this topic focuses on training large neural networks, but the authors highlight that inference is done much more frequently and on a larger scale than training once a model is deployed.

# Go Wide: Article Summaries

### Facebook's 'Red Team' Hacks Its Own AI Programs
(Original *Wired* article by Tom Simonite)

We are increasingly relying on content moderation by automated means because of the rise in the amount of content produced and the detrimental mental health effects faced by human content moderators. Given the current limitations of AI systems, we have settled into an adversarial setting between content moderation systems and malicious actors who try to slip content by the moderation system.

Facebook found that as they tried to curb nudity in their platform families, content creators came up with creative ways to elude censorship. When the team created new fixes to address those "attacks", they found that the creators came up with other novel ways of evading filters.

The CTO of Facebook mentioned that as AI systems become more integral to their production systems, it has become even more critical to ensure that they are robust against such circumvention attempts. In the same vein, to understand deepfakes in the wild and how to limit their spread, they launched the DeepFake Detection Challenge, which was an attempt at surfacing techniques that might be effective in countering the rising ubiquity of these tools. Building on the work of some academic researchers, it is crucial to develop tooling similar to the ones used in cybersecurity that can be integrated into existing developer workflows to enhance the robustness of AI systems.

### Why Asking an AI to Explain Itself Can Make Things Worse
(Original *MIT Tech Review* article by Will Douglas Heaven)

Of late, there have been a lot of calls to have explainable AI systems. This article goes into detail on why that might be problematic. One of the reasons to ask for explanations is so that people have the chance to understand why a system made a decision and be empowered to agree or disagree with it. The article talks about glassbox models that are simplified versions of neural networks that enable easy tracking of how data gets used and how decisions get made. But, for more complicated problems where simplistic approaches don't work well, there is a need to rely on more complex models. There is still the potential to use glassbox models such that they get used initially to identify problems with the datasets followed by a more complex model used on top of that. Visualizations are also known to aid in explainability.

But, the researchers point out pitfalls with this demand for explainability. Specifically, they emphasize how humans tend to overtrust the system when provided explanations leading to the token human problem. As an example, the researchers talk about visualizations and how the people they surveyed didn't even understand what the visuals indicated. A phrase commonly used to describe this behavior dubbed "mathwashing" is where humans trust machines because of their numerical nature. One of the goals for these explanations is to provoke critical reflection, and not to elicit blind trust from humans. Tailoring the explanation to the audiences will also enhance their utility.

## Unpredictability of Artificial Intelligence
([Original *Hacker Noon* article](#) by Roman Yampolskiy)

AI governance and safety literature aims to implement intermediary controls to enhance security and produce leading indicators to predict outcomes from the system. This article dives into how complex AI systems are inherently unpredictable through examples of Rice's theorem and Wolfram's Computational Irreducibility. It states that we can't accurately comment on the transitory steps that the AI system will take even if we have full knowledge of the end goals of the system. It can be formally measured by Bayesian surprise that calculates the difference between the prior and posterior beliefs of an agent.

The author further motivates this by expounding on Vinge's principle and instrumental convergence. Specifically, in designing an advanced agent, we might have to approve the design of the system even if we don't know all the decisions that it might take. We can surmise the potential purposes for which it was built based on the designs which we observe even if we don't know the precise goal.

Cognitive uncontainability further expands on this idea by pointing out that a human mind cannot perceive all decisions that an agent might take, especially one that might have access to facts and knowledge that are invisible to us. A couple of examples that illustrate that are partial knowledge of an inherently rich domain (such as human psychology) and projections of the future (such as 10th-century humans not knowing the capabilities of humans in the 21st century). Alonzo Church and Alan Turing had mathematically proven that it is impossible to ascertain that an algorithm fulfills several properties without executing it. Finally, AI systems can

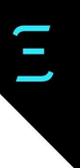
become complex enough to be potentially threatening to human safety, hence the domain of AI safety warrants investigation.

## Autonomous Cars: Five Reasons They Still Aren't on Our Roads
(Original *The Conversation* article by John McDermid)

The pandemic has certainly thrown a wrench in terms of testing of autonomous vehicles (AV) that, at present, require the presence of human overseers. This article highlights some of the main challenges that the industry will have to overcome in making the presence of AVs in everyday settings a reality.

The current crop of sensors used on AVs is susceptible to failure in bad weather, heavy traffic, unexpected situations, graffiti on traffic signs, and other naturally occurring adversarial items. These sensors will need to be universal in their performance so that systems trained in one place might be amenable to deployment in another with different circumstances. The underlying machine learning systems that do object detection, path planning, and other operations don't yet have standardized approaches for training, validation, and testing, which is essential for benchmarking and safety. In an online learning context where these systems will learn on the open road, we need mechanisms that we can certify in terms of their safety even after updates to them through the learning that they have post-deployment.

Recognized regulations and standards across the industry that square with existing vehicle safety checks need implementation before there is widespread deployment. Finally, people need convincing from a social acceptance standpoint that the systems are safe for use. Without that, we risk abandonment and distrust that will kill the industry.

## Picking Locks with Audio Technology
(Original *CACM* article by Paul Marks)

The article highlights some recent research work that demonstrates an attack on how to break physical locks using captured audio samples from the insertion of a key into a slot to unlock the door. The metallic clicks as the key goes into the lock can be used to compute the depth of the ridges on a key that would be required so that a duplicate key can be 3D-printed to give someone access to a secured space. The way pin-tumbler locks work is that the ridges on the key alter the height of the pins attached to springs inside the lock and once they are all in the right position it allows the tumbler to turn and unlocks the door. Typically for the most common kinds of these locks there are around 330,000 combinations possible for the patterns on the key. The research approach presented in the article narrows that down to the 3 most likely key patterns.

The research group that produced this work works on harvesting audio and other signals in the environment that are typically of little value and translate them into attacks on the security of systems. As an example, they mention how the gyroscope sensors on a smartwatch and tracking the rotatory movements can help unearth the combination of a safe lock that works through rotation. One of the things that presents a limitation to their current approach is in the quality of audio samples that they are able to capture. Additionally, there are also limitations in terms of interference emerging from other sounds around in the environment, for example, the jingling of other keys, traffic, and other sounds that are typically found in the real-world.

Some of the limitations can be overcome through the attack vectors that were identified in the paper, including the use of machine learning over repeated samples that can help to eliminate noise, and provide a higher degree of accuracy. Yet, it remains a pipe dream at the moment in terms of mounting realistic attacks but as the article mentions, it sounds like a fun idea to insert into a movie plot.

# 8. The Future of AI Ethics

**Opening Remarks** by Rumman Chowdhury (Global Lead for Responsible AI, Accenture)

Nearly four years ago, when I started my job as Accenture's Global Lead for Responsible AI, I used to begin my talks with a slide that stated: "There are three things I don't talk about: Terminator, Hal, and Silicon Valley entrepreneurs saving the world." It was not only meant to get a laugh out of the crowd but to get them thinking about something quite important. While "AI" sounds like a future technology plucked from movies, like most things that matter, it's the seemingly mundane applications of this technology that have the most impact.

It's been a wild ride, these past few years, and our little community has gone from largely unnoticed positions at companies, small meetings in forgotten back rooms of universities, and social media fringes, to invitations into some of the most powerful rooms in the world. We did so not with flashy promises that we're the saviors who will fix the world, but by discussing real human problems and encouraging constructive dissent from a wide range of disciplines to ground technological promises in reality.

The backlash to our efforts has been to paint us as tech-ignorant and pessimistic, versus the "tech-forward" folks who imagine a 'Star Trek future.' As Spock would say, "Fascinating". Star Trek is one of the few tech-utopian visions that openly encouraged discussion of the impact of technology on society, culture, class, the environment, and more. The most frightening enemies in Star Trek lore are the Borg — part-robots who were once vibrant living creatures, but forcibly merged with technological 'improvements' and integrated into a subservient collective in a manner that stripped them of individuality, personality, and meaning in their lives, all to serve a never-ending pursuit of technological greatness. The Borg act as a warning that developing technology for technology's sake is purposeless, stark, and empty.

Painting a false adversarial binary (ethics 'pessimists' vs the tech-forward 'optimists') creates a narrative that's not conducive to positive change. The tech ethics community are also optimists. Instead of simply building technology that reinforces existing broken institutions, we want technology that embraces the complexities of the current state of humanity and gets us from where things are, to where things ought to be.

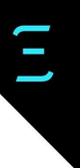

The subsequent articles in this section discuss the future of AI ethics in a manner that reflects this mission. Articles on pressing social issues such as mass incarceration, contact tracing, and labor mingle with business imperatives such as risk management, organizational governance, and applied research on algorithmic transparency. Encapsulated in this discussion is the future I hope we see in technology and in general - substantive, useful, and forward-thinking, respectful of humanity, and constructively critical of misuse with a willingness to collaborate and improve.

The future of AI ethics is, simply put, the future of us.

Rumman Chowdhury (**@ruchowdh**)
Global Lead for Responsible AI, Accenture

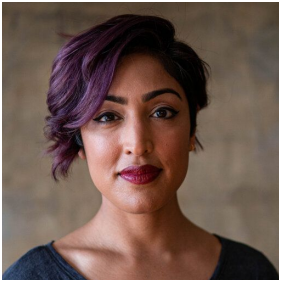

Rumman Chowdhury's passion lies at the intersection of artificial intelligence and humanity. As the Global Lead for Responsible AI at Accenture Applied Intelligence, she works with C-suite clients to create cutting-edge technical solutions for ethical, explainable, and transparent AI. In addition, she is a CADE Ethical Entrepreneurship Fellow at the University of San Francisco. Dr. Chowdhury is also the founder and CEO of Parity, a startup dedicated to identifying and mitigating the risk and harm introduced by AI systems.

# Go Deep: Research Summaries

## AI Governance in 2019, A Year in Review: Observations of 50 Global Experts
([Original report](#) by Shanghai Institute of Science of Science)

2019 has seen a sharp rise in interest surrounding AI Governance. This is a welcome addition to the lasting buzz surrounding AI and AI Ethics, especially if we are to collectively build AI that enriches people's lives.

The AI Governance in 2019 report presents 44 short articles written by 50 international experts in the fields of AI, AI Ethics, and AI Policy. Each article highlights, from its author's or authors' point of view, the salient events in the field of AI Governance in 2019. Apart from the thought-provoking insights it contains, this report also offers a great way for individuals to familiarize themselves with the experts contributing to AI governance internationally, as well as with the numerous research centers, think tanks, and organizations involved.

Throughout the report, many experts mention the large amount of AI Ethics principles published in the past few years by organizations and governments attempting to frame how AI should be developed for good. Experts also highlight how, in 2019, governments were slowly moving from these previously established ethical principles towards more rigid, policy measures. This, of course, is far from accomplished. Currently, many governments are [holding consultations](#) and [partnering with organizations](#) like MAIEI to help them develop their AI strategy. Authors of the articles featured in this report also suggest considerations they deem necessary to getting AI governance right. For one, Steve Hoffman (pp. 51-52) suggests policymakers take advantage of market forces in regulating AI. FU Ying (pp. 81-82) stresses the importance of a China-US partnership regarding AI, for which better relations between both governments are necessary.

On another note, the release of gradually larger versions of OpenAI's GPT-2 language model and the risks around its publication are mentioned by many authors as a salient event of 2019. For many, this brought up issues surrounding responsible publishing in AI, as well as more general concerns around how AI may be used to do harm. The report even features an article written by four members of OpenAI discussing the event and its impact on the discussion concerning publishing norms in AI (pp. 43-44).

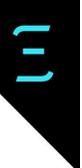

One expert, Prof. Yang Qiang, also mentions the importance of new advances like federated learning, differential privacy, and homomorphic encryption, and their importance in ensuring that AI is used to the benefit of humanity (pp. 11-12). In his article, Prof. Colin Allen, highlights a crucial but oft forgotten element of good AI governance: strong AI journalism (pp. 29-30). He writes: "The most important progress related to AI governance during the year 2019 has been the result of increased attention by journalists to the issues surrounding AI" (p. 29). It is necessary for policymakers, politicians, business leaders, and the general public to have a proper understanding of the technical aspects of AI, and journalists play a large role in building public competence in this area.

It's interesting to note that the report was released by the Shanghai Institute of Science for Science. Its editor-in-chief (Prof. SHI Qian) and one of its executive editors (Prof. Li Hui) are affiliated with this Institute, and the report features numerous Chinese AI experts. In light of this, it is particularly refreshing to see such a collaboration not only between Chinese and American or British experts, but also with other scholars from around the world. Efforts in AI governance can easily become siloed due to politics and national allegiances. This report, thankfully, does away with these to privilege an international and collaborative approach. In addition, twenty of the fifty experts featured are women, and many of them are at the beginning of their careers. This is commendable, considering the field of AI tends to be male-dominated. However, none of the fifty experts featured in the report are Black. This is unacceptable. There are numerous Black individuals doing innovative and crucial work in AI, and their voices are central to developing beneficial AI. I encourage our readers to engage with the work of Black AI experts. For one, start by listening to this playlist of interviews from the TWIML podcast, which features Black AI experts talking about their work. If a similar report on AI governance is put together next year, it must include the perspectives of Black AI experts.

## Classical Ethics in A/IS
([Original document](#) by the IEEE)

The ethical implications of autonomous and intelligent systems (A/IS) are, by now, notably numerous and complex. This chapter of Ethically Aligned Design: A Vision for Prioritizing Human Well-being with Autonomous and Intelligent Systems by The IEEE Global Initiative on Ethics of Autonomous and Intelligent Systems simultaneously adds some definition to the issues surrounding the ethics of autonomous systems by providing clear analysis and recommendations. The topics of inquiry covered in the paper are wide, and each is given between two and four pages of background and subsequent recommendations. At the end of each section, readers will be happy to find a list of further readings if they wish to dive deeper into a specific topic.

The topics covered include, for instance: making ethical concepts and philosophical vocabulary accessible to programmers, policymakers, companies, and other stakeholders; the importance of considering Buddhist, Ubuntu, and Shinto ethics along with typically Western ethical traditions like consequentialism, deontology, and virtue ethics; an overview of what Buddhist, Ubuntu, and Shinto ethical perspectives can contribute to the discourse surrounding A/IS; the impact of automation in the workplace; the importance of maintaining human autonomy; and the implications of cultural migration for A/IS. The overarching theme that unites these specific topics and the others presented in the paper is the role of classical ethics in the creation, implementation, and use of A/IS. The insights provided on these topics and others covered in the paper can be understood as offering meta-analysis of how ethics and A/IS should interact.

Three features of this piece standout: its use of "A/IS" instead of "AI", its balanced view concerning ethical implications, and that it features Buddhist, Ubuntu, and Shinto value systems.

First, the authors make a point of using "A/IS" (autonomous and intelligent systems), instead of AI (artificial intelligence) throughout their analysis. This may seem innocuous, but it highlights the authors' commitment to taking a critical stance towards applying "classical concepts of anthropomorphic autonomy to machines" (p. 37). Using A/IS instead of AI limits these potentially misleading connotations, and gives a more nuanced description of these technologies. This is significant, according to the authors, as it helps orient ethical concerns with regards to A/IS towards established ethical issues, instead of conjuring up insubstantial ones because of misleading terminology.

Second, the paper presents a balanced view of the expected ethical implication of A/IS. While it makes clear that A/IS carry important ethical risks, it also emphasizes that A/IS have had and can continue to have very positive effects on societies. In addition, the authors make tangible recommendations to help make ethics more accessible, more representative, and better-implemented into A/IS.

Third, the authors include Buddhist, Ubuntu, and Shinto ethics in their analysis, and advocate for these traditions' inclusion in the wider debate about the ethical implications of A/IS. They highlight the possible importance of the Buddhist view on privacy in the context of A/IS. For instance, from a Buddhist perspective, privacy is not merely an individual protection, but a necessity "for a well-functioning society to prosper in the globalized world." In addition, the paper emphasizes how

Buddhism, Ubuntu and Shinto traditions are marked by relationships – not only with other human beings, but with oneself, too, as well as with A/IS.

The authors also explain why the hegemony of typical Western ethics is cause for concern. One central impact of Western ethics' monopoly on the A/IS ethics discourse is related to standardization as highlighted by Pak-Hang Wong, and which the authors behind the IEEE Global Initiative on Ethics of Autonomous and Intelligent Systems cite. According to Wong, making Western ethics the standard (in this case, specifically to evaluate A/IS systems) is a problem because it indirectly assigns greater value to Western ethics than other traditions, establishing the former as "the normative criteria for inclusion to the global network" (Wong, 2016). Following this, the authors note that, in light of this standardization, it falls on those who work outside the accepted standard to devise ways to include other ethical traditions and value systems. They then mention that "liberal values arose out of conflicts of cultural and subcultural differences and are designed to be accommodating enough to include a rather wide range of differences." (pp. 50-51)

While this may be true, it sidesteps the fact that even though liberal values may be flexible enough to eventually accommodate different value systems, such accommodating is unlikely to come without at least some pushback. What is more, it appears this pushback may overwhelmingly place the burden on individuals from outside the West and/or are already marginalized to advocate for the inclusion of Buddhist, Ubuntu, Shinto, and other perspectives. Considering the power and dominance Western ethics and values possess, it is clear that the struggle to get Eastern and other ethical traditions and value systems (like Indigenous ones) "accommodated" by liberal values will be an unequal one. Later on in the text, the authors nonetheless highlight that "intentionally making space for ethical pluralism is one potential antidote to dominance of the conversation by liberal thought, with its legacy of Western colonialism." Indeed, if we are to include ethical traditions other than the Western one and make the struggle to do so fairer, the ethics of A/IS community will have to work to create space for these non-Western value systems.

## Principles Alone Cannot Guarantee Ethical AI
([Original paper](#) by Brent Mittelstadt)

AI Ethics has been approached from a principled angle since the dawn of the practice, drawing great inspiration from the medical ethics field. However, this paper advocates how AI Ethics cannot be tackled in the same principled way as the medical ethics profession. The paper bases this argument on features of the medical ethics field that its AI counterpart lacks, and then aims to suggest ways

The State of AI Ethics, October 2020                                                                                          121

forward. Taking this into account, I will split this post into 3 sections to demonstrate how this is the case. Section 1 will show what the paper believes the AI Ethics field lacks compared to the medical field, section 2 will be how this is the case, and section 3 will be how this is to be resolved. I will then end with my thoughts on the discussion.

**Section 1: What the AI Ethics field lacks**

Firstly, the practitioners in the AI Ethics field all lack a common aim or 'patient' that can align all the differing interests of the different institutions involved. The field is filled with different practitioners of diverse backgrounds, and private companies all with varying interests. Hence, a principled approach here would have to unite these differing views under the maxims it proposes. However, in order to accommodate all the different viewpoints, the principles start to become more and more abstract. Proposals such as 'fair' and 'equal' end up being the point of agreement for all parties, which this paper highlights as hiding the "fundamental normative and political tensions embedded" in these concepts (Mittelstadt, 2019, p. 1). For example, there are deep disagreements over what equality actually means, such as whether it purports to egalitarianism or complete equality for all (such as wage distribution). Medical ethics instead can unite on the subject of a patient, and prioritise their interest in their methods, forming a focus point for the differing views within the field. This is then further enforced by medical bodies being rigorously reviewed by legally backed institutions to make sure this prioritisation is taking place, with no such body existing in the AI Ethics field yet. Hence, a principled approach to said field may not be the most fruitful path to undertake.

**Section 2: Why is this the case?**

The paper then proposes that a principled approach as such is hindered by the field not having an established history. There are no previous lessons to draw on in order to demonstrate what "good" AI is. There is no 'AI Hippocratic Oath' to undertake for behaviour to be modelled on, and the unpredictability of AI means that any one single method can't be guaranteed to always produce a 'good' result. Instead, each company is almost left to forge their own practice, tailored to their own company values. Resultantly, each company produces their own exemplars of how 'good' AI is deployed, leaving little scope for principled practical advice to influence how to implement ethical AI.

Such lack of advice then emphasises the importance of accountability when deploying AI, as there is no regulation apparent to signify what is seen as 'bad' AI. Even then, the AI Ethics field lacks the accountability framework to counterbalance the lack of regulation. With many different actors involved in processes that are hard to trace all the way back, it would be difficult to pin the responsibility on any one person. Whereas, the medical ethics field has a fixed team of actors at any one time, making a stronger case for the presence of accountability. Thus, approaching the AI Ethics field in the same way as the medical ethics arena may in fact be like mixing oil and water.

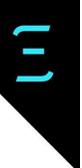

**Section 3: Potential solutions**

The paper then concludes by offering some ways forward for the AI Ethics field. Defining clear pathways that are most likely to end up in ethical AI will then help foster support for more emphasis on a "bottoms-up" (Mittelstadt, 2019, p. 9) approach to AI deployment. Such an approach will help generate novel problems that repeatedly face the field of AI Ethics, generating methods on how to tackle them rather than seeing similar problems surfacing from the companies at the top. This may then lead to AI deployment being crafted as a licensed profession, which can be utilised by both large and small corporations. Such a licensing can then smooth the approach away from individual AI Ethics, and more towards organisational ethics being considered. Individuals corrupting the use of AI will be held accountable as well as the corporations who allowed it to happen, with their role being previously left unquestioned. In this way, a principled approach to AI Ethics as seen in medical ethics will be better able to take form.

# Virtues Not Principles
**(Op-ed by Ryan Khurana of the Montreal AI Ethics Institute)**

There has been a recent explosion of interest in the field of Responsible Artificial Intelligence (aka AI Ethics). It is well understood that AI is having (and will continue to have) significant social implications as the technology currently exists. Concerns over bias, fairness, privacy, and labour market impacts do not require the arrival of some mythical Artificial General Intelligence (AGI) to be socially destructive. Organisations from around the world are publishing frameworks for the ethical development and deployment of AI, corporations are affirming their commitment to social responsibility by funding responsible AI research and creating review boards, and a political interest in AI has led to state guidelines from numerous countries. Despite this, to many observers, little feels to have changed. Algorithms with suspected bias continue to be used, privacy continues to be a major concern, and accountability is near non-existent.

The failure of progress in responsible AI stems from a mistaken approach which prioritises good outcomes over good behaviour. It is uncontroversial to say that bias is bad and that privacy is good, but what this means in practice is more contentious. By attempting to simplify the work of achieving good outcomes to "frameworks" or "principles" the work being done in the field risks bearing little fruit. Our understanding of how AI systems can lead to problematic social outcomes is inherently reactive, in that we respond to problems that can be documented. The goal of responsible AI, however, is to be proactive by anticipating potential harms and mitigating their impact. Checklists on what ought to be done can never achieve the full range of potential risks that responsible AI seeks to address, and as a result are inherently limited. Proactive concern with socially beneficial outcomes requires not just work on frameworks for ethical use, but the cultivation of virtuous technologists and managers, who are motivated to take the concerns of responsible AI seriously.

The importance of virtue is clear when we consider the failure modes of principles-driven approaches to Responsible AI: non-adoption and recuperation. Either principles will be ignored, or they will be contorted to serve the interests of the status quo.

The risk of non-adoption is currently the more pervasive problem facing responsible AI work. In the rush to develop principles and guidelines competing approaches fail to generate consensus, and as a result have low impetus for adoption. Where a set of guidelines have low implementation costs, they may be championed to little or no effect. Where the burden of principles is too high, nobody cares to use them. If work is too technically focussed, it's hard to communicate what has been achieved. If work is too value driven, it's hard to audit whether anything has been done and it has limited practical applicability.

Even if consensus were to exist and the ideal set of principles formulated, this would not in and of itself motivate adoption. A lot of work has gone into consensus building among ethicists and commoditising the various implementations of responsible AI tools such as differential privacy, but this alone does not make technologists or business leaders care. Regulatory approaches seek to create the right incentives for following ethical guidelines by penalising bad behaviour, but these run into huge cost barriers to enforce and are slow to develop and diffuse. Even if ideal enforcement mechanisms were discovered, they would fail to create the proactive concern with social outcomes that responsible AI practitioners desire. Good incentives are no substitute for good citizens.

The degradation of responsibility and civic duty that incentive driven adoption creates leads to the second failure mode of current responsible AI approaches, recuperation. Recuperation is the risk of sincere work being co-opted by those who are in power. The common cries of "ethics washing" and "ethics theatre" that dominate high profile efforts by corporations to respond to the concerns of responsible AI practitioners reveal that without a sincere motivation to care, principles can be reduced to PR gimmicks.

Technical jargon can bury sincere conversations about social consequences, making it seem as if lots is being done. Review boards that hold no actual power over decisions become poster children for corporate efforts. Internal memos may normalise a certain sort of discourse around responsibility, with little to no actual understanding of the content. Regulatory compliance may become a way to capture markets by increasing barriers for new entrants. Even without the malice associated with naked self-interest, recuperation can result from laziness. If ethical concerns are reduced to checklists one simply has to tick off, the letter of the law may be followed, but its spirit completely lost. This would require the field of responsible AI to remain in constant vigilance to identify new risks and formulate new approaches for everyone to follow. In doing so, the law would get larger, but adherence would decline.

None of this is to say that work on principles is inherently useless, but rather that good behaviour relies on good people. Principles are put into place by people, and they embody the character of those who implement them. In order to achieve the goals of the responsible AI community, making people care about being good needs to be the primary goal. This requires a cultivation of virtue.

Dating back to Ancient Greece the concept of the Cardinal Virtues has been a pillar in moral philosophy as prerequisites for living a good life. A virtuous person is one capable of doing good, and they provide a bedrock upon which the internal motivation to do the right thing rests. Traditionally, the Cardinal Virtues are Prudence, Fortitude, Temperance, and Justice. By focussing on cultivating these virtues among the practitioners who develop and deploy AIs, from researchers to ML engineers to data scientists to project managers to executives, the responsible AI community would be more successful in ensuring proactive social concern. Each of these virtues is essential, and none on their own is sufficient to guarantee prosocial outcomes.

The first of these virtues is Prudence, which can be roughly equated to foresight or practical wisdom. It is probably the only virtue that is emphasised in education and work environments today. A prudent person is able to judge what the right thing to do is at the right time. The ability to make rational judgements about what one can best spend their time doing, what is the best use of the resources at their disposal, and whether a risk is worth taking are the fruits of prudence.

In the AI context, prudent practitioners are able to understand the consequences of what they are building, accurately describe the limits of their programs, and evaluate the opportunity costs of solving different problems using AI or solving a problem using AI rather than more conventional methods (such as human labour or rules-based programs). All of these decisions are critical for evaluating whether

The State of AI Ethics, October 2020                                                              125

the outcomes of a system are socially beneficial or harmful, and the process by which this understanding comes about cannot be reduced to a checklist.

The next virtue is Fortitude, which can be equated with moral courage or perseverance. Fortitude enables one to do the right thing no matter the personal consequences or the stigma associated with a course of action. There are fields where fortitude is emphasised, such as in the military and in nuclear power management, where there is a pervasiveness of uncertainty and the consequences of error are disastrous. A lack of fortitude among AI practitioners poses many risks as without fortitude those who notice a problem may be afraid to speak up, the voice of one's conscience may be drowned out by a desire to go with the flow or to protect one's career, and doubt may confuse one's ability to make effective judgements. Cultivating fortitude goes beyond the education of what the right thing to do is that currently grabs the attention of the responsible AI community by providing practitioners with the tools to act on the right thing.

Then there is Temperance, which can be equated with self-restraint or moderation. Plato himself considered this the most important of all the virtues as it enabled one to be humble and avoid acting rashly. Temperance enables people to understand their own failings and critically examine their own whys for acting. This virtue has played a foundational role in most of the world's great religions and cultures, though its popular emphasis ebbs and flows. Organisations such as the Boy Scouts seek to cultivate this virtue, as do philosophy and meditation. The risks of intemperance for AI are severe as it would allow personal desires, for recognition or power or profit or even pure intellectual curiosity, to cloud judgement. Without temperance practitioners would be less willing to acknowledge their own biases and limitations, and as a result may refuse to acknowledge the harms that could be caused by what they develop.

Finally, there is Justice, which encompasses concepts of fairness and charity. The scales of justice in classical iconography sum up the balance between selfishness and selflessness that defines the just person. The golden rule that one ought to treat others as one ought to be treated cuts to the heart of just behaviour. The concerns of justice are front and centre in the work of most responsible AI researchers, as they seek to ensure that AIs do not discriminate against the disenfranchised and that diversity among practitioners provides representation to all voices. The risks of being unjust are evident and well understood, that the development of AIs will prioritise the people building it or people like them at the expense of the whole. Despite this understanding, the development of the moral sense of justice is typically absent from most schools and workplaces, with a carrot and stick approach often being used to enforce concern with social justice.

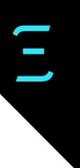

The value of these virtues I hope have been made clear, not only in their necessity in creating the motivation to develop and deploy responsible AI, but also in their ability to help foster a good society filled with good people. Formation in the virtues, however, is quite uneven, with prudence and to a lesser extent justice being the primary focus, with fortitude and temperance being more niche concerns. Calls to make ethical reasoning core to AI education should make personal virtue their aim, providing the tools for students to reason about the particulars on their own. At the Rotman School of Management, my alma mater, the business ethics curriculum was recently redesigned, moving away from a more legalistic compliance curriculum to the study of Aristotle's Nicomachean Ethics and a focus on the personal moral inquiry of students and their character. Should this be a more popular approach, the benefits would be immeasurable. But the development of virtue does not end in school, it is rather a life-long task.

To be trained in virtue is to be given the opportunity to develop the right habits and frames of mind. This often comes though having good routines and being faced with challenges that exercise one's moral faculties. Some ideas for how workplaces can cultivate this include:

- Regular workplace exercises that have individuals and teams solve complex moral problems from day to day life would allow for critical inquiry of values and enable character development. Michael Sandel's Harvard Justice course provides a lot of insight on how to organise such problems and engage people with sound ethical reasoning.

- Encouraging time for reflection and philosophical reading would allow workers to develop their moral faculties. Scheduled unstructured time is critical for children's moral development, but it is often lacking in schools and homes. There is no reason to believe that providing it for adults would fail to bring about the same results in allowing for improved communication, empathy, and social concern.

- Interdisciplinary work environments, which not only have diverse teams, but have individuals go out of their comfort zone to work on tasks on which they are less skilled, would aid in character building. When the Apollo missions had infighting about what the ideal approach to landing on the moon would be, each competing team was asked to complete a research document arguing for the option they disagreed with. This approach enabled greater understanding of others and a concern for full knowledge of the circumstance. Adapting these lessons to more workplaces can help take individuals outside of themselves.

This is a non-exhaustive list as the ways to build virtue are myriad and there is no checklist for how to do it. The particular environment and the circumstances of the people in that environment call for tailored strategies. Helping managers and human resources staff understand the value of virtue development in the workplace and working with them to create easily implementable solutions for their organisations would lead to larger returns to responsible AI work than focussing solely on reactive solutions to identified problems.

There has been significant valuable research in the responsible AI community in developing tools to address important problems posed by AI, clarifying ethical questions, and circulating frameworks that allow for prosocial development and deployment. The major hurdles faced now are in adoption and workplace culture. To address these, attention needs to be paid on the character of people involved in AI, and the moral virtues that they possess.

## Social Work Thinking for UX and AI Design
([Original paper](#) by Desmond Upton Patton)

What if tech companies dedicated as much energy and resources to hiring a Chief Social Work Officer as they did technical AI talent (e.g. engineers, computer scientists, etc.)? If that was the case, argues Desmond Upton Patton (associate professor of social work, sociology, and data science at Columbia University, and director of SAFElab), they would more often ask: Who should be in the room when considering "why or if AI should be created or integrated into society?"

By integrating "social work thinking" into their process of developing AI systems and ethos, these companies would be better equipped to anticipate how technological solutions would impact various communities. To genuinely and effectively pursue "AI for good," there are significant questions that need to be asked and contradictions that need to be examined, which social workers are generally trained to do. For example, Google recently hired individuals experiencing homelessness on a temporary basis to help collect facial scans to diversity Google's dataset for developing facial recognition systems. Although on the surface this was touted as an act of "AI for good," the company didn't leverage their AI systems to actually help end homelessness. Instead, these efforts were for the sole purpose of creating AI systems for "capitalist gain." It's likely this contradiction would have been noticed and addressed if social work thinking was integrated from the very beginning.

It's especially difficult to effectively pursue "AI for good" when the field itself (and tech more broadly) remains largely racially homogenous, male, and socioeconomically privileged; as well as restricted to those with "technical" expertise while other expertise is largely devalued. Patton asks, "How might AI impact society in more positive ways if these communities [e.g., social workers, counselors, nurses, outreach workers, etc.] were consulted often, paid, and recognized as integral to the development and integration of these technologies…?"

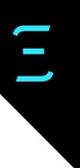

Patton argues that systems and tools can be used to both help a community, and hurt it. "I haven't identified an ethical AI framework," he wrote, "that wrestles with the complexities and realities of safety and security within an inherently unequal society." Thus, an AI technology shouldn't be deployed in a community unless a "more reflective framework" can be created that "privileges community input." When developing these systems, it's important to admit, as Patton does, that the technical solution may not be what's needed to solve the problem.

Through his work at SAFElab, Patton has nurtured collaboration between natural language processing (NLP) and social work researchers to "study the role of social media in gun violence," and create an AI system that predicts aggression and loss. Their approach was to first collect qualitative data by social workers trained in annotation who provided an analysis that then influenced the development of the "computational approach for analyzing social media content and automatically identifying relevant posts." By working closely together, the social workers and the computer scientists were able to develop a more contextualized technical solution to a problem that was cognizant of the "real-world consequences of AI."

In order to effectively ask the right questions and deal with the inherent complexities, problems, and contradictions with developing "AI for good," we need to change who we view as "domain experts." For the project at SAFElab, for example, they developed an "ethical annotation process" and hired youth from the communities they were researching in order to center "context and community voices in the preprocessing of training data." They called this approach Contextual Analysis of Social Media (CASM). This approach includes collecting a baseline interpretation of a social media post from an annotator who provides a contextualized assessment; and then debriefing, evaluating, and reconciling disagreements on the labeled post with the community expert and the social work researcher. Once done, the labeled dataset is then given to the data science team to use in training the system. This approach eliminates the "cultural vacuum" that can exist in training datasets from the beginning and throughout the entire development process.

The code of ethics that guides social workers, argues Patton, should be used to guide the development of AI systems—leading companies to create systems that actually help people in need, address social problems, and are informed by conversations with the communities most impacted by the system. In particular, before looking for a technical solution to a problem, the problem must be fully understood first, especially as it's "defined by the community." These communities should be given the power to influence, change, or veto a solution. To integrate this social work thinking into UX and AI design, we must value individuals "beyond academic and domain experts." Essentially, we must center humanity and acknowledge that in the process of doing so, we may end up devaluing the power and role of the technology itself.

## Overcoming Barriers to Cross-Cultural Cooperation in AI Ethics and Governance
([Original paper](#) by Seán S. ÓhÉigeartaigh, Jess Whittlestone, Yang Liu, Yi Zeng, Zhe Liu)

As AI development continues to expand rapidly across the globe, reaping its full potential and benefits will require international cooperation in the areas of AI ethics and governance. Cross-cultural cooperation can help ensure that positive advances and expertise in one part of the globe are shared with the rest and that no region is left disproportionately negatively impacted by the development of AI. At present, there are a series of barriers that limit the capacity that states have to conduct cross-cultural cooperation, ranging from the challenges of coordination to cultural mistrust. For the authors, misunderstandings and mistrust between cultures are often more of a barrier to cross-cultural cooperation rather than fundamental differences in ethical principles. Other barriers include language, a lack of physical proximity, and immigration restrictions which hamper on possibilities for collaboration. The authors argue that despite these barriers, it is still possible for states to reach a consensus on principles and standards for areas of AI.

The researchers Seán S. ÓhÉigeartaigh, Jess Whittlestone, Yang Liu, Yi Zeng and Zhe Liu define cross-cultural cooperation as different populations and cultural groups working together to ensure that AI is developed, deployed and governed in societally beneficial ways. They make the distinction that cross-cultural cooperation on AI does not entail that all parts of the world are to follow or be imposed the same standards. Rather, it involves identifying areas where global agreement is needed and necessary and others where cultural variation and a plurality of approaches is required and desirable.

The authors focus their areas of study on the misunderstandings between the "West", Europe and North America, and in the "East", East Asia. Both areas of the globe have been recognized for their fast-developing AI, but also their active steps in the development of ethical principles and governance recommendations for AI. The authors argue that the competitive lens in which technological progress is framed, one common example being discourses of an AI race between the US and China, creates greater cross-cultural misunderstandings and mistrust between the two regions. They posit that these misunderstandings and mistrust are some of the biggest barriers to international cooperation rather than fundamental disagreements on ethical and governance issues. These misunderstandings must be corrected before they become entrenched in intellectual and public discussions.

The history of political tensions between the US and China and their different founding philosophical traditions have led to the perception that Western and Eastern ethical traditions are fundamentally in conflict. This idea of an East/West divide, despite being oversimplistic and ignoring the differences in values within the regions, is repeatedly manifested in discourses surrounding the development of ethical AI. Claims of differences between the two regions, according to ÓhÉigeartaigh et al., often rest on unexamined concepts and a lack of empirical evidence.

One example is data privacy, where it is supposed that China's laws are laxer compared to Europe and the US. The authors point to how this view may now be outdated, as Beijing recently banned 100 apps for data privacy infringements. They argue that those working in AI ethics must achieve a more nuanced understanding of how privacy may be prioritized differently when it comes to conflict with other key values, such as security. Another example is China's much demonized social credit score system (SCS). The authors argue that Western publications fail to underscore how the measures in the SCS are largely aimed at tackling fraud and corruption in local governments, and how blacklisting and mass surveillance already exists in the US. Researchers in both regions will need to work reversing assumptions on building greater mutual understandings.

Part of the misunderstandings between the two regions is also due to the language barrier which limits the opportunities for shared knowledge. Researchers in China, Japan and Korea tend to have a greater knowledge of English, while only a small fraction of North American and European researchers know Mandarin, Japanese or Korean. Even in cases where key documents are translated, misunderstandings can arise due to subtleties in language. One case is the Beijing principles where China's goal of AI leadership was misinterpreted as a claim of AI dominance. Commentators concentrated on this miswording instead of focusing on the overlapping principles of human privacy, dignity, and AI for the good of humankind listed in the document.

For ÓhÉigeartaigh et al., constructive progress in AI ethics and governance can be achieved without finding consensus on philosophical issues or needing to resolve decades of political tension between nations. The dialogue should shift to areas where cooperation between states is crucial, like in military technology and arms development, over areas where it may be more appropriate to respect a plurality of approaches, such as healthcare. The delineation of where global standards for AI ethics and governance are needed should be informed by diverse cross-cultural perspectives that consider the needs and desires of different populations. As in the case of the previous nuclear weapons ban treaty, overlapping consensus on norms and practical guidelines can also be achieved even when countries have different political or ethical considerations for justifying these principles.

For the authors, academia has a large role to play in facilitating cross-cultural cooperation on AI ethics and governance and identifying further areas where it is possible. Research initiatives that promote the free-flowing and intercultural exchange of ideas can help foster greater mutual understandings. Diverse academic expertise will also be needed to outline where fundamental differences do exist and whether value alignment is possible.

ÓhÉigeartaigh et al., are optimistic that academia and wider civil society can actively shape the principles behind binding international regulations. They refer to cases where both groups successfully intervened to shape issues of global importance, such as campaigns for the ban of lethal autonomous weapons and the abandonment of Google's Project Maven.

To achieve greater cross-cultural cooperation, the authors offer a series of further recommendation and calls to action which include:

– Developing AI ethics and governance research agendas requiring cross-cultural cooperation. This is aimed at a global research community that can support international policy cooperation.

– Translating key papers and reports. This includes higher quality and several translations that explore the nuances and context of language.

– Alternate continents for major AI research conferences and ethics and governance conferences. This can allow for greater international and multilingual participation as well as reduce the cost and time commitment for scholars and AI experts to take part.

– Establish joint/or exchange programs for PhD students and postdocs. International fellowships will allow researchers to be exposed to different cultures early on in their careers and give the capacity to reach mutual understandings.

Greater efforts aimed at achieving a more nuanced understanding between AI superpowers will help reduce the mistrust and correct assumptions of fundamental differences. As encouraged by the authors, cross-cultural cooperation is possible even among countries with divergent ethical principles by delineating the areas that necessitate cooperation and those that can support a diversity of values as well as by concentrating on practical issues. In the absence of cross-cultural cooperation, the competitive pressures and tensions between states will lead to underinvestment in safe, ethical and socially beneficial AI. It will also be cause for concern when ensuring that applications of AI are set to cross-national and regional boundaries. The authors conclude that as AI systems become more capable and ubiquitous, cultivating deep cooperative relationships on AI ethics and governance should be an immediate and pressing challenge for the global community.

## Decision Points in AI Governance
([Original *UC Berkeley* white paper](#) by Jessica Cussins Newman)

The inspiration behind Newman's paper lies within her observation that governance on AI principles focuses too much on the what, and not enough on the how. As a result, her paper aims to denote examples to act as suggestions to how to best operationalise the AI principles being discussed. To do this, she illustrates 3 different case studies, which I will now illustrate in turn.

**Case study 1: Can an AI Ethics advisory committee help advance responsible AI (Microsoft AETHER committee)?**

A very current debate topic is of whether a company's ethics boards can actually impact the work done by its engineers. Newman refers to Microsoft's AETHER committee attempt to do just that.

As a big company, Microsoft's moves in the AI world will have a significantly larger impact than other smaller businesses, putting even more emphasis on making this known to the key stakeholders. To act on this, Microsoft organised their principles based on the engineering processes involved, including guidance on privacy, and accountability. The committee (comprising 7 working groups, with about 23 members from each major department) would then write reports on any AI concerns had by different employees raised through their Ask-AETHER phone-line. This was made available to all departments within Microsoft, and allowed the reports compiled to represent each concern raised. These reports would then be sent to senior management for review, keeping those at the top connected with what goes on elsewhere.

Qualms were nonetheless raised about the council's impact, with Microsoft winning the $10 billion contract in 2019 to restructure the Department of Defense's

cloud system. Their response was that there was no objection to being involved with the military within the company's AI principles, so long as the system was safe, reliable, and accountable. No official objection was ever published from AETHER, but they apparently did raise a policy concern on an executive retreat that same year.

Newman's takeaways were resultantly based on the welcomed move of establishing the AETHER call line, and involving the executives at the top. For the principles to be truly representative, all concerns must be taken into account, and inter-disciplinary departments are to be involved. Microsoft did exactly that, but AETHER's true impact is still to be seen.

**Case study 2: Does shifting publication norms of AI reduce its risk?**

Here, Newman considers the staged-release publication process of AI systems, in complete contrast to the norm in the AI field of an all at once release. The staged process has been examined as a possible way to prevent the use of the AI software by malicious actors, as well as being able to give time to policy makers and human actors involved. Such a process gives policy-makers time to consider how best to approach the software and its societal effects, while human actors have time to reflect on their own usage of the product.

However, the process has been criticised for potentially stifling the speed and growth of the AI field through having a more delayed process. Admittedly, such a process can prevent potential harms, but it can also prevent potential benefits. Here, Newman utilises OpenAI's GPT-2 language model as an example. Committed to releasing it in stages, models with greater parameters and specs were released before GPT-2 had fully been made available. Furthermore, once released, a doctor from Imperial College London repurposed GPT-2 to write accurate scientific abstracts in solely 24 hours, something which could have occurred much earlier had the model been fully released.

Newman believes that open source AI information is key to the field progressing, whether released in stages or fully. Releasing in stages can help prevent certain harms, but can also make it harder for independent researchers to properly evaluate the model without its full release. Altering publication norms can potentially help prevent malicious usage of the product, but can also prevent its proper evaluation in the first place.

**Case study 3: Can a global focus point provide for international coordination on AI policy and implementation?**

Newman takes advantage of the monumental OECD principles as her example of (and one of the only) points of international agreement on AI principles. On may 22nd 2019, 42 countries signed up to the OECD's intergovernmental principles on AI, ranging from Asia, South America, Europe and Africa. The language utilised in the principles that stand out to me are words such as stewardship, plain easy-to-understand information, human-centeredness, and underrepresented. Strong and powerful language contained within principles agreed upon by 42 countries was never anticipated, and proved an extremely positive step in the right direction.

Unfortunately, Newman acknowledges that the implementation of these principles in each country will be different. Cultural considerations, the presence of infrastructure and the economic situation will impact which principles can be adopted in which way. Bodies such as the AI observatory have been established to try and link practical instantiations of the principles with their desired goal, but how each country develops its AI strategy remains to be seen.

Newman's paper has provided us with real life examples of how AI principles are trying to be implemented in the real world. Involving leaders at large corporations like AETHER has done can help to move towards a great cognizance of the implications of decisions made on AI. Such a cognizance can then help influence the publication norms to prevent evil-use of AI products, helping international governments do the same. While there are many challenges ahead, turning talk into action is certainly the way to overcome them.

## Changing My Mind About AI, Universal Basic Income, and the Value of Data
([Original long-form blog post](#) by Vi Hart)

Artificial Intelligence may soon become powerful enough to change the landscape of work. When it does, will it devastate the job market and widen the wealth gap, or will it lay the foundation for a technological utopia where human labor is no longer required? A potential intersection between these seemingly opposed theories has developed into an increasingly popular idea in the past 5 years: the idea that human work may become obsolete, but that AI will generate such excess wealth that redistribution in the form of Universal Basic Income is possible. In the article "Changing my Mind about AI, Universal Basic Income, and the Value of Data", author Vi Hart explores the attractive idea of UBI and AI – long prophesied by tech industry leaders – and weighs its practicality and pitfalls.

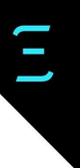

Universal Basic Income is a program that provides every individual with a standardized unconditional income. It has been presented as a salve to the existential problem of massive unemployment as AI replaces human workers. It could reduce financial dependence on traditional jobs, freeing individuals to pursue meaningful (rather than market-driven) skill development. And although UBI may appear costly, the relative cheapness of AI labor could generate capital for redistribution.

While it might seem an ideal solution at first glance, UBI doesn't address the most dangerous threat presented by AI: the devaluation of the human labor that makes AI programs work.

For the past 5 years, the tech elite have justified the devaluation of the human worker by claiming artificial intelligence will be orders of magnitude more productive than manual work. They extend this line of reasoning by idealizing "pure" AI, which will move beyond the need for human participation at all.

But this rhetoric is untrue: human contributions are necessary inputs for AI to make decisions. AI is only as useful as the "collective intelligence" it draws upon – human-generated data collected knowingly or unknowingly. The gig economy of producing data through online marketplaces like MTurk is unregulated and can pay less than a living wage. This is, in part, because the value of data is set by an unbalanced data market (a monopsony), as many data are collected freely in exchange for use of online services.

In addition to their role in data creation, human workers participate in customer service, delivery, and other on-demand tasks under the guise of full automation. Call center workers, content moderators, and other humans invisibly fill in the "last mile" of decisions that AI systems cannot make. This illusion helps justify the artificially low value of data labor, even though that labor will generate massive wealth for corporations.

In sum, a marketplace radically transformed by AI will likely drive workers' perceived worth down – and UBI may not reverse the harmful results. The utopian vision for AI and UBI, touted by the tech elite, deflects responsibility from corporations to pay for the data labor that is so valuable to them. The author proposes a solution that goes beyond UBI to establish "data dignity": fair compensation for data labor in a balanced marketplace. Above all else, individuals must be recognized and valued for their data. They must be able to reason on the value of their contributions and make the choice to contribute.

# **Go Wide: Article Summaries**

### **Is The Business World Ready For A Chief Data Ethics Officer?**
**(**[**Original *Forbes* article**](#) **by Randy Bean)**

The article highlights some of primary concerns that an organization might face as it utilizes data to make inferences about their customers or users. Going into details about how privacy risks abound in the use of proxy data from which private details like sexual orientation or political preferences might be inferred, the article highlights how there is a need for data ethics officers who can help address or mitigate some of these concerns. Most people are familiar with how such inferred details can be used to subvert the integrity of democratic institutions by manipulating people and persuading them for political gains. What is essential here is to understand how such consequences can be mitigated through the appointment of someone who would have a fiduciary responsibility to data subjects, something that could potentially be folded into this role.

A position like this would empower the organization to enact top-down change where there is a central authority that is responsible for organization-wide practices to ensure the ethical use of data. Discussions around the bias in these systems have been hotly debated and the opacity surrounding the operations of these systems only exacerbates the problem. A Chief Data Ethics Officer can potentially help to align the existing business practices with these responsible data principles which would be crucial for adoption and implementation rather than just the discussion of this.

Finally, in a highly competitive space, this can become a strong differentiator for an organization in terms of minimizing reputational risk, enhancing employee retention and recruitment, and increasing compliance with regulation. The important thing would be to empower this role in a way to lead to successful, measurable outcomes rather than just paying lip-service and checking off a box.

### **An Understanding of AI's Limitations Is Starting to Sink in**
**(**[**Original *The Economist* article**](#)**)**

Whispers in terms of limitations of the approaches to achieving the "intelligence" in AI have been abounding for a while. Specifically, over the last few years AI has become a buzzword and more companies claim to use AI than actually do, trying to garner investment and media interest. We acknowledge here that the definition of intelligence itself is quite open to interpretation and everyone has different expectations in terms of what these systems should be capable of achieving to be called intelligent. Yet, in some scenarios where we have seen stellar progress, decades ahead of what experts had thought possible, for example with AlphaGo, there are still scenarios where the intelligence exhibited by these systems is quite limited even within a narrow domain where the system starts to fall apart when presented with data that falls outside of the distribution it expects.

The fundamental promise of AI is that it is great at identifying patterns which makes it a general-purpose technology that has wide applicability across every possible domain. As an example, with the ongoing pandemic, there is a strong case being made for using AI in every part of the viral management lifecycle from drug discovery, inventory management, allocating resources, and more. Especially as it relates to the use of AI in contact-tracing, a lot of promises have been made, but pulling open the hood points to a host of problems that don't yet have clear answers. This is supplemented also by the views from Eric Topol who says that advances in the hype of AI have far outpaced that in the science of AI. In the past this has led to "AI Winters" which had a tremendously negative impact on the field. While this time around there has been a lot of actual and positive deployment of AI systems, it hasn't been without a great deal of ethics, safety, and inclusivity issues.

Given that a lot of consulting firms make predictions that the widespread use of AI will add trillions of dollars worth of economic value and output to the global economy. Yet, a more realistic view given the current capabilities and speaking with technical experts working in the field brings a dampener on such extravagant claims made by firms that put out such reports.

### The Loss Of Public Goods To Big Tech
([Original *Noema Magazine* article](#) by Safiya Noble)

An article that elucidates how there is a massive asymmetry between private and public goods - it highlights essentially how private corporations are able to benefit from the vast public infrastructure that helps to support their activities while they eschew their responsibility to contribute back into the pool from which they draw. Pulling our attention to "charity theatre", the author asks us to examine how philanthropic efforts from these corporations are minimal drops in the bucket compared to the benefits that they are able to extract from the publicly funded facilities. In terms of the tax benefits and offshoring practiced by the firms and how their expenditures on technology that enables malpractices like the suppressive use of facial recognition technology on people in the BLM movement, the author

The State of AI Ethics, October 2020    138

advocates for rerouting that money towards other initiatives that have the potential to bring prosperity for people.

The platforms are also able to polarize society through the passive role that they play in the propagation of misinformation and other harmful content while simultaneously profiting from it. The incentives from the public welfare and business standpoint are so misaligned that they are poised in an almost zero-sum game where they are encouraged to take from one to benefit the other.

Instead of helping existing public infrastructure, the firms actively profit from its erosion as it drives more users and customers to them looking for products and services that might have been provided elsewhere by the public sector. There is also the asymmetry in how crises like the current pandemic exacerbate the impact faced by those who have too little while those who are powerful have to give up close to nothing, often even gaining in the times of such crises.

Finally, making a call for addressing problems like the current pandemic, at least in part, the author appeals to us to make a collective effort to help each other through these times, citing that it would be unjust and unethical to not do so. The pandemic has the potential to reshape society significantly, now is the time that we grab the opportunity to shape it into something that does benefit us all.

## How Deepfakes Could Actually Do Some Good
([Original *Vox* article](#) by Rebecca Heilweil)

In breaking away from the oft-cited examples of how "deepfakes" can cause harm, and there are all the reasons to do that given the abuses that have resulted from the use of this technology, this article sheds a new light on how this technology could be used for good. Such a dual-use for this technology, something that the AI domain is quite familiar with, poses new challenges in assessing the regulatory stance around how to think about the public use of this technology.

The article mentions the upcoming documentary about Welcome to Chechnya on HBO where members of the LGBTQ community who are heavily persecuted there share their experiences in the documentary without having to give up their identities. They are essentially anonymized by digitally grafting on faces of volunteers who the documentary makers call "activists" in an attempt to humanize them and better convey the experiences of those people rather than the typical techniques used in such efforts where the face of the person is just blotted out. The creators of the documentary consulted with researchers experienced in neuroscience and psychology to make sure that they minimized the "Uncanny

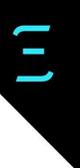

Valley" effect to prevent the experience of the audiences from being too jarring. The results, while admittedly slightly off, do a great job of making it quite realistic.

There are ethical concerns even when this sort of technology is used explicitly for the good, as it might backfire against the "activists", it nonetheless has great potential to empower people to share their experiences without being traumatized by having to reveal their identities. There are similar instances of using this technology in Snapchat filters for victims of sexual abuse to share their experiences more freely. Ultimately, what separates the good use of "deepfakes" from the bad uses is clear consent from those whose likeness is being used, from those on whom it is used, and clear articulation of the purpose for which this is being done.

## Is It Time for a 'Digital New Deal' to Rein in Big Tech?
(Original *Protocol* article by Emily Birnbaum)

Advocating for a significant overhaul in regulation and legislation around the power that Big Tech holds, this article talks about the "Digital New Deal" to evoke in people the desire to move towards broad changes rather than incrementalist reforms that currently plague the system. The four pillars of this deal are more robust antitrust enforcement, nondiscrimination principles, transparency, and public utility regulation.

Going into some details on each, the article mentions how we need a higher degree of scrutiny and reflection in how the power dynamics are framed when thinking about the required regulations. It especially discusses the current ecosystem in the US whereby the calls for regulating Big Tech are mired in political motivations as well, which have the potential of weakening some of the calls to action because of people's specific political alignments.

As we've mentioned in many past editions of the MAIEI newsletter, a bold move towards action is required and that needs steps to be outlined in a clear and concise fashion so that there is impetus to move and act on them rather than be paralyzed by over-analysis. This proposal is a positive move in that direction.

## China and AI: What the World Can Learn and What It Should Be Wary of
(Original *The Conversation* article by Hessy Elliott)

A brief but much required article that offers a balanced view on the developments in AI coming out of China. Specifically, we require a critical look into the trope of pitting the US against China in an "AI arms race" which already sets the stage in an adversarial manner. Instead, we need to weigh the pros and cons from a more holistic standpoint rather than taking a reductionist view of the whole debate.

The article highlights some of the positive developments coming out of China in the use of AI, keeping in line with the current pandemic, some of the spotlighted solutions are the use of AI in medicine, geared towards combating COVID-19. There is emerging evidence on how this can benefit those who don't live in regions with sufficient healthcare resources.

In the negative developments segment, most readers of the MAIEI newsletter are already familiar with a lot of the problems that arise from the unmitigated use of facial recognition technology applied indiscriminately even in the face of overwhelming evidence that there are many false positives and errors. A key requirement in making better use of this technology is transparency and accountability as fundamental tenets incorporated into how these systems are deployed.

Lastly, another outsider perspective which is clarified in this article is how many perceive that the national AI strategy in China is a key driver for how the ecosystem is shaped, yet there are ample grassroots and municipally driven initiatives that are reimagining and interpreting the national AI strategy tailoring it to the local context and economy. Ultimately, the article calls for more informed discussions on the subject to evade the crutch of reductionism which will ultimately harm the quality of debate on this subject.

## An Update on [Google's] Work on AI and Responsible Innovation
(Original *Google Blog* article by Kent Walker)

This post penned by Jeff Dean from Google highlights some of the work that they have done in putting responsible AI principles in practice within their organization. It serves as a useful guide for others who are aiming to do the same. One of the initiatives that caught our eye was the deployment of a mandatory technology ethics course trialed with several employees. It is akin to secure coding practices that new hires have to undertake before they are allowed to write a single line of code.  It is essential in creating a groundswell of interest and adoption across the organization.

Though they highlight some of the work done internally in operationalizing ethics, the deployment rates appear to be quite low, which means there is still a lot of work to be done. They provide some details on their internal review process: it is an iterative approach that factors in internal domain expertise. When relevant, they have also chosen to engage external experts to understand the societal impacts of their solutions. Illustrating their learning from previous work, they mention how Google TTS and Google Lens have incorporated feedback to make these technologies more ethical. As an example, for Google TTS, they added additional layers of protection to prevent misuse by bad actors. They also limited open-sourcing the TTS solution to prohibit the creation of deepfakes.

The final idea that was quite exciting for us at MAIEI was the use of community and civil-society based bodies to consult on ethical AI implications. They used insights from the discussion with this body to shape their operational efforts and decision-making frameworks. Adopting a transparent approach in how responsible AI principles are being developed and deployed will be crucial in evoking a high degree of trust from users.

### Not Just The Sprinkles On Top: Baking Ethics Into AI Design
([Original *Forbes* article](#) by James Freeze)

An interesting article that rehashes notions from the field of cybersecurity and presents them in a new light (unfortunately, it doesn't make that attribution). Both the efficacy and cost of mitigation measures are better in the earlier stages of development. Our founder Abhishek Gupta has referenced this principle in many of his talks about "baking in, rather than bolting on" ethics to AI. The use of interdisciplinary teams has the potential to improve current processes. Specifically, it allows unearthing blindspots and making the entire pipeline of design, development, and deployment more ethical.

There is also a need to think of the guidelines as being tailored for different use cases. For example, Roombas evoke different reactions compared to those from Alexa. An application should convey the right signals to match the expectations and sensibilities of the users.

The State of AI Ethics, October 2020                                                                                          142

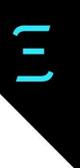

Another interesting consideration is to think about the high-impact scenarios that have the potential to impact human lives. For example, self-driving vehicles should be able to provide explanations as do medical systems in how they arrive at particular decisions. To reach ubiquitous deployment, we need to evoke a high degree of trust from consumers. The onus to make that lies on the shoulders of the organizations developing these systems.

## Four Steps for Drafting an Ethical Data Practices Blueprint
(**Original *TechCrunch* article** by Joel Shapiro, Reid Blackman)

Prompting the need for having an ethical data science blueprint, the article starts by mentioning how a healthcare provider that relied on AI ended up prioritizing white patients over black patients. The article advocates for four different steps to building responsible AI within the organization. The first step is the integration of ethical AI practices within the existing work in the organization around the assessment of privacy risks and legal compliance. Leveraging existing bodies reduces friction in adoption. The second step is maintaining an appropriate level of transparency, which can be decided by a C-suite executive who takes on this role. They can analyze the ethics against business requirements and then articulate them to product managers who are responsible for implementing them in practice.

The importance of a robust and clear blueprint is that it leads to consistency in implementation rather than delegating decisions to individual data scientists. From a technical standpoint, since there are numerous definitions for fairness, a body of experts must be consulted to select those that are most appropriate for the problem at hand.

Finally, education and training on data ethics will equip and empower front-line workers in making the right decisions in their everyday work and help them align with the ethical blueprint for the organization.

## How Open Data Could Tame Big Tech's Power and Avoid a Breakup
(**Original *The Conversation* article** by Patrick Leblond)

With the recent antitrust hearings where some of the biggest tech companies were put on the stand to justify their behaviour, this article proposes an interesting alternative to the traditional call for breaking up Big Tech. Specifically, it highlights how the problem of breaking up Big Tech is fraught with problems of assessing what constitutes Big Tech. In addition, given the virtuous data cycles and network effects of these platforms, the effects of which have been exacerbated by automation, only serves to repeat the cycle even if Big Tech is broken up into smaller pieces. The smaller pieces have the potential to again metastasize into something large with concentrated market power that would be hard to control. In other cases, the merger of certain smaller players might suddenly create an entity that has unfettered market power.

Instead, a focus on creating public data commons whereby data is made available openly for anyone to use and develop on while protecting the privacy of individuals and keeping the market operations information of firms out of the picture could be a way to distribute market power. While the article is scant in terms of details on how to address the privacy challenges and (in our opinion) mistakenly mentions anonymization as a sufficient methodology for protecting the privacy of individuals, it is nonetheless an interesting idea. Additionally, the article also proposes the creation of a market regulatory authority that could operate in a way similar to the regulation of financial data and how individuals and institutions trade securities for profits and the restrictions that are made in using insider information. Such ideas provide a fresh perspective on the current paradigm and are welcome additions to the debate surrounding the best way to act in the interest of the users.

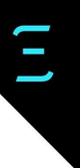

# 9. Outside The Boxes

**Opening Remarks** by Abhishek Gupta (Founder, Montreal AI Ethics Institute)

While there have certainly been a lot of developments in the domain of AI ethics from the (now) standard lenses of privacy, bias, fairness, governance, among other areas, there are still a few more perspectives that caught our attention when compiling the research and developments over this past quarter. One of the things revolved around an increasing focus on the environmental impacts of AI and how very large compute requirements can have externalities.

My [work at MAIEI](#) aims to address some of these problems through the extension of the work mentioned in this report by advocating for a more end-to-end and standardized, frictionless solution that enables comparisons for researchers and consumers to make environmentally-conscious decisions in picking AI models.

As the pandemic continues to rage, it is no surprise that there needs to be a greater emphasis on being able to effectively evaluate the contact-tracing applications to ensure that they respect our rights. In one of the pieces of work in this report, we get a glimpse of what questions we should ask as consumers on whether we should download and use these applications.

Failure of autonomous systems has been on our mind and we believe that addressing issues like the "moral crumple zone" mentioned in this report are essential so we don't offload the unintended consequences onto consumers of the system and operate in a more ethical manner allocating responsibility onto those who have the resources to address concerns effectively. Especially with the case of the A-level exams and the subsequent uproar as the lives of students were upended, one needs to critically question the transparency requirements around the public use of automated systems and accountability for those who are creating and deploying the systems. In other areas like healthcare where such systems can nudge doctors towards initiating uncomfortable end-of-life conversations with patients, there is a stark need for design philosophies that can bring out the best in terms of the use of the technology while still doing so in a manner that is aligned with the best interests of the patients at heart.

Finally, science wouldn't be science unless we adhere to strict principles of what constitutes good science: reproducibility is certainly one of them. As more and more fields integrate complex data and machine learning workflows within their research work, it is essential that we are able to implement the right measures in making the entire experiment reproducible to adhere to principles of doing good scientific research.

I believe that this is a great opportunity to expand our thinking beyond just the core principles in the domain of AI ethics and make broader considerations in evaluating the impacts of technology, especially from a societal perspective since these systems are inherently socio-technical. The efficacy of any and all of these measures relies critically on this realization and it would behoove both researchers and practitioners in the domain to lay a greater emphasis on actionable guidance rather than just paying lip-service to principles or engage in these measures as a box-ticking exercise. I hope that you enjoy this chapter and it prompts critical reflections for you and your team in assessing the impacts of your research and applied work.

Abhishek Gupta (**@atg_abhishek**)
Founder & Principal Researcher, Montreal AI Ethics Institute

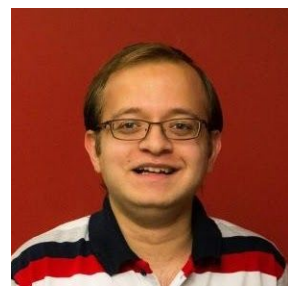

Abhishek Gupta is the founder and principal researcher at the Montreal AI Ethics Institute, seeking to define humanity's place in a world increasingly characterized and driven by algorithms. He is also a machine learning engineer at Microsoft, where he serves on the CSE Responsible AI Board. His book '[Actionable AI Ethics](#)' will be published by Manning in 2021.

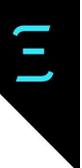

# Go Deep: Research Summaries

## Towards the Systematic Reporting of the Energy and Carbon Footprints of Machine Learning
([Original paper](#) by Peter Henderson, Jieru Hu, Joshua Romoff, Emma Brunskill, Dan Jurafsky, Joelle Pineau)

Climate change and environmental destruction are well-documented. Most people are aware that mitigating the risks caused by these is crucial and will be nothing less than a Herculean undertaking. On the bright side, AI can be of great use in this endeavour. For example, it can help us optimize resource use, or help us visualize the devastating effects of floods caused by climate change.

However, AI models can have excessively large carbon footprints. Henderson et al.'s paper details how the metrics needed to calculate environmental impact are severely underreported. To highlight this, the authors randomly sampled one-hundred NeurIPS 2019 papers. They found that none reported carbon impacts, only one reported some energy use metrics, and seventeen reported at least some metrics related to compute-use. Close to half of the papers reported experiment run time and the type of hardware used. The authors suggest that the environmental impact of AI and relevant metrics are hardly reported by researchers because the necessary metrics can be difficult to collect, while subsequent calculations can be time-consuming.

Taking this challenge head-on, the authors make a significant contribution by performing a meta-analysis of the very few frameworks proposed to evaluate the carbon footprint of AI systems through compute- and energy-intensity. In light of this meta-analysis, the paper outlines a standardized framework called experiment-impact-tracker to measure carbon emissions. The authors use 13 metrics to quantify compute and energy use. These include when an experiment starts and ends, CPU and GPU power draw, and information on a specific energy grid's efficiency.

The authors describe their motivations as threefold. First, experiment-impact-tracker is meant to spread awareness among AI researchers about how environmentally-harmful AI can be. They highlight that "[w]ithout consistent and accurate accounting, many researchers will simply be unaware of the impacts their models might have and will not pursue mitigating strategies". Second, the framework could help align incentives. While it is clear that lowering one's environmental impact is generally valued in society, this is not currently the case in the field of AI. Experiment-impact tracker, the authors believe, could help bridge this gap, and make energy efficiency and carbon-impact curtailment valuable objectives for researchers, along with model accuracy and complexity. Third, experiment-impact-tracker can help perform cost-benefit analyses on one's AI model by comparing electricity cost and expected revenue, or the carbon emissions saved as opposed to those produced. This can partially inform decisions on whether training a model or improving its accuracy is worth the associated costs.

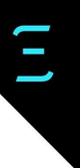

To help experiment-impact-tracker become widely used among researchers, the framework emphasizes usability. It aims to make it easy and quick to calculate the carbon impact of an AI model. Through a short modification of one's code, experiment-impact-tracker collects information that allows it to determine the energy and compute required as well as, ultimately, the carbon impact of the AI model. Experiment-impact-tracker also addresses the interpretability of the results by including a dollar amount that represents the harm caused by the project. This may be more tangible for some than emissions expressed in the weight of greenhouse gases released or even in CO2 equivalent emissions (CO2eq). In addition, the authors strive to: allow other ML researchers to add to experiment-impact-tracker to suit their needs, increase reproducibility in the field by making metrics collection more thorough, and make the framework robust enough to withstand internal mistakes and subsequent corrections without compromising comparability.

Moreover, the paper includes further initiatives and recommendations to push AI researchers to curtail their energy use and environmental impact. For one, the authors take advantage of the already widespread use of leaderboards in the AI community. While existing leaderboards are largely targeted towards model accuracy, Henderson et al. instead put in place an energy efficiency leaderboard for deep reinforcement learning models. They assert that a leaderboard of this kind, that tracks performance in areas indicative of potential environmental impact, "can also help spread information about the most energy and climate-friendly combinations of hardware, software, and algorithms such that new work can be built on top of these systems instead of more energy-hungry configurations".

The authors also suggest AI practitioners can take an immediate and significant step in lowering the carbon emissions of their work: run experiments on energy grids located in carbon-efficient cloud regions like Quebec, the least carbon-intensive cloud region. Especially when compared to very carbon-intensive cloud regions like Estonia, the difference in CO2eq emitted can be considerable: running an experiment in Estonia produces up to thirty times as much emissions as running the same experiment in Quebec. The important reduction in carbon

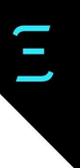
emissions that follows from switching to energy-efficient cloud regions, according to Henderson et al., means there is no need to fully forego building compute-intensive AI as some believe.

In terms of systemic changes that accompany experiment-impact-tracker, the paper lists seven. The authors suggest the implementation of a "green default" for both software and hardware. This would make the default setting for researchers' tools the most environmentally-friendly one. The authors also insist on weighing costs and benefits to using compute- and energy-hungry AI models. Small increases in accuracy, for instance, can come at a high environmental cost. They hope to see the AI community use efficient testing environments for their models, as well as standardized reporting of a model's carbon impact with the help of experiment-impact-tracker.

Additionally, the authors ask those developing AI models to be conscious of the environmental costs of reproducing their work, and act as to minimize unnecessary reproduction. While being able to reproduce other researchers' work is crucial in maintaining sound scientific discourse, it is merely wasteful for two departments in the same business to build the same model from scratch. The paper also presents the possibility of developing a badge identifying AI research papers that show considerable effort in mitigating carbon impact when these papers are presented at conferences. Lastly, the authors highlight important lacunas in relation to driver support and implementation. Systems that would allow data on energy use to be collected are unavailable for certain hardware, or the data is difficult for users to obtain. Addressing these barriers would allow for more widespread collection of energy use data, and contribute to making carbon impact measurement more mainstream in the AI community.

### Trust and Transparency in Contact Tracing Applications
([Original paper](#) by Stacy Hobson, Michael Hind, Aleksandra Mojsilovic, Kush R. Varshney)

It is estimated that 50-60% of the population must use the contact tracing app deployed in Ontario in order for it to work as intended, warning individuals of exposure to COVID-19. But, how much do we really know about this technology? Of course, automatic contact tracing can be more accurate, efficient and comprehensive when identifying and notifying individuals who have been exposed to the virus relative to manual contact tracing; but, what are the trade-offs of this solution? To guide our thinking, authors of "Trust and Transparency in Contact Tracing Applications" have developed FactSheets, a list of questions users should consider before downloading a contact tracing application.

According to the article, users should begin by asking which technology the app uses to track users' location. If the app uses GPS, it works by identifying a user's geographical location and pairing that data with a timestamp. In terms of efficacy, the technology is impeded when users are indoors or in a building with different stories (e.g. an apartment building). Bluetooth, on the other hand, establishes contact events through proximity detection.

However, Bluetooth's signal strength can be obstructed by the orientation of the device as well as the signal's absorption into the human body, radio signals or in buildings and trains. Neither GPS nor Bluetooth capture variables such as ventilation or the use of masks and gloves, which also impact the likelihood of transmission. Not to mention, both technologies rest on assumptions that the device is in possession of one individual and stays with them at all times. Both of these assumptions can result in a false determination of exposure.

In addition to accuracy concerns, users should consider:

- Privacy: sensitive data users are asked to share with the application (health status, location details, social interactions, name, gender, age, health history)

- Security: the vulnerability of the application to attack

- Coverage: the number of users that will opt into the use of the application

- Accessibility: whether the technology is accessible to the entire population (consider that 47% of people aged 65 and older do not have smartphones)

- Accuracy: whether the limitations of Bluetooth and GPS location tracking will undermine the accuracy of the app

- Asynchronous Contact Events: whether the app will capture risk of exposure from transmission in circumstances other than proximity to others (i.e. infected surfaces)

- Device Impacts: the app's impact on the users' devices (battery life etc.)
- Ability: users' capacity to use the app as intended

- Ability: interoperability between contact tracing applications downloaded by the rest of the population

- Reluctance in Disclosure: whether users will submit information about their positive COVID-19 diagnosis

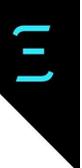

# Go Wide: Article Summaries

### Who Is Responsible When Autonomous Systems Fail?
([Original *CIGI* article](#) by Madeleine Clare Elish)

In the infamous case of the Uber self-driving vehicle accident in 2018 that brought forth a dark cloud over testing of autonomous vehicles (AVs) on public roads for a while, it was interesting to note that they were back on the streets a few months later with almost no reprimands. The safety drive in the AV though still continues to grapple with legal action. The allocation of responsibility in this context has been uneven, whereby the manufacturers of the vehicle, the developer of the software for the AV, and the state of Arizona that allowed for testing to take place held no responsibility in the accident.

Often the human-in-the-loop is pushed forth as a safety mechanism that is going to be the failsafe in case the automation doesn't work as intended, yet one has to carefully analyze the position of the human in that loop, especially as it relates to whether they are empowered or disempowered to take action. We can end up wrongly estimating the capabilities of both the humans and the machines such that the gap that is meant to be covered in case of failures in automation are in fact left even more widely exposed. Specifically, there is an argument to be made that automation can help to fill over the smaller errors but leave room for even larger errors to occur, especially as the humans who are supposed to be in the loop atrophy in skills and become tokenized.

The author points out the concept of the moral crumple zone, drawing from the examples in the aviation industry where there is an increasing amount of automation but when it comes to allocating blame for things going wrong, that is put squarely on the shoulders of the pilots / the nearest human around with little to no scrutiny on the automated systems themselves. Any of these systems always operate in a complicated environment where they have a variety of feedback loops with the humans that are a part of the system and disregarding those loops and the interactions between the humans and the machines takes a very narrow look at the problems. The term crumple zone comes from vehicle parts that are meant to absorb the bulk of the damage and impact during an accident to protect the human. In the case of highly complex and automated systems, the humans become the scapegoat or crumple zone for taking on the moral liabilities when it comes to the failure of these systems.

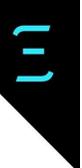

The author also points out the concept of the irony of automation whereby automation doesn't fully eliminate human errors, it just creates opportunities for new kinds of errors. This also brings to light the handoff problem whereby we still don't have effective ways to transfer control from a machine to a human quickly and meaningfully in the case of failures on the part of the automated systems. As a closing argument, it is important to consider, as the author points out, that we don't doom these essential workers who form the human infrastructure to give us the illusion of automation running smoothly to become sacrificial workers without adequate protections. Governance in AI needs to address these challenges as a priority before more people end up in a situation where they take on a disproportionate burden of the fallout from the failure of these systems.

## Complex Data Workflows Contribute to Reproducibility Crisis in Science, Stanford Scientists Say

([Original *Stanford News* article](#) by Adam Hadhazy)

The field of machine learning is no stranger to complex data workflows. It relies on crunching large amounts of data that are taken from their raw form and carefully processed and transformed by engineers and data scientists to arrive into a format and shape that is amenable to being processed by machine learning algorithms. Reproducibility has been a huge concern off late in the field and this experiment mentioned in the article highlights this in the context of life sciences where research teams from across the world were provided with the same dataset, the same hypotheses to test, and they came up with differing results on 5 out of the 9 hypotheses.

Diving deeper into the issues that gave rise to these discrepancies, the investigators behind the study surfaced that there was a lack of consistency in how the data was preprocessed, what libraries were used to do this transformation, what activations and thresholds were used in analyzing the fMRI data among other differences which ultimately led to differences in the final results. One of core tenets for good science revolves around being able to take the data, methods, and other information and reproduce the results independently. Barring which we end up just relying on the word of the scientists who originally conducted the experiment for the validity of the results.

While this challenge has plagued medicine, psychology, and other fields, especially in the domain of AI where there are a lot of resources being invested and we have a new generation of researchers utilizing fora like arXiv to grab the latest research and build from it, we must be cautious, accountable, and transparent in the conduction of our experiments and research so that we do right by the research community around us.

## An Experiment in End-of-Life Care: Tapping AI's Cold Calculus to Nudge the Most Human of Conversations
([Original *STAT* article](#) by Rebecca Robbins)

Bringing the worst sort of dystopias alive from the series *Black Mirror*, medical institutions are beginning to use AI systems to nudge doctors to have end-of-life conversations with patients that the system deems are at risk of dying. Doctors are put in awkward positions bringing up the most intimate and sensitive thing up to a patient based on the recommendation of a machine.

In some cases, doctors agree with the recommendations coming from the system in terms of which patients they should broach these subjects, the article points to the case of one doctor who mentions that this system has helped to make her judgement sharper. Other doctors make sure to exclude mentioning that it was a machine that prompted them to have that conversation because it is inherently cold and often unexplainable in why it arrived at a certain decision which makes it even more challenging for patients to grasp why this is being discussed.

A particular challenge arises when doctors disagree with the recommendation from the system. In this case, even though the doctor does have the final say, they are weighed down by the consideration of whether or not they are making the right decision in not bringing up these advanced care options with the patient in case they are wrong and the system is indeed right. Another problem to be highlighted is a potential over-reliance on the system for making these decisions and reducing the autonomy that doctors would have, related to the token human problem.

The designers of the systems have taken into consideration many different design choices, especially around the number of notifications and alerts to provide the doctors to avoid "notification fatigue". Another design consideration is to explicitly not include the probability rating with the patient list, given the understanding that humans are terrible at discerning differences between percentage figures unless they are on the extremes or dead-center. The labeling around the recommendations coming from the system are also framed as those requiring "palliative care" rather than talking about "will die" which can subtly create different expectations. One of the benefits of a system like this, as documented in an associated study, is that it has helped to boost the number of conversations around this subject that are being had with the patients which are essential and are sometimes ignored due to competing priorities and time burdens on the doctors.

## Meet the Secret Algorithm That's Keeping Students Out of College

([Original *Wired* article](#) by Tom Simonite)

The IB board adopted an algorithmic approach to providing scores to students because of the ongoing pandemic disrupting in-person exams. As we have discussed many times in past editions of this newsletter, such solutions are not free of potentially harmful consequences. Students, parents, and educators alike are questioning the underlying mechanisms of this algorithmic system. It has become a hotly debated issue in popular media because of the life-altering consequences that this has for the students in the current cohort.

The article points to cases where students who had received conditional admission offers and scholarships have seen those rescinded because of inadequate performance in the IB evaluation.

One of the issues pointed out is how for schools where there was limited enrolment, the model would have flaws because it utilizes information from other schools to infer the results. The IB foundation defended the system by pointing out how it crafts a bespoke equation for each school. But, for schools that don't have much of a track record, this can be a red flag where scores are calculated differently for different students. For small classes, since there are fewer data points, the system is bound to generate noisier results. Ultimately, when operating automated systems that don't have adequate transparency, the burden for existence should fall on the shoulders of the organization.

# Conclusion

Congratulations on making it all the way to the end!

You are not alone in feeling a sense of dread and hopelessness when it comes to the challenges that face us if we are to build a more just society, one that uses technology to empower us to reach the nadir of our potential.

I am optimistic though. We are much better equipped than prior technological revolutions. We are connected and have the ability to navigate and connect bodies of knowledge from different parts of the world, and rapidly iterate on our findings, and cross disciplines in an intrepid manner. Ideology has its place and has certainly brought many benefits to the research and development of more ethical, safe, and inclusive technology but we have also seen efforts in operationalizing some of these ideas including frameworks to overcome barriers in cross-cultural cooperation, reporting mechanisms for the environmental impacts of AI, a roadmap for ensuring fair outcomes for people with disabilities, and more.

We do have the potential to use AI to foster more good in the world, for example by boosting the efforts of human rights activists when they are trying to prove war crimes in courts by parsing satellite imagery with AI to garner larger amounts of evidence and utilizing computing as a diagnostic mechanism for surfacing areas of injustice and providing blueprints for us to improve.

We have the tools, we have the ability, we have the will, we have the networks, we have the incentives, we have the conscience, we have the belief, we have the knowledge, and I believe we have the willingness to utilize AI for the betterment of society. Let's come together and use the learnings from this quarter's report to move the conversation forward to a place where we are not just talking about problems but are actively experimenting with solutions and sharing what we learn with each other in the true spirit of science to bring about human flourishing — *eudaimonia!*

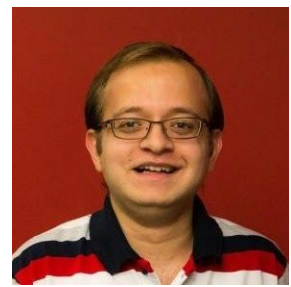

Abhishek Gupta (**@atg_abhishek**)
Founder & Principal Researcher, Montreal AI Ethics Institute

Abhishek Gupta is the founder and principal researcher at the Montreal AI Ethics Institute, seeking to define humanity's place in a world increasingly characterized and driven by algorithms. He is also a machine learning engineer at Microsoft, where he serves on the CSE Responsible AI Board. His book 'Actionable AI Ethics' will be published by Manning in 2021.

# Further Reading

- **[Our AI Ethics Newsletter (weekly, ongoing) — subscribe here](#)**

Subscribe to get full access to the newsletter and have the latest from the field of AI ethics delivered right to your inbox every week — no spam, we promise. Every week we summarize research papers as well as provide constructive commentary on how that research can be implemented. We also share brief thoughts on interesting articles and other developments in the field.

- **[The State of AI Ethics Report (June 2020)](#)**

This pulse-check for the state of discourse, research, and development is geared towards researchers and practitioners alike who are making decisions on behalf of their organizations in considering the societal impacts of AI-enabled solutions. We cover a wide set of areas in this report spanning Agency and Responsibility, Security and Risk, Disinformation, Jobs and Labor, the Future of AI Ethics, and more.

- **[Publication Norms for Responsible AI](#)**

In order to ensure that the science and technology of AI is developed in a humane manner, we must develop research publication norms that are informed by our growing understanding of AI's potential threats and use cases. To examine this challenge and find solutions, the Montreal AI Ethics Institute (MAIEI) collaborated with the Partnership on AI in May 2020 to host two public consultation meetups. These meetups examined potential publication norms for responsible AI, with the goal of creating a clear set of recommendations and ways forward for publishers.

- **[The Unnoticed Cognitive Bias Secretly Shaping the AI Agenda](#)**

This explainer was originally written in response to colleagues' requests to know more about temporal bias, especially as it relates to AI ethics. It explains how humans understand time, time preferences, present-day preference, confidence changes, planning fallacies, and hindsight bias.

- **[The Short Anthropological Guide to the Study of Ethical AI](#)**

The State of AI Ethics, October 2020     156

To encourage social scientists, in particular anthropologists, to play a part in orienting the future of AI, we created the Short Anthropological Guide to Ethical AI. This guide serves as an introduction to the field of AI ethics and offers new avenues for research by social science practitioners. By looking beyond the algorithm and turning to the humans behind it, we can start to critically examine the broader social, economic and political forces at play and ensure that innovation does not come at the cost of harming lives.

- **[Green Lighting ML: Confidentiality, Integrity, and Availability of Machine Learning Systems in Deployment](#)**

Automated systems for validating privacy and security of models need to be developed, which will help to lower the burden of implementing hand-offs from those building a model to those deploying the model, and increasing the ubiquity of their adoption.

- **[SECure: A Social and Environmental Certificate for AI Systems](#)**

This work proposes an ESG-inspired framework combining socio-technical measures to build eco-socially responsible AI systems. The framework has four pillars: compute-efficient machine learning, federated learning, data sovereignty, and a LEEDesque certificate.

- **[Report prepared for the Santa Clara Principles for Content Moderation](#)**

The Electronic Frontier Foundation publicly called for comments on expanding the Santa Clara Principles on Transparency and Accountability (SCP). The Montreal AI Ethics Institute (MAIEI) responded to this call by drafting a set of recommendations based on insights and analysis by the MAIEI staff, supplemented by workshop contributions from the AI Ethics community.